\documentclass[reprint,superscriptaddress,amsmath,amssymb,aps,pre]{revtex4-2}
\usepackage{graphicx}
\usepackage{dcolumn}
\usepackage{bm}
\usepackage{color}
\usepackage{units}
\usepackage{wasysym}
\usepackage{MnSymbol}
\usepackage{comment}
\usepackage{times}
\usepackage{amsmath}
\usepackage{footnote}
\usepackage[unicode=true,pdfusetitle,bookmarks=false,colorlinks=true,citecolor=blue,urlcolor=blue,linkcolor=blue]{hyperref}

\begin{document}

\title{Theory of Nonequilibrium Multicomponent Coexistence}

\author{Yu-Jen Chiu}
\thanks{These authors contributed equally to this work}
\affiliation{Department of Materials Science and Engineering, University of California, Berkeley, California 94720, USA}
\author{Daniel Evans}
\thanks{These authors contributed equally to this work}
\affiliation{Department of Materials Science and Engineering, University of California, Berkeley, California 94720, USA}
\affiliation{Materials Sciences Division, Lawrence Berkeley National Laboratory, Berkeley, California 94720, USA}

\author{Ahmad K. Omar}
\email{aomar@berkeley.edu}
\affiliation{Department of Materials Science and Engineering, University of California, Berkeley, California 94720, USA}
\affiliation{Materials Sciences Division, Lawrence Berkeley National Laboratory, Berkeley, California 94720, USA}

\begin{abstract}
Multicomponent phase separation is a routine occurrence in both living and synthetic systems. 
Thermodynamics provides a straightforward path to determine the phase boundaries that characterize these transitions for systems at equilibrium.
The prevalence of phase separation in complex systems outside the confines of equilibrium motivates the need for a genuinely nonequilibrium theory of multicomponent phase coexistence. 
Here, we develop a mechanical theory for coexistence that casts coexistence criteria into the familiar form of equality of state functions. 
Our theory generalizes traditional equilibrium notions such as the species chemical potential and thermodynamic pressure to systems out of equilibrium. 
Crucially, while these notions may not be identifiable for all nonequilibrium systems, we numerically verify their existence for a variety of systems by introducing the phenomenological Multicomponent Active Model B+. 
Our work establishes an initial framework for understanding multicomponent coexistence that we hope can serve as the basis for a comprehensive theory for high-dimensional nonequilibrium phase transitions.

\end{abstract}
\maketitle
\section{Introduction}
\begin{figure*}
    \centering
    \includegraphics[width=0.8\linewidth]{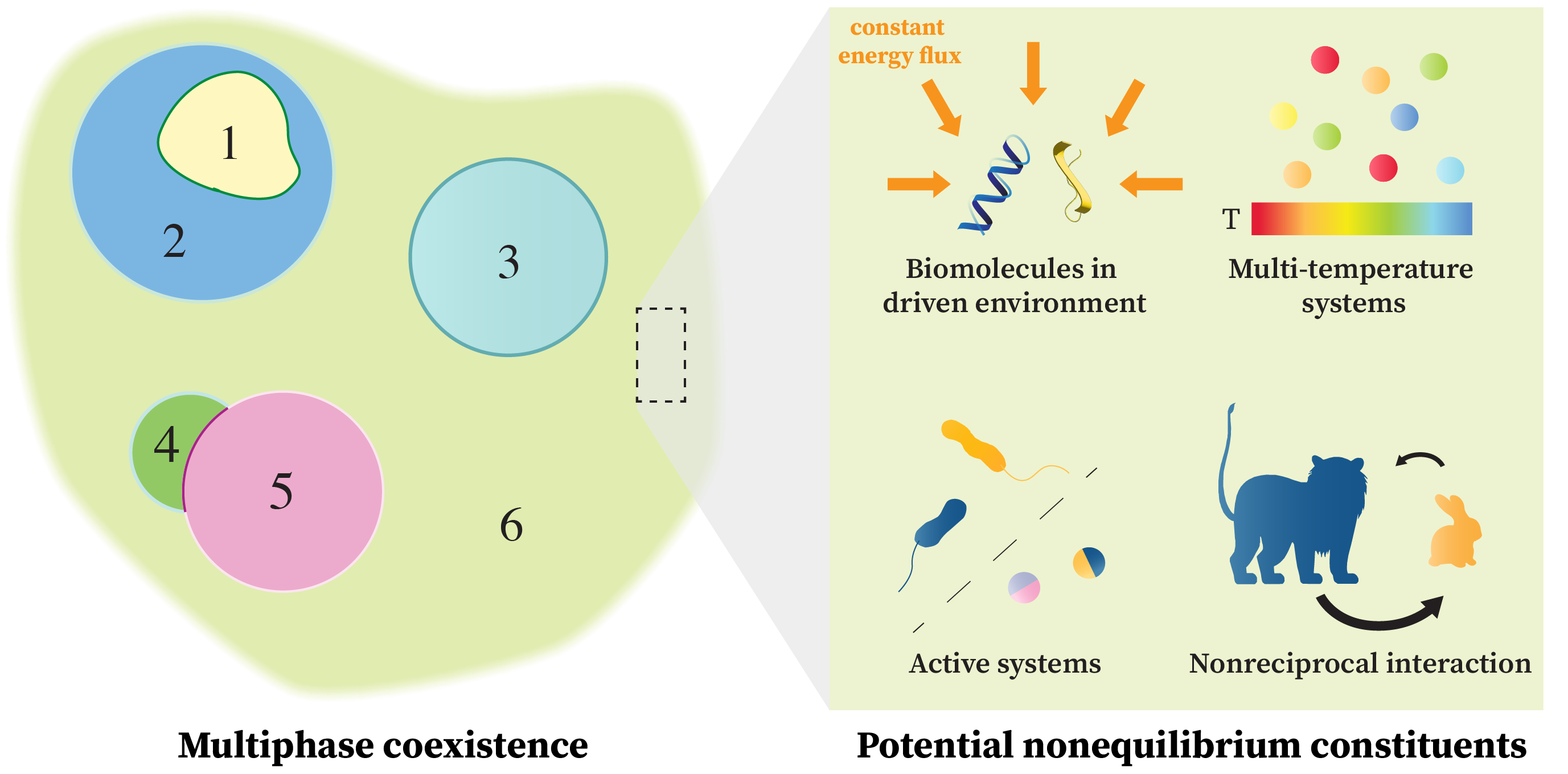}
    \caption{A schematic of multiphase coexistence (six phases shown) and possible nonequilibrium constituents comprising the phases (including biomolecules in a driven environment~\cite{Doan2024}, multi-temperature particle mixtures~\cite{Grosberg2015, Weber2016, Han2017}, active-passive particle mixtures~\cite{Angelani2011,Stenhammar2015, Wittkowski2017, Omar2019, Batton2024, Manson2024, Kreienkamp2024}, and nonreciprocally interacting systems~\cite{You2020,Fruchart2021,Chiu2023, Dinelli2023}).
    While curved interfaces are displayed, our theory assumes that the sizes of the phases are such that the radii of curvature far exceed any length scale associated with the constituents and thus the interfaces can be approximated as planar.
    }
    \label{fig:bulk_coexist}
\end{figure*}

From membrane-less organelles~\cite{Hyman2014, Banani2017, Berry2018, Lee2020} to the segregation of cell types during differentiation~\cite{Trinkaus1955, Weiss1960, Steinberg1962, Teixeira2024, Mccarthy2024}, multicomponent phase separation has been implicated as the origin of a number of processes that are essential to living, functional matter.
Theoretical treatments of these scenarios have largely borrowed from the equilibrium literature, yet it remains to be seen whether a local equilibrium is achieved in living systems that are decidedly out of equilibrium~\cite{Harris1976, Doan2024}. 
In parallel with these efforts, many studies in the field of active matter have been dedicated to generalizing thermodynamic theories of phase coexistence to systems out of equilibrium~\cite{Fily2012, Redner2013, Wittkowski2014, Takatori2015, Speck2016, Solon2018, Hermann2019, Agudo2019, Fruchart2020, Hermann2021, Omar2023, Evans2023}.
Much of these investigations have aimed to describe nonequilibrium steady states of phase coexistence: states where two (or more) macroscopic phases coexist (e.g.,~liquid-liquid coexistence) but energy is constantly consumed at the microscale [see Fig.~\ref{fig:bulk_coexist}]. 
The length scales associated with the entities that comprise the phases can differ dramatically and need not be molecular. 
Both synthetic and biological colloidal systems and even macroscopic living matter can exhibit states of coexistence that are beyond the description of equilibrium thermodynamics~\cite{Petroff2015, Natan2022, Tan2022, Kang2024}. 

While direct measurements of heterogeneous states from experiments or computer simulations can allow for the construction of phase diagrams (both in and out of equilibrium), in equilibrium, thermodynamics allows us to circumvent these approaches and construct phase diagrams by simply \textit{equating bulk equations of state}, which are properties of homogeneous states.
These coexistence criteria, famously derived by Maxwell, van der Waals, and Gibbs, are obtained variationally from the system free energy~\cite{Maxwell1875}. 
The absence of this variational principle out of equilibrium has obfuscated which (or, if any) bulk state functions are equal between coexisting nonequilibrium phases. 

For the above reasons, a genuinely nonequilibrium theory for multicomponent phase coexistence is an outstanding challenge, although much progress has been made towards a theory of \textit{single-component} coexistence. 
Recently, an entirely mechanical approach for single-component systems has been developed to describe two-phase coexistence while making no appeals to equilibrium notions~\cite{Aifantis1983, Solon2018, Omar2023}.
The coexistence criteria derived for these systems can be equivalently expressed as a weighted-area construction~\cite{Aifantis1983}, a generalized common tangent construction~\cite{Solon2018}, or a generalized Maxwell equal-area construction~\cite{Omar2023}.
This approach, however, makes use of several simplifications that are not applicable to multicomponent systems. 
While system-specific criteria have been developed for some nonequilibrium multicomponent systems~\cite{Dinelli2023, Saha2024,Greve2024}, there remains an absence of a general theoretical treatment of nonequilibrium multicomponent coexistence.

In this Article, we construct an entirely mechanical theory of phase coexistence applicable to both equilibrium and nonequilibrium systems \textit{in terms of bulk equations of state}. 
While such criteria were found in single-component nonequilibrium systems, it is not clear \textit{a priori} if equality of state functions is to be generally expected for multicomponent nonequilibrium systems. 
Indeed, our work allows us to identify systems in which: (i) the criteria are exact and take the form of equality of bulk state functions (equilibrium and select nonequilibrium systems); (ii) the criteria are approximate and take the form of a combination of equality of state functions and a generalized equal-area Maxwell construction on state functions; (iii) no coexistence criteria in terms of bulk state functions exist, exact or approximate.
The utility of our approach is demonstrated through its application to both equilibrium systems, in which we precisely recover the equilibrium coexistence criteria, and a phenomenological nonequilibrium field theory which we term Multicomponent Active Model B+ (MAMB+). 
Numerical simulations of the full spatial density profiles of a two-component MAMB+ system reveal excellent agreement with our coexistence theory. 
It is our hope that this approach can provide a practical means by which nonequilibrium phase diagrams with multiple conserved order parameters can be constructed from first principles.

\section{Mechanical Theory of Nonequilibrium Multicomponent Coexistence}
In this Section, we describe a theory for the bulk coexistence criteria of phase-separated multicomponent systems arbitrarily far from equilibrium.
Our perspective is rooted in mechanics and makes no appeals to equilibrium notions. 
We briefly review the canonical thermodynamic arguments used to derive the familiar equilibrium multicomponent coexistence criteria of equality of chemical potentials and pressure in Section~\ref{sec:eq_coex}. 
Beginning from the evolution equation of the species densities, we derive generalized coexistence criteria in Section~\ref{sec:noneq_coex} in terms of bulk mechanical equations of state. 
Importantly, while coexistence criteria resembling those of equilibrium systems may not always be identifiable within this framework, in select cases we are able to find the form of species pseudopotentials, analogous to the equilibrium chemical potentials, and a generalized Gibbs-Duhem relation that connects these species pseudopotentials to a global quantity that is analogous to the equilibrium pressure.

\subsection{Equilibrium Multicomponent Coexistence Criteria}
\label{sec:eq_coex}
Consider a macroscopic system with $n_c$ number of components, $n_p$ number of coexisting phases, and a fixed overall system volume $V$.
We define two sets: $\mathcal{C}$, containing all $n_c$ species in the system, and $\mathcal{P}$, containing all $n_p$ coexisting phases in the system.
We use Roman letters to index different species and Greek letters to index different phases. 
The order parameters that distinguish the coexisting phases are the species densities. 
We therefore define a vector of species number densities $\boldsymbol{\rho} \equiv \begin{bmatrix} \rho_{i} & \rho_{j} & \cdots & \rho_{n_c} \end{bmatrix}^{\rm T}$ with length $n_c$.
The overall number density of each species is constrained to a constant, $\boldsymbol{\rho}^{\rm sys}$, where the $i$th element is defined as $\rho_i^{\rm sys} \equiv N_{i}/V$, where $N_i$ is the total number of particles of species $i$.

In equilibrium, for a system with uniform temperature $T$ below the critical temperature $T_c$, the system can reduce its overall free energy through phase separation. 
The coexistence curve (or binodal) describes the values of the coexisting densities in each phase as a function of the global system parameters, e.g.,~$T$ and $\boldsymbol{\rho}^{\rm sys}$.
The coexisting densities can be extracted from the complete spatial species density profiles, $\boldsymbol{\rho}(\mathbf{x})$.
Thermodynamics allows us to circumvent solving for these spatially resolved density profiles and, instead, \textit{entirely determine the phase diagram from bulk equations of state.}

With $n_p$ coexisting phases and $n_c$ species densities in each phase, we have $n_cn_p$ unknown species densities. 
Using the extensivity of the free energy, we can express the total free energy density of a system as ${f = \sum_\alpha^{n_p} \phi^\alpha f^{\rm bulk}(\boldsymbol{\rho}^\alpha)}$ where ${f^{\rm bulk}(\boldsymbol{\rho})}$ is the bulk free energy density of a homogeneous system with densities $\boldsymbol{\rho}$ and $\phi^\alpha$ is the fraction of the system volume occupied by phase $\alpha$.
Here, we have neglected the interfacial free energy which is negligible for phases of macroscopic extent. 
With the free energy in hand, we can obtain the coexistence criteria by minimizing $f$ with respect to the species densities in each coexisting phase, $\boldsymbol{\rho}^{\alpha}$, and each phase fraction, $\phi^\alpha$, subject to the constraints of volume (${\sum_\alpha^{n_p} \phi^\alpha = 1}$) and particle number (${\sum_\alpha^{n_p} \boldsymbol{\rho}^\alpha \phi^\alpha= \boldsymbol{\rho}^{\rm sys}}$ ) conservation. 
This results in the familiar multicomponent equilibrium coexistence criteria:
\begin{subequations}
\label{eq:equilibrium_coex}
\begin{align}
    \boldsymbol{\mu}^{\rm bulk}(\boldsymbol{\rho}^\alpha) = \boldsymbol{\mu}^{\rm coexist} \ \forall \alpha \in \mathcal{P}, \label{eq:coex_equilibrium_chemical_potential}\\
    P^{\rm bulk}(\boldsymbol{\rho}^\alpha) = P^{\rm coexist} \ \forall \alpha \in \mathcal{P}, \label{eq:coex_equilibrium_pressure}
\end{align}
\end{subequations}
where $\boldsymbol{\mu}^{\rm bulk} \equiv \begin{bmatrix}\mu_i^{\rm bulk} & \mu_j^{\rm bulk} & \cdots & \mu_{n_c}^{\rm bulk} \end{bmatrix}^{\rm T}$ is a vector of bulk species chemical potentials with ${\mu_i^{\rm bulk} = \partial f^{\rm bulk}/\partial\rho_i}$, ${P^{\rm bulk} = -f^{\rm bulk}+ \boldsymbol{\rho}\cdot\boldsymbol{\mu}^{\rm bulk}}$ is the thermodynamic pressure, and $\boldsymbol{\mu}^{\rm coexist}$ and $P^{\rm coexist}$ are coexistence values for the chemical potential vector and pressure, respectively.
Simply stated, the chemical potential of like species in different phases must be equal and the pressure of each phase must also be equal. 
The values of these chemical potentials and pressures at coexistence will generally depend on the temperature, $T$, and overall system composition, $\boldsymbol{\rho}^{\rm sys}$.
Equations~\eqref{eq:coex_equilibrium_chemical_potential}~and~\eqref{eq:coex_equilibrium_pressure}  constitute $n_c(n_p-1)$ and $n_p-1$ equations, respectively. 
Subtracting the number of equations provided by the coexistence criteria from the number of unknown species densities results in $n_c - n_p + 1$ degrees of freedom, the well-known Gibbs phase rule (with constant temperature)~\cite{Gibbs1878}. 

Equation~\eqref{eq:equilibrium_coex} can be equivalently recast as a Maxwell equal-area construction~\cite{Maxwell1875}:
\begin{subequations}
\label{eq:eq_maxwell}
\begin{align}
    & \boldsymbol{\mu}^{\rm bulk}(\boldsymbol{\rho}^\alpha) = \boldsymbol{\mu}^{\rm coexist} \ \forall \alpha \in \mathcal{P},\\
    & \int_{\boldsymbol{\rho}^\alpha}^{\boldsymbol{\rho}^\beta}[\boldsymbol{\mu}^{\rm bulk}(\boldsymbol{\rho}) - \boldsymbol{\mu}^{\rm coexist}] \cdot d \boldsymbol{\rho}= 0 \ \forall \alpha, \beta \in \mathcal{P}, \label{eq:equilibrium_maxwell}
\end{align}
\end{subequations}
which can be straightforwardly recovered by the Gibbs-Duhem relation~\cite{Plischke1994}:
\begin{equation}
\label{eq:multi_gibbs_duhem}
    dP^{\rm bulk} = \boldsymbol{\rho} \cdot d\boldsymbol{\mu}^{\rm bulk}.
\end{equation}
Equation~\eqref{eq:eq_maxwell} no longer explicitly contains the thermodynamic pressure and solely contains the species chemical potentials. 
We note that for \textit{one-component} systems, the coexistence criteria can be expressed entirely with \textit{either} the chemical potential \textit{or} pressure as chemical potential equality can be expressed through the appropriate Maxwell construction on the pressure. 
For multicomponent systems, this symmetry is lost: there is no way to express the coexistence criteria without $n_c-1$ species chemical potentials. 
Next, we seek to determine the nonequilibrium analogs of the species chemical potentials and Gibbs-Duhem relation in order to express our coexistence criteria solely in terms of bulk equations of state.

\subsection{Nonequilibrium Multicomponent Coexistence Criteria}
\label{sec:noneq_coex}
The phase diagram, for both equilibrium and nonequilibrium systems, can always be obtained by determining (e.g., through experiment, simulation, or theoretically) the spatially resolved species density profiles, $\boldsymbol{\rho}(\mathbf{x})$, during coexistence.
We aim to develop a framework that, much like equilibrium thermodynamics, allows us to circumvent the often difficult task of obtaining the complete spatial $\boldsymbol{\rho}(\mathbf{x})$ and, instead, obtain nonequilibrium phase diagrams through a procedure that solely requires bulk equations of state. 
We take the following approach:
\begin{itemize}
    \item [1.] Develop the mechanical conditions necessary for stationary multicomponent phase coexistence.
    \item [2.] Perform a formal expansion of the terms appearing in the mechanical conditions with respect to species density gradients.
    \item [3.] Perform a series of linear operations on the expanded mechanical conditions to recover coexistence criteria in terms of bulk equations of state.
\end{itemize}
Our approach is able to reproduce the criteria derived in the previous section for equilibrium systems while also describing a large class of nonequilibrium systems.

We begin by considering the spatial and temporal dynamics of each species density field.
We aim to describe states of macroscopic phase coexistence and therefore consider a quasi-1D geometry where spatial variations only occur in the $z$-dimension with translational invariance in the other dimensions.
This choice will ensure that $z$ will be the normal direction to the interface for phase separated states and amounts to neglecting the effects of interfacial curvature.
Such an approximation is justified for macroscopic phases where the radius of curvature far exceeds any other length scale.
The evolution equations for the species densities, in the absence of chemical reactions, satisfy the continuity equation:
\begin{equation}
\label{eq:species_continuity_eq}
    \partial_t \boldsymbol{\rho} = -\partial_z\mathbf{J},
\end{equation}
where $\partial_t \equiv \partial / \partial t$, $\partial_z \equiv \partial / \partial z$, and the $i$th element of $\mathbf{J}$ is the $z$-component of the number density flux of species $i$.
The fluxes in the directions orthogonal to $z$ are neglected, consistent with our quasi-1D geometry and translational invariance in these directions.
We now require a description of the species density flux, $J_i =\rho_iv_i$ (where $v_i$ is the average velocity of species $i$ in the $z$-direction).
We can appreciate that the flux of species $i$ is directly proportional to the momentum density of species $i$, differing only by a factor of the species mass, $m_i$~\cite{Omar2023}. 
The evolution equation for the flux of each species $i$ is governed by linear \textit{species momentum balance}:
\begin{equation}
\label{eq:species_linear_momentum}
    m_i \partial_t J_i + m_i \partial_z (J_iJ_i/\rho_i) = \partial_z{\sigma}_i + b_i \ \forall i \in \mathcal{C},
\end{equation}
where $\sigma_i (z;t)$ and $b_i (z;t)$ are the stresses ($zz$-component of the full spatial stress tensor) and body forces ($z$-component of the spatial body force vector) acting on species $i$, respectively.
We emphasize that even for passive systems, the species momentum density is not a conserved quantity, and body forces (sources/sinks of momentum) are thus anticipated to be present. 
It is only the total momentum density (i.e.,~$\sum_i^{n_c} m_i J_i)$ that is conserved for passive systems (in the absence of external forces).

The coupled continuity equations and species momentum balances, along with \textit{constitutive relations} for the stresses and body forces, can in principle be used to describe the spatial and temporal evolution of the density fields in 1D.
Here, as we are interested in stationary states of phase coexistence, we can simplify the continuity equation to $\partial_z\mathbf{J} = \mathbf{0}$.
With zero-flux boundary conditions, the flux of each species must vanish everywhere, $\mathbf{J} = \mathbf{0}$, for the coexistence to be stationary: macroscopic phase coexistence is a flux-free stationary state in both equilibrium and nonequilibrium systems. 
This reduction enables us to simplify the species linear momentum balance [Eq.~\eqref{eq:species_linear_momentum}] to a static species mechanical force balance:
\begin{equation}
    \label{eq:species_linear_momentum_0}
    \mathbf{0} = \partial_z\boldsymbol{\sigma} + \mathbf{b}.
\end{equation}
Equation~\eqref{eq:species_linear_momentum_0} represents the \textit{species-level mechanical conditions} for flux-free stationary states, including macroscopic phase coexistence.

While the distinction between stresses and body forces can be essential~\cite{Omar2020}, here we define an \textit{effective body force} and restate the static mechanical balance:
\begin{equation}
\label{eq:species_linear_momentum_1}
    \mathbf{b}^{\rm eff} \equiv \partial_z\boldsymbol{\sigma} + \mathbf{b} = \mathbf{0},
\end{equation}
which are $n_c$ exact mechanical conditions for any planar coexistence scenario.
Any further analysis requires a form for $\mathbf{b}^{\rm eff}$.
In detail, the exact microscopic form can be obtained from an Irving-Kirkwood procedure~\cite{Irving1950}, or, in equilibrium, variationally using the methods of classical density functional theory~\cite{Hansen2013}.
The exact form of $\mathbf{b}^{\rm eff}$ often results in integro-differential equations, with the non-local contributions arising from, for example, long-range particle interactions and correlations.
Solving these equations for the complete spatial profiles can be an arduous challenge that we aim to circumvent. 

To make analytical progress, we first propose performing a general expansion of the effective body forces with respect to spatial gradients of the species densities. 
Such an expansion can be formally performed (as will later be done for passive systems).
Here, we retain terms up to third-order in spatial gradients and neglect even gradients as the body force must be odd with respect to spatial inversion. 
The accuracy of this expansion will depend on the degree to which the non-local terms can be approximated with local terms.
Our body force now takes the form: 
\begin{subequations}
\label{eq:body_force_expansion}
\begin{equation}
    \mathbf{b}^{\rm eff}
    \approx \mathbf{b}^{(1)} + \mathbf{b}^{(3)},
\end{equation}
where:
\begin{align}
    \mathbf{b}^{(1)} =&  \mathbf{b}^{(1,1)} \cdot\partial_z \boldsymbol{\rho}, \\
    \mathbf{b}^{(3)} =& - \mathbf{b}^{(3,1)} \vdots \partial_z\boldsymbol{\rho} \partial_z\boldsymbol{\rho} \partial_z\boldsymbol{\rho} \nonumber \\
    & - \mathbf{b}^{(3,2)} : \partial^2_{zz}\boldsymbol{\rho} \partial_z\boldsymbol{\rho} - \mathbf{b}^{(3,3)}\cdot \partial^3_{zzz}\boldsymbol{\rho},
\end{align}
\end{subequations}
with definition $\partial^2_{zz} \equiv \partial^2 / \partial z^2$ and $\partial^3_{zzz} \equiv \partial^3 / \partial z^3$.
Our proposed form of the body force now contains a number of equations of state. 
$\mathbf{b}^{(1,1)}$, $\mathbf{b}^{(3,1)}$, $\mathbf{b}^{(3,2)}$, and $\mathbf{b}^{(3,3)}$ are tensors of rank 2, 4, 3, and 2, respectively, where each element of these tensors is a state function dependent on $\boldsymbol{\mathbf{\rho}}$ but not its spatial gradients.
We further note that not every component of $\mathbf{b}^{(3,1)}$ is unique as certain anti-symmetric components will vanish upon contraction with ${\partial_z\boldsymbol{\rho} \partial_z\boldsymbol{\rho} \partial_z\boldsymbol{\rho}}$. 
This is more readily appreciated by adopting Einstein indicial notation, which makes clear that, when triple contracting into the tensor with components ${\partial_z \rho_i \partial_z \rho_j \partial_z \rho_k}$, we only retain the contributions to $b^{(3,1)}_{nijk}$ that are symmetric under the exchange of any pair of the $i$, $j$, and $k$ indices.
We define $\mathbf{b}^{(3,1)}$ to satisfy this symmetry.
In total, the expanded body force contains $2n_c^2+n_c^3+\frac{1}{6}n_c^2(n_c+1)(n_c+2)$ equations of states.
For a one-component system, only four state functions are required but this number grows quartically with the number of components.

It was demonstrated that in one-component systems, the mechanical balance always and immediately results in one coexistence criterion for both equilibrium and nonequilibrium systems: equality of pressure~\cite{Omar2023}. 
For one-component phase separation, there is a single order parameter (the density), which \textit{always} allows the $z$-component of the static momentum balance to be expressed as ${\partial_z\sigma + b = b^{\rm eff} \equiv \partial_z \Sigma = 0}$ where $\Sigma$ is the relevant component of the dynamic (or effective) stress tensor.
Simple integration of the one-component mechanical balance results in the condition of uniform dynamic stress.
The uniform density within the macroscopic coexisting phases coupled with the uniform stress condition results in the coexistence criterion of equality of the bulk contribution to the stress (i.e., pressure) between phases.

Multicomponent systems stand in contrast to one-component systems as the mechanical balance of each species \textit{generally} does not result in a condition of a spatially uniform mechanical property. 
All (true and effective) body forces will vanish within the phases where no density gradients are present, precluding the immediate identification of a bulk state function that is equal between phases. 
We seek to perform the following \textit{transformation} of the mechanical balance:
\begin{equation}
\label{eq:b_to_u_transform}
    \mathbf{b}^{\rm eff} = \mathbf{0} \rightarrow \partial_z \mathbf{u} = \mathbf{0},
\end{equation}
where we have \textit{defined} the vector of species pseudopotentials, $\mathbf{u}$.
Consistent with the effective body forces being odd with respect to spatial inversion, each species pseudopotential must be even.
Before proposing a specific transformation relating the species pseudopotentials to the effective body forces, let us consider the expanded form of $\mathbf{u}$:
\begin{subequations}
\label{eq:pseudo_pot_expansion}
\begin{equation}
    \mathbf{u} \approx \mathbf{u}^{\rm bulk} + \mathbf{u}^{\rm int},
\end{equation}
where:
\begin{equation}
    \mathbf{u}^{\rm int} = - \mathbf{u}^{(2,1)} : \partial_z\boldsymbol{\rho} \partial_z\boldsymbol{\rho}
    -  \mathbf{u}^{(2,2)} \cdot \partial^2_{zz} \boldsymbol{\rho}.
\end{equation}
\end{subequations}
Here, we explicitly delineate the bulk (i.e.,~gradient-independent) and interfacial contributions to the species pseudopotentials as $\mathbf{u}^{\rm bulk}$ and $\mathbf{u}^{\rm int}$, respectively.
$\mathbf{u}^{\rm bulk}$, $\mathbf{u}^{(2,1)}$, and $\mathbf{u}^{(2,2)}$ are tensors of rank 1, 3, and 2, respectively, where each element of these tensors is a state function dependent on $\boldsymbol{\rho}$ but not its spatial gradients. 
Not every element of $\mathbf{u}^{(2,1)}$ is unique as certain anti-symmetric components vanish upon contraction with ${\partial_z \boldsymbol{\rho} \partial_z \boldsymbol{\rho}}$.
In indicial notation, considering $u^{(2,1)}_{nij}$, only contributions symmetric under the exchange of the $i$, $j$ indices are retained when doubled contracted into the tensor with components ${\partial_z \rho_i \partial_z \rho_j}$.
We define $\mathbf{u}^{(2,1)}$ to satisfy this symmetry.
In total, the expanded species pseudopotentials contain ${n_c+n_c^2+\frac{1}{2}n_c^2(n_c+1)}$ equations of state that each generally depend on $\boldsymbol{\rho}$.

If the form of the species pseudopotentials can be identified, we immediately recover the majority of our coexistence criteria.
Integration of the steady-state condition, ${\partial_z \mathbf{u} = \mathbf{0}}$, results in the condition of spatially uniform species pseudopotentials.
The absence of density gradients in the bulk phases then allows us to identify that the bulk contribution of each species pseudopotential is equal in each of the coexisting phases:
\begin{equation}
\label{eq:pseudopotential_bulk}
    \mathbf{u}^{\rm bulk} (\boldsymbol{\rho}^\alpha) = \mathbf{u}^{\rm coexist} \ \forall \alpha \in \mathcal{P}.
\end{equation}
As we will later show, these coexistence criteria reduce to equality of species chemical potentials for systems at equilibrium. 

The central task is now to determine the precise form of the transformation symbolically proposed in Eq.~\eqref{eq:b_to_u_transform}. 
We introduce a \textit{transformation tensor}, $\boldsymbol{\mathcal{T}}$, defined such that:
\begin{equation}
\label{eq:T_intro}
    \boldsymbol{\mathcal{T}} \cdot \mathbf{b}^{\rm eff}  = \partial_z \mathbf{u}.
\end{equation}
Here, $\boldsymbol{\mathcal{T}}$ is an invertible second-rank tensor, where each of its elements is a state function. 
Substituting the expanded forms of $\mathbf{b}^{\rm eff}$ [Eq.~\eqref{eq:body_force_expansion}] and $\mathbf{u}$ [Eq.~\eqref{eq:pseudo_pot_expansion}] into Eq.~\eqref{eq:T_intro} (detailed in Appendix~\ref{ap:transformation_tensor}) results in:
\begin{subequations}
\label{eq:pseudo_sys_eq}
    \begin{align}
        & \boldsymbol{\mathcal{T}} \cdot \mathbf{b}^{(1,1)}  = \partial_{\boldsymbol{\rho}} \mathbf{u}^{\rm bulk},\label{eq:pseudo_bulk} \\
        & \boldsymbol{\mathcal{T}} \cdot \mathbf{b}^{(3,1)} = \left[ \partial_{\boldsymbol{\rho}} \mathbf{u}^{(2,1)}\right]^{\rm S},  \\
        & \boldsymbol{\mathcal{T}} \cdot \mathbf{b}^{(3,2)} = \left( \partial_{\boldsymbol{\rho}} \mathbf{u}^{(2,2)} + 2 \mathbf{u}^{(2,1)} \right), \\
        & \partial_{\boldsymbol{\rho}}(\boldsymbol{\mathcal{T}} \cdot \mathbf{b}^{(3,3)}) = \partial_{\boldsymbol{\rho}} \mathbf{u}^{(2,2)},
    \end{align}
\end{subequations}
where $\partial_{\boldsymbol{\rho}} \equiv \partial / \partial \boldsymbol{\rho}$ and $[\partial u^{(2, 1)}_{nij} / \partial \rho_k]^{\rm S} \equiv \frac{1}{6}(\partial u^{(2, 1)}_{nij} / \partial \rho_k +\partial u^{(2, 1)}_{nik} / \partial \rho_j +\partial u^{(2, 1)}_{nji} / \partial \rho_k +\partial u^{(2, 1)}_{njk} / \partial \rho_i +\partial u^{(2, 1)}_{nki} / \partial \rho_j +\partial u^{(2, 1)}_{nkj} / \partial \rho_i)$ extracts the symmetric (with respect to exchange of the $i$, $j$, and $k$ indices) portion of the fourth rank tensor $\partial_{\boldsymbol{\rho}} \mathbf{u}^{(2,1)}$ with components $\partial u^{(2, 1)}_{nij} / \partial \rho_k$.

Crucially, the system of equations presented in Eq.~\eqref{eq:pseudo_sys_eq} is overdetermined, with $n_c^2+2n_c^3+\frac{1}{6}n_c^2(n_c+1)(n_c+2)$ linear, first-order, homogeneous PDEs, and $n_c+2n_c^2+\frac{1}{2}n_c^2(n_c+1)$ unknown functions (the elements of both $\boldsymbol{\mathcal{T}}$ and $\mathbf{u}$). 
The number of equations [$\mathcal{O}(n_c^4)$] exceeds the number of unknown functions [$\mathcal{O}(n_c^3)$] and hence Eq.~\eqref{eq:pseudo_sys_eq} is overdetermined when $n_c > 1$, meaning \textit{a solution is not guaranteed to exist}.
The single-component coexistence theories~\cite{Aifantis1983, Solon2018, Omar2023} thus represent an exceptional case where the number of unknowns is precisely the number of equations.

To find solutions for both $\boldsymbol{\mathcal{T}}$ and $\mathbf{u}$, one approach may be to eliminate $\mathbf{u}$ from Eq.~\eqref{eq:pseudo_sys_eq} through a series of linear operations, obtaining a set of equations solely in terms of $\boldsymbol{\mathcal{T}}$ and the known body force coefficients.
Proceeding this way, we must ensure that certain conditions are met, namely, that the order of partial differentiation with respect to the species densities of the elements of both $\boldsymbol{\mathcal{T}}$ and $\mathbf{u}$ is inconsequential (this is more formally referred to as involutivity~\cite{Seiler2010}).
These conditions ensure that the resulting state functions, i.e.~the components of $\boldsymbol{\mathcal{T}}$ and $\mathbf{u}$, are continuous and differentiable functions of the species densities.
The general relations that emerge from stipulating these conditions are difficult to determine for an arbitrary number of components. 
Nevertheless, Eq.~\eqref{eq:pseudo_sys_eq} is the starting point for any approach to determine solutions for $\boldsymbol{\mathcal{T}}$ and $\mathbf{u}$.
Any solutions that satisfy Eq.~\eqref{eq:pseudo_sys_eq} can subsequently be checked to ensure that partial differentiation with respect to species density is indeed commutative -- we will later show that the form of the body force coefficients for passive equilibrium systems results in solutions [$\mathcal{T}_{ij} = -\rho_i^{-1}\delta_{ij}$ (where $\delta_{ij}$ is the Kronecker delta) and $u_i$ is simply the equilibrium chemical potential of species $i$] that are clearly consistent with this property.

It is important to mention that many phenomenological models for nonequilibrium mixtures \textit{take as a starting point} the existence of species pseudopotentials (sometimes defined as nonequilibrium chemical potentials) whose gradients drive the species flux~\cite{Saha2020ScalarModel}.
The present work makes clear that the existence of species-level chemical potential-like quantities is not at all guaranteed for multicomponent systems. 
This undoubtedly limits the number of microscopic systems that can be mapped to species-potential based phenomenological theories for multicomponent systems.
In the absence of finding a full-rank solution (where all $n_c$ columns/rows of $\boldsymbol{\mathcal{T}}$ are linearly independent) for $\boldsymbol{\mathcal{T}}$, the species pseudopotentials cannot be rigorously defined using the transformation proposed in this work.
Importantly, the absence of a well-defined $\mathbf{u}$ for a particular system does not necessarily signify the absence of phase coexistence scenarios.
However, if a full-rank solution of $\boldsymbol{\mathcal{T}}$ can be found, we can use Eq.~\eqref{eq:pseudo_sys_eq} to identify the form of $\mathbf{u}$ as:
\begin{subequations}
\label{eq:u_form}
    \begin{align}
        & \mathbf{u}^{\rm bulk} = \int \boldsymbol{\mathcal{T}} \cdot \mathbf{b}^{(1,1)} \cdot d\boldsymbol{\rho},\\
        & \mathbf{u}^{(2,1)} = \frac{1}{2}\left(\boldsymbol{\mathcal{T}} \cdot \mathbf{b}^{(3,2)} - \partial_{\boldsymbol{\rho}}(\boldsymbol{\mathcal{T}} \cdot \mathbf{b}^{(3,3)})\right), \\
        & \mathbf{u}^{(2,2)} = \boldsymbol{\mathcal{T}}\cdot \mathbf{b}^{(3,3)}.
    \end{align}
\end{subequations}

For systems in which the species pseudopotentials can be found, we still require additional coexistence criteria that are analogous to the equality of pressure between phases in equilibrium systems. 
To identify the remaining coexistence criteria, we make an ansatz of a generalized multicomponent Gibbs-Duhem relation, mapping the species pseudopotentials, $\mathbf{u}$, to a global quantity (i.e.,~a quantity that is not on a per-species basis), $\mathcal{G}$:
\begin{equation}
\label{eq:generalized_GD}
    \boldsymbol{\mathcal{E}} \cdot d\mathbf{u} = d\mathcal{G},
\end{equation}
where we define a generalized Maxwell construction vector, $\boldsymbol{\mathcal{E}}$, containing $n_c$ elements, each of which is a state function.
The connection of this vector to a Maxwell construction will be made evident in Eq.~\eqref{eq:G_maxwell_dE_form}.
Consistent with our second-order expansion of $\mathbf{u}$, we expand $\mathcal{G}$:
\begin{subequations}
\label{eq:G_expansion}
\begin{equation}
    \mathcal{G} \approx \mathcal{G}^{\rm bulk} + \mathcal{G}^{\rm int},
\end{equation}
where:
\begin{equation}
    \mathcal{G}^{\rm int} = - \boldsymbol{\mathcal{G}}^{(2,1)} : \partial_z\boldsymbol{\rho} \partial_z\boldsymbol{\rho}
    -  \boldsymbol{\mathcal{G}}^{(2,2)} \cdot \partial^2_{zz}\boldsymbol{\rho}.
\end{equation}
\end{subequations}
Here, we again explicitly delineate the bulk and interfacial contributions to the global quantity as $\mathcal{G}^{\rm bulk}$ and $\mathcal{G}^{\rm int}$, respectively. 
Note that $\mathcal{G}^{\rm bulk}$ is a scalar function while $\boldsymbol{\mathcal{G}}^{(2,1)}$ and $\boldsymbol{\mathcal{G}}^{(2,2)}$ are tensors of rank 2 and 1, respectively, where each element is a state function dependent on $\boldsymbol{\rho}$ but not its spatial gradient. 
Not every element of $\boldsymbol{\mathcal{G}}^{(2,1)}$ is unique as anti-symmetric components vanish upon contraction with ${\partial_z \boldsymbol{\rho} \partial_z \boldsymbol{\rho}}$. 
In indicial notation, $\mathcal{G}^{(2,1)}_{ij}$ is symmetric under the exchange of the $i$, $j$ indices, just as $u^{(2,1)}_{nij} \ \forall n \in \mathcal{C}$.
We define $\boldsymbol{\mathcal{G}}^{(2,1)}$ to only include terms that satisfy this symmetry.
In total, the expanded global quantity contains ${1+n_c+\frac{1}{2}n_c(n_c+1)}$ equations of state that each generally depend on $\boldsymbol{\rho}$.

If a global quantity can be found, we identify our remaining coexistence criteria by substituting the spatial uniformity of species pseudopotentials into the generalized Gibbs-Duhem relation [Eq.~\eqref{eq:generalized_GD}], finding $\partial_z\mathcal{G} = 0$ and consequently $\mathcal{G}$ must also be spatially constant. 
As the species densities are homogeneous within the bulk phases, we recognize that $\mathcal{G}^{\rm bulk}$ must be equal across phases:
\begin{equation}
    \mathcal{G}^{\rm bulk} \left( \boldsymbol{\rho}^{\alpha} \right) = \mathcal{G}^{\rm coexist} \ \forall \alpha \in \mathcal{P},
\end{equation}
providing our final $n_p - 1$ coexistence criteria.
In equilibrium, this set of criteria reduces to equality of pressures [Eq.~\eqref{eq:coex_equilibrium_pressure}] between phases.

To determine the elements of the Maxwell construction vector, we insert the expanded forms of both $\mathbf{u}$ [Eq.~\eqref{eq:pseudo_pot_expansion}] and $\mathcal{G}$ [Eq.~\eqref{eq:G_expansion}] into our proposed generalized Gibbs-Duhem relation [Eq.~\eqref{eq:generalized_GD}] and identify the following system of equations (see Appendix~\ref{ap:variable_vector} for details):
\begin{subequations}
\label{eq:E_system_eq}
    \begin{align}
        & \boldsymbol{\mathcal{E}} \cdot \partial_{\boldsymbol{\rho}} \mathbf{u}^{\rm bulk} = \partial_{\boldsymbol{\rho}} \mathcal{G}^{\rm bulk}, \label{eq:G_bulk}\\
        & \partial_{\boldsymbol{\rho}}(\boldsymbol{\mathcal{E}} \cdot \mathbf{u}^{(2, 2)}) = \partial_{\boldsymbol{\rho}} \boldsymbol{\mathcal{G}}^{(2,2)},  \label{eq:G22}\\
        & \boldsymbol{\mathcal{E}} \cdot \left(2 \mathbf{u}^{(2, 1)}+ \partial_{\boldsymbol{\rho}} \mathbf{u}^{(2, 2)}\right) = 2 \boldsymbol{\mathcal{G}}^{(2,1)} + \partial_{\boldsymbol{\rho}} \boldsymbol{\mathcal{G}}^{(2,2)}, \label{eq:G22_G21}\\
        & \boldsymbol{\mathcal{E}} \cdot \left[\partial_{\boldsymbol{\rho}} \mathbf{u}^{(2, 1)}\right]^{\rm S}= \left[\partial_{\boldsymbol{\rho}} \boldsymbol{\mathcal{G}}^{(2,1)}\right]^{\rm S'}, \label{eq:G21}
    \end{align}
\end{subequations}
where $\left[\partial \mathcal{G}_{i j}^{(2,1)} / \partial {\rho_k}\right]^{\rm S'} \equiv \frac{1}{6} \big( \partial \mathcal{G}_{i j}^{(2,1)} / \partial {\rho_k} + \partial \mathcal{G}_{j i}^{(2,1)} / \partial {\rho_k} + \partial \mathcal{G}_{i k}^{(2,1)} / \partial {\rho_j} + \partial \mathcal{G}_{k i}^{(2,1)} / \partial {\rho_j} + \partial \mathcal{G}_{j k}^{(2,1)} / \partial {\rho_i} + \partial \mathcal{G}_{k j}^{(2,1)} / \partial {\rho_i} \big)$ extracts the components, $\partial \mathcal{G}_{i j}^{(2,1)} / \partial {\rho_k}$, of $\partial_{\boldsymbol{\rho}} \boldsymbol{\mathcal{G}}^{(2, 1)}$ that are symmetric with respect to exchanging $i$, $j$, and $k$.
The system of PDEs in Eq.~\eqref{eq:E_system_eq} is again overdetermined, with $n_c+2n_c^2+\frac{1}{6}n_c(n_c+1)(n_c+2)$ linear, first-order, homogeneous PDEs and $1+2n_c+\frac{1}{2}n_c(n_c+1)$ unknown functions, as described in Appendix~\ref{ap:variable_vector}.
Again, just as was the case with Eq.~\eqref{eq:pseudo_sys_eq}, we have an excess of equations in comparison to our unknowns for $n_c>1$.
Solutions for $\mathcal{G}$ and $\boldsymbol{\mathcal{E}}$ that satisfy Eq.~\eqref{eq:E_system_eq} will again need to be checked to ensure that partial differentiation with respect to species densities is commutative for every state function in $\mathcal{G}$ and $\boldsymbol{\mathcal{E}}$.
As shown in the SI~\footnote{Supporting information [URL].}, for a two-component system with $n_c=2$, we can differentiate (or, formally, prolong~\cite{Seiler2010}) Eq.~\eqref{eq:E_system_eq} twice to obtain a system of PDEs where the commutativity of partial differentiation is guaranteed.
However, since this system is overdetermined, a consistent solution to all equations may not exist, including those found after differentiating Eq.~\eqref{eq:E_system_eq}.
Importantly, this means a solution for $\boldsymbol{\mathcal{E}}$ generally may not exist, even when a full-rank solution for $\boldsymbol{\mathcal{T}}$ can be found. 
If a solution of $\boldsymbol{\mathcal{E}}$ can be found, we can use Eq.~\eqref{eq:E_system_eq} to identify the form of $\mathcal{G}$ as:
\begin{subequations}
\label{eq:G_functional_form}
\begin{align}
    & \mathcal{G}^{\rm bulk} = \boldsymbol{\mathcal{E}} \cdot \mathbf{u}^{\rm bulk} - \int \mathbf{u}^{\rm bulk} \cdot d\boldsymbol{\mathcal{E}} , \\
    & \boldsymbol{\mathcal{G}}^{(2,1)} = \left(\boldsymbol{\mathcal{E}} \cdot \mathbf{u}^{(2,1)}  - \frac{1}{2} \mathbf{E}^{\rm T} \cdot \mathbf{u}^{(2,2)} \right),\\
    & \boldsymbol{\mathcal{G}}^{(2,2)} = \boldsymbol{\mathcal{E}}\cdot\mathbf{u}^{(2,2)},
\end{align}
\end{subequations}
where $\mathbf{E} \equiv \partial_{\boldsymbol{\rho}} \boldsymbol{\mathcal{E}}$.

For systems at equilibrium, $\mathbf{u}$ is simply the species chemical potentials and Eq.~\eqref{eq:generalized_GD} reduces to the equilibrium Gibbs-Duhem relation with the Maxwell construction vector as the species densities, $\boldsymbol{\mathcal{E}} = \boldsymbol{\rho}$, and the pressure as the global quantity, $\mathcal{G} = P$.
These solutions are indeed shown to be consistent with Eq.~\eqref{eq:E_system_eq} in the SI~\cite{Note1}. 
We emphasize that while the mechanical interpretation of the global quantity is clear for systems at equilibrium, this may not always be the case out of equilibrium. 

Our nonequilibrium multicomponent coexistence criteria are thus:
\begin{subequations}
\label{eq:final_coex_criteria}
\begin{align}
    & \mathbf{u}^{\rm bulk} (\boldsymbol{\rho}^\alpha) = \mathbf{u}^{\rm coexist} \ \forall \alpha \in \mathcal{P}, \label{eq:u_coex_criteria} \\
    & \mathcal{G}^{\rm bulk} (\boldsymbol{\rho}^\alpha) = \mathcal{G}^{\rm coexist} \ \forall \alpha \in \mathcal{P}, \label{eq:G_coex_criteria}
\end{align}
\end{subequations}
which can be expressed in alternate forms. 
We can recast Eq.~\eqref{eq:G_coex_criteria} as a generalized Maxwell equal-area construction with respect to $\boldsymbol{\mathcal{E}}$:
\begin{equation}
\label{eq:G_maxwell_dE_form}
\int_{\boldsymbol{\mathcal{E}}^\alpha}^{\boldsymbol{\mathcal{E}}^\beta} \left[\mathbf{u}^{\rm bulk}(\boldsymbol{\mathcal{E}}) - \mathbf{u}^{\rm coexist} \right]
    \cdot d\boldsymbol{\mathcal{E}} = 0 \ \forall \alpha, \beta \in \mathcal{P}.
\end{equation}
We can further replace $d\boldsymbol{\mathcal{E}}$ with $\mathbf{E} \cdot d\boldsymbol{\rho}$ to obtain the ``weighted-area'' construction with respect to the species densities:
\begin{equation}
    \label{eq:G_maxwell_drho_form} \int_{\boldsymbol{\rho}^\alpha}^{\boldsymbol{\rho}^\beta} \left[\mathbf{u}^{\rm bulk}(\boldsymbol{\rho}) - \mathbf{u}^{\rm coexist} \right]\cdot \mathbf{E}(\boldsymbol{\rho})\cdot d\boldsymbol{\rho} = 0 \ \forall \alpha, \beta \in \mathcal{P}.
\end{equation}

As shown in Section~\ref{sec:eq_coex}, one can derive the equilibrium coexistence criteria [Eq.~\eqref{eq:equilibrium_coex}] by minimizing a free energy subject to the constraints of particle number and volume conservation.
This results in the well-known common tangent construction.
We can similarly express our criteria as a common tangent construction on an effective bulk free energy defined as $w^{\rm bulk} \equiv \int \mathbf{u}^{\rm bulk} \cdot d\boldsymbol{\mathcal{E}}$.
The common tangent construction on $w^{\rm bulk}$ is with respect to $\boldsymbol{\mathcal{E}}$ and takes the form:
\begin{subequations}
    \label{eq:common_tangent}
    \begin{align}
        \mathbf{u}^{\rm coexist} &= \frac{\partial w^{\rm bulk}}{\partial \boldsymbol{\mathcal{E}}} \bigg |_{\boldsymbol{\rho}=\boldsymbol{\rho}^{\alpha}} \ \forall \alpha \in \mathcal{P}, \label{eq:common_tangent_u}\\
        \mathcal{G}^{\rm coexist} &= \mathbf{u}^{\rm coexist} \cdot \boldsymbol{\mathcal{E}}(\boldsymbol{\rho}^{\alpha}) - w^{\rm bulk}(\boldsymbol{\rho}^{\alpha}) \ \forall \alpha \in \mathcal{P}. \label{eq:common_tangent_G}
    \end{align}
\end{subequations}
 
\begin{figure*}
	\centering
	\includegraphics[width=.95\textwidth]{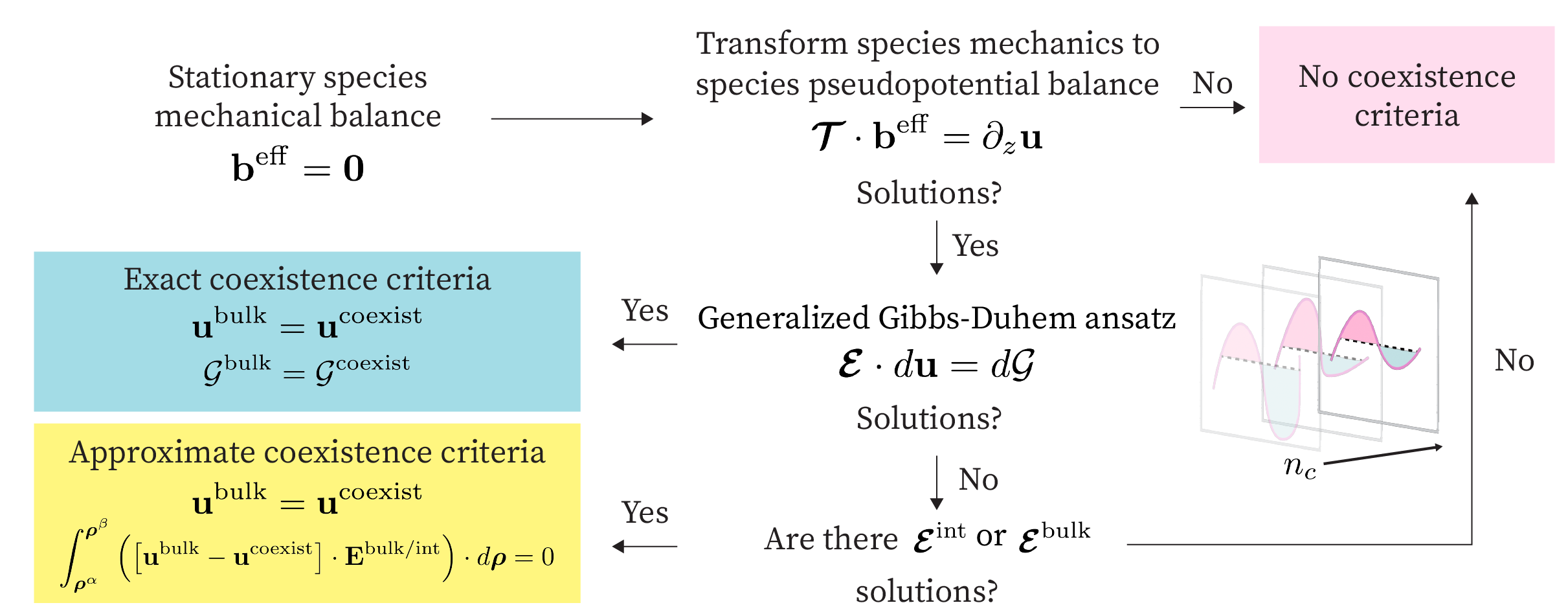}
	\caption{ \label{fig:procedure} 
 A visual procedure to obtain the coexistence criteria for multicomponent nonequilibrium systems.
 First, determine the stationary species mechanical balance [Eq.~\eqref{eq:species_linear_momentum_1}]. 
 Next, transform the effective body forces to species pseudopotentials with the transformation tensor $\boldsymbol{\mathcal{T}}$ [Eq.~\eqref{eq:T_intro}].
 Only proceed if solutions for $\boldsymbol{\mathcal{T}}$ and $\mathbf{u}$ can be found. 
 If so, identify the Maxwell construction vector, $\boldsymbol{\mathcal{E}}$, used in the generalized Gibbs-Duhem relation [Eq.~\eqref{eq:generalized_GD}].
 If solutions for $\boldsymbol{\mathcal{E}}$ and $\mathcal{G}$ can be found, we have exact coexistence criteria [Eq.~\eqref{eq:final_coex_criteria}].
 If $\boldsymbol{\mathcal{E}}^{\rm bulk} \neq \boldsymbol{\mathcal{E}}^{\rm int}$ but solutions for $\boldsymbol{\mathcal{E}}^{\rm bulk}$ and/or $\boldsymbol{\mathcal{E}}^{\rm int}$ can be found, we can approximate the coexistence criteria.
 }	
\end{figure*}

Before concluding this Section, it is important to note that there are cases in which a full solution to $\boldsymbol{\mathcal{E}}$ [from Eq.~\eqref{eq:E_system_eq}] does not exist but approximate coexistence criteria can still be formulated.
To see this, we first separately consider the bulk and interfacial contributions to the generalized Gibbs-Duhem relation and distinguish the Maxwell construction vectors that satisfy each relation with:
\begin{subequations}
\begin{align}
    & \boldsymbol{\mathcal{E}}^{\rm bulk} \cdot d\mathbf{u}^{\rm bulk} = d\mathcal{G}^{\rm bulk}, \label{eq:E_bulk}\\
    & \boldsymbol{\mathcal{E}}^{\rm int} \cdot d\mathbf{u}^{\rm int} = d\mathcal{G}^{\rm int}. \label{eq:E_int}
\end{align}
\end{subequations}
The global quantity is rigorously defined with Eq.~\eqref{eq:E_system_eq} when 
$\boldsymbol{\mathcal{E}} = \boldsymbol{\mathcal{E}}^{\rm bulk} = \boldsymbol{\mathcal{E}}^{\rm int}$.
Then, the generalized Gibbs-Duhem relation holds and equality of $\mathcal{G}^{\rm bulk}$ between coexisting phases is recovered.
We now consider two approximations to equality of $\mathcal{G}^{\rm bulk}$ [Eq.~\eqref{eq:G_coex_criteria}] when $\boldsymbol{\mathcal{E}}^{\rm bulk} \neq \boldsymbol{\mathcal{E}}^{\rm int}$, both of which rely on the existence of a $\boldsymbol{\mathcal{E}}^{\rm int}$ that satisfies Eq.~\eqref{eq:E_int}.
When this is the case, as shown in Appendix~\ref{ap:partial}, the weighted-area construction (with $\mathbf{E}=\mathbf{E}^{\rm int}$) [Eq.~\eqref{eq:G_maxwell_drho_form}] is no longer equal to the difference in $\mathcal{G}^{\rm bulk}$ between phases and hence generally depends on the selected integration path.
Notably, this construction vanishes if the integrals are performed along the path where the relationships between the species densities coincides with the full spatial solution of the coexisting densities, $\boldsymbol{\rho}^{\rm c}(z)$, which we initially set out to avoid determining.
This constitutes the first approximation for the coexistence criteria corresponding to equality of $\mathcal{G}^{\rm bulk}$: the weighted-area construction [Eq.~\eqref{eq:G_maxwell_drho_form}] with $\mathbf{E}=\mathbf{E}^{\rm int}$ allows one to approximate the coexistence densities, with the approximation gaining accuracy as the selected integration path approaches the actual coexistence profiles.

We can also introduce a second approximate criteria when $\boldsymbol{\mathcal{E}}^{\rm bulk} \neq \boldsymbol{\mathcal{E}}^{\rm int}$, now using the $\boldsymbol{\mathcal{E}}^{\rm bulk}$ that satisfies Eq.~\eqref{eq:E_bulk}. 
Defining the bulk contribution to the global quantity as $\mathcal{G}^{\rm bulk} \equiv \boldsymbol{\mathcal{E}}^{\rm bulk} \cdot \mathbf{u}^{\rm bulk} - \int \mathbf{u}^{\rm bulk} \cdot d \boldsymbol{\mathcal{E}}^{\rm bulk}$, we arrive at our second approximation: equality of this $\mathcal{G}^{\rm bulk}$ across phases [Eq.~\eqref{eq:G_coex_criteria}], or equivalently the weighted-area construction [Eq.~\eqref{eq:G_maxwell_drho_form}] with $\mathbf{E} = \mathbf{E}^{\rm bulk}$.
Equality of this $\mathcal{G}^{\rm bulk}$ across phases is not guaranteed in these scenarios, as the weighted-area construction only vanishes along the path corresponding to the coexistence profiles when using $\mathbf{E}=\mathbf{E}^{\rm int}$.
However, the weighted-area construction using $\mathbf{E}=\mathbf{E}^{\rm bulk}$ is path-independent, as its value is the difference in $\mathcal{G}^{\rm bulk}$ across the phases being considered.
This difference is generally unknown but can expressed as (detailed in Appendix~\ref{ap:partial}):
\begin{multline}
    \label{eq:deltaG}
    \Delta^{\alpha \beta} \mathcal{G}^{\rm bulk} \equiv \mathcal{G}^{\alpha} - \mathcal{G}^{\beta} \\ = \int_{\boldsymbol{\rho}^{\alpha}}^{\boldsymbol{\rho}^{\beta}} \left[ \mathbf{u}^{\rm bulk}(\boldsymbol{\rho}^{\rm c}) - \mathbf{u}^{\rm coexist} \right] \cdot \left( \mathbf{E}^{\rm bulk}(\boldsymbol{\rho}^{\rm c}) - \mathbf{E}^{\rm int}(\boldsymbol{\rho}^{\rm c}) \right) \cdot d \boldsymbol{\rho}^{\rm c},
\end{multline}
where the integrals are evaluated along the coexistence profiles, $\boldsymbol{\rho}^{\rm c}(z)$.
Clearly, the approximation of equality of $\mathcal{G}^{\rm bulk}$ across phases improves as the integrals in Eq.~\eqref{eq:deltaG} decrease.
Notably, the validity of this approximation is unalterable for a given system, as it depends on the value of the right-hand side of Eq.~\eqref{eq:deltaG}.
This contrasts with the first approximation (the weighted-area construction with $\mathbf{E} = \mathbf{E}^{\rm int}$), as its accuracy can be tuned by altering the selected integration path.
A schematic of the above framework to obtain either the exact or approximate coexistence criteria is shown in Fig.~\ref{fig:procedure}.

\section{Generalized Gibbs Phase Rule}
For systems with exact coexistence criteria [Eq.~\eqref{eq:final_coex_criteria}], it can be straightforwardly shown that the Gibbs phase rule holds. 
For a system with $n_c$ species and $n_p$ coexisting phases, we have $n_c n_p$ unknown species densities ($n_c$ in each phase) and $n_p$ unknown phase fractions.
Our generalized coexistence criteria [Eq.~\eqref{eq:final_coex_criteria}] result in a total of $(n_c+1)(n_p-1)$ equations that solely contain the $n_pn_c$ unknown densities (i.e.,~the phase fractions do not appear in these equations).
We always have an additional equation, arising from volume conservation, that solely contains the phase fractions:
\begin{subequations}
\label{eq:conservation_constraints}
\begin{equation}
\label{eq:volume_constraints}
    \sum_\alpha^{n_p} \phi^\alpha = 1,
\end{equation}
and an additional $n_c$ equations that contain both the unknown densities and phase fractions, arising from particle number conservation:
\begin{equation}
\label{eq:particle_constraints}
    \sum_\alpha^{n_p} \boldsymbol{\rho}^\alpha \phi^\alpha = \boldsymbol{\rho}^{\rm sys}.
\end{equation}
\end{subequations}

While at first glance, we have precisely the same number of equations as unknowns, we can gain insight by examining the precise dependencies of the state function criteria [Eq.~\eqref{eq:final_coex_criteria}], volume conservation equation [Eq.~\eqref{eq:volume_constraints}], and particle conservation equations [Eq.~\eqref{eq:particle_constraints}].
We begin by subtracting the number of unknown species densities from the number of coexistence criteria that solely contain the species densities. 
We denote this quantity as the ``degrees of freedom''~\footnote{Our definition differs from the often presented, ``textbook'' definition of degrees of freedom, which includes an additional degree of freedom corresponding to the system temperature. 
In principle, an arbitrary number of microscopic or external parameters can impact $\mathbf{u}$ and $\mathcal{G}$ and subsequently the phase behavior. 
Here, we avoid a subjective accounting of the dependencies of our state functions.}, $n_{\rm dof} = n_c-n_p+1$.
When $n_{\rm dof}=0$, the number of coexistence criteria is the same as the number of unknown species densities, allowing for the determination of the coexisting densities \textit{without} referring to the conservation constraints [Eq.~\eqref{eq:conservation_constraints}].
The conservation constraints can then be used to determine the remaining unknowns, the phase fractions.
This is known as the lever rule.
In these scenarios, the independence of the constraints and the coexistence criteria indicates that the coexisting densities are \textit{fixed and cannot be adjusted by altering the global constraints}.
It is in this sense that the system has zero degrees of freedom.

The picture changes when $n_{\rm dof} > 0$. 
Here, there are fewer coexistence criteria than the number of unknown densities, and the conservation constraints must be solved simultaneously with the coexistence criteria to obtain the unknown phase densities and phase fractions. 
In these scenarios, in contrast to $n_{\rm dof}=0$, the constraints on the global species densities \textit{do} impact the coexisting densities.

Finally, when $n_{\rm dof} < 0$, the number of coexistence criteria exceeds the number of unknown species densities while there are fewer conservation constraints than unknown phase fractions.
Only in pathological cases can one solve the coexistence criteria for the species densities in each of the $n_p$ phases, however, these cases can generally be accessed by tuning system-level control parameters.
While the coexisting densities can be solved for in these pathological scenarios, the conservation constraints are underspecified with respect to the phase fractions, and thus $|n_{\rm dof}|$ phase fractions can be chosen arbitrarily while satisfying our coexistence criteria.
This has the consequence that, in practice, one will always observe fewer than $n_p$ phases (with finite phase fractions) simultaneously coexist as this reduces the overall interfacial area of the system.
For example, in a single-component system, while a state of three phase coexistence can be predicted (a ``triple point''),  one of the phase fractions will always be zero to minimize the interfacial area, and two-phase coexistence will be the dominant form of coexistence that is observed.
While these surface energy arguments may appear to not be extendable to active or nonequilibrium systems, it has been recently demonstrated that many active interfaces continue to exhibit interfacial area-minimizing statistics~\cite{Langford2024}.
Indeed, systems that display macroscopic phase separation are invariably those with area-minimizing interfacial statistics!
Therefore, in practice, the maximum number of coexisting phases that will be observed is simply $n_p^{\rm max} = n_c + 1$ 

In the event that an ensemble can be constructed in which any of the species pseudopotentials or the global quantity are fixed (corresponding to the semi-grand canonical ensemble or Gibbs ensemble in equilibrium, respectively), elements of $\boldsymbol{\rho}^{\rm sys}$ are no longer fixed and the mean value of these elements must be solved for.
At most, an additional $n_c$ (the total number of elements of $\boldsymbol{\rho}^{\rm sys}$) number of unknowns can be generated through this hypothetical change of ensemble. 
If we denote the number of fixed potentials (which again consists of the species pseudopotentials and global quantity) as $n_{\rm fix}$, the total number of degrees of freedom is now $n_{\rm dof} = n_c - n_p +1 - n_{\rm fix}$.
Thus, in equilibrium, fixing the pressure for a one-component system would result in $n_{\rm dof} = 1 - n_p$, and hence it is practically impossible to observe two phase coexistence with non-zero phase fractions.
If one fixes the pressure precisely at the coexistence pressure, one will observe either of the two phases (with equal probability) but not both coexisting, as the presence of an interface is generally disfavored.

\section{Mechanics of Passive Pairwise Interacting Systems}
\label{sec:micro_passive}
All of our coexistence criteria follow from the species force densities.
To demonstrate how to derive the form of the effective body force from microscopic considerations and that our theory recovers the known equilibrium criteria, we consider multicomponent passive systems with pairwise interactions. 
The static momentum balance for these systems is obtained through an Irving-Kirkwood~\cite{Irving1950} procedure with the z-component of the effective body force acting on species $i$ taking the following \textit{exact} form~\cite{Zwanzig2001,Note1}: 
\begin{multline}
    \label{eq:beff_eq}
    b_i^{\rm eff} (z)
    = - k_BT \partial_z \rho_i (z) \\
    - \rho_i(z) \sum_{j}^{n_c} \int_{V} d \mathbf{x}' \rho_j(z') g_{ij}(z,z', \Delta x, \Delta y) F_{ij}(\Delta x, \Delta y, \Delta z).
\end{multline} 
The first term on the right-hand side arises from the single-body translational degrees of freedom of the particles. 
If expressed as a stress, this would be the ``ideal gas'' contribution to the stress.
The second term arises from the pairwise particle interactions and, in principle,  requires an integration over the entire system volume, $V$.
The difference between the position we are integrating over and the position we seek to evaluate the body force at is represented by $\Delta x = x'-x$, $\Delta y = y'-y$, and $\Delta z = z'-z$.
Here, $g_{ij}(z,z', \Delta x, \Delta y)$ is the pair distribution function which provides a measure of the likelihood of finding a particle of species $j$ located at $z'$ conditioned on a particle of species $i$ is located at $z$. 
We note that the translational invariance in the $x$ and $y$ directions results in the pair distribution function only depending on $\Delta x$ and $\Delta y$ and hence the choice of $x$ and $y$ are arbitrary. 
${F_{ij}(\Delta x, \Delta y, \Delta z)}$ is the $z$-component of the pairwise interaction force between particles $i$ and $j$ and solely depends on the relative positions.
The integral in Eq.~\eqref{eq:beff_eq} thus represents the $z$-component of all interaction forces exerted on particles of species $i$ located at $z$ by all particles of species $j$.

The effective body force above is exact for pairwise interacting passive particles. 
Our task is now to formally expand this expression up to third-order in spatial density gradients and explicitly identify the coefficients [see Eq.~\eqref{eq:body_force_expansion}] for this class of systems.
We can then use the machinery developed in the previous Section to extract the multicomponent coexistence criteria and confirm that they are indeed identical to those obtained from bulk thermodynamics. 
The expansion of Eq.~\eqref{eq:beff_eq}, formally performed in the SI~\cite{Lindell1993,Note1}, involves a gradient expansion of the non-local densities [i.e.,~$\rho_j(z')$ appearing in Eq.~\eqref{eq:beff_eq}] and approximates the pair distribution functions of inhomogeneous systems using the pair distribution functions of homogeneous systems.
The latter approximation is likely a source of error in our higher order expansion coefficients as there is no widely accepted microscopic theory for the pair distribution function of inhomogeneous fluids~\cite{Lurie2014}. 
We note that we truncate the expansion (with respect to species densities) of the inhomogeneous pair distribution function at first-order, whereas we truncate the spatial expansions of the non-local species densities at third-order.
Using our expansion, we obtain explicit forms for the coefficients of each $b^{\rm eff}_i$ in a passive pairwise interacting system:
\begin{subequations}
\label{eq:body_force_coeffs_eq}
\begin{align}
    & b_{ij}^{(1,1)} = -k_BT\delta_{ij} \nonumber\\
    & - \rho_i \sum_k^{n_c}\int d\mathbf{x}' \Delta z \left(F_{ij}g^{0}_{ij} + \frac{1}{2} \sum_k^{n_c} F_{ik}\rho_k \frac{\partial g_{ik}^0}{\partial\rho_j}\right),\\
    & b_{ijkl}^{(3,1)} \approx 0, \label{eq:bijkl_def}\\
    & b_{ijk}^{(3,2)} \approx \rho_i \int d\mathbf{x}' \frac{1}{4} (\Delta z)^3 \left(F_{ij} \frac{\partial g_{ij}^0}{\partial \rho_k} + F_{ik} \frac{\partial g_{ik}^0}{\partial \rho_j}\right),\\
    & b_{ij}^{(3,3)} \approx \rho_i \int d\mathbf{x}' \frac{1}{6} (\Delta z)^3 \left(F_{ij} g_{ij}^0 +\frac{1}{2} \sum_k^{n_c} F_{ik}\rho_k \frac{\partial g_{ik}^0}{\partial\rho_j}\right),
\end{align}
\end{subequations}
where $\rho_i \equiv \rho_i (z)$, $F_{ij} \equiv F_{ij} (\Delta x, \Delta y, \Delta z)$, and $g_{ij}^0 \equiv g_{ij}^0 (\Delta x,\Delta y,\Delta z)$ is the pair distribution function of species $i$ and $j$ when the species densities are spatially uniform with values $\boldsymbol{\rho}(z)$.
We again emphasize that the nature of our approximation for our inhomogeneous pair distribution function likely introduces error in the third-order coefficients above. 

The coefficients in Eq.~\eqref{eq:body_force_coeffs_eq} can be evaluated provided expressions for the pairwise interaction forces and a theory for the pair distribution function of homogeneous systems. 
While the precise values of these coefficients can vary dramatically for different systems, we recognize that \textit{all passive systems} must have precisely the same coexistence criteria (see Section~\ref{sec:eq_coex}). 
These criteria must emerge from the specific \textit{structure} of these coefficients.
The nature of this structure can be best identified by noting that the body force coefficients are entirely consistent with the following effective body forces:
\begin{subequations}
\label{eq:variational_b}
    \begin{equation}
        \label{eq:body_force_mu_def}
        b_i^{\rm eff} = - \rho_i \partial_z \mu_i,
    \end{equation}
    where:
    \begin{equation}
    \label{eq:equil_chemical_potential_kappa}
        \mu_i = \mu_i^{\rm bulk} - \sum_{j}^{n_c} \sum_{k}^{n_c}\frac{1}{2} \frac{\partial \kappa_{ij}}{\partial \rho_{k}}\partial_z \rho_j \partial_z \rho_k - \sum_{j}^{n_c}  \kappa_{ij} \partial^2_{zz}\rho_j.
    \end{equation}
    This second-order expansion of the chemical potential is obtained from $\mu_i = \delta \mathcal{F} / \delta \rho_i$ with:
    \begin{equation}
        \label{eq:variational_b_free_energy}
        \mathcal{F}[\boldsymbol{\rho}(z)] = \int_{V} d\mathbf{x}' \left( f^{\rm bulk} + \frac{1}{2} \sum_{i}^{n_c} \sum_{j}^{n_c} \kappa_{ij} \partial_z \rho_i \partial_z \rho_j \right).
    \end{equation}
\end{subequations}
Here, $\kappa_{i j}$ is symmetric with respect to the exchange of indices $i$ and $j$ (as any antisymmetric contributions are inconsequential).
Furthermore, derivatives of the interfacial coefficient ${\partial\kappa_{ij}/\partial\rho_k}$ are symmetric under exchange of any pairs of the $i$, $j$, and $k$ indices~\cite{Hansen2013}. 
This variational form of the body force [Eq.~\eqref{eq:body_force_mu_def}] is consistent with the perspective of linear irreversible thermodynamics which posits that the direct force acting on particles of species $i$ is simply $-\partial_z \mu_i$~\cite{DeGroot2013}.
Moreover, this form of the body force is entirely consistent with obtaining the thermodynamic pressure $P$ upon summing all body forces with ${\sum_i^{n_c}b_i^{\rm eff} = -\partial_z P}$ in accordance with both mechanics and the equilibrium Gibbs-Duhem relation~\cite{Note1}.

Comparison of Eqs.~\eqref{eq:body_force_coeffs_eq} and \eqref{eq:variational_b} allows us to identify the following relations:
\begin{subequations}
\label{eq:kappa_relation}
    \begin{align}
         & \frac{\partial\mu_i^{\rm bulk}}{\partial\rho_j} = \frac{k_BT}{\rho_i }\delta_{ij} \nonumber\\
         & + \sum_{k}^{n_c} \int d\mathbf{x}' (\Delta z) \left( F_{ij}g_{ij}^0 +\frac{1}{2} F_{ik}\rho_k\frac{\partial g_{ik}^0}{\partial \rho_j} \right), \label{eq:hessian_expression}\\
         & \kappa_{ij} = - \int d\mathbf{x}' (\Delta z)^3 \frac{1}{6}  \left( F_{ij} g_{ij}^0 +\frac{1}{2}\sum_{k}^{n_c} F_{ik}\rho_k\frac{\partial g_{ik}^0}{\partial \rho_j} \right), \label{eq:kappa_expression}\\
        & \frac{\partial\kappa_{ij}}{\partial\rho_k} = - \int d\mathbf{x}'(\Delta z)^3  \frac{1}{8}\left( F_{ij}\frac{\partial g_{ij}^0}{\partial \rho_k}+ F_{ik}\frac{\partial g_{ik}^0}{\partial \rho_j} \right). 
    \end{align}
\end{subequations}
We can compactly express the form of the body force coefficients of passive pairwise interacting systems as:
\begin{subequations}
    \begin{align}
        & b_{ij}^{(1,1)} = -\rho_i \frac{\partial\mu_i^{\rm bulk}}{\partial\rho_j},\\
        & b_{ijkl}^{(3,1)} = - \frac{1}{2} \rho_i \frac{\partial^2\kappa_{ij}}{\partial\rho_k\partial\rho_l}, \label{eq:bijkl_def2}\\
        & b_{ijk}^{(3,2)} = -2 \rho_i \frac{\partial\kappa_{ij}}{\partial\rho_k},\\
        & b_{ij}^{(3,3)} = -\rho_i\kappa_{ij}.
    \end{align}
\end{subequations}
Interestingly, comparison of Eqs.~\eqref{eq:bijkl_def} and \eqref{eq:bijkl_def2} indicates every component of the Hessian of every $\kappa_{i j}$ must vanish.
Truncating our expansion of the inhomogeneous pair distribution functions at first-order in derivatives with respect to the species densities appears to limit our theory to $\kappa_{i j}$ coefficients that are constant or linear in the species densities.

While our theory cannot describe nonlinear $\kappa_{i j}$ coefficients, the nature of our approximation does not appear to have impacted Eq.~\eqref{eq:hessian_expression}.
We recognize that $\partial\mu_i^{\rm bulk}/\partial\rho_j$ comprises the Hessian matrix of the bulk free energy density with elements $H_{ij}^{\rm bulk} = \partial\mu_i^{\rm bulk}/\partial\rho_j = \partial^2 f^{\rm bulk}/\partial \rho_i \partial \rho_j $.
Equation~\eqref{eq:hessian_expression} thus offers \textit{microscopic expressions} for bulk thermodynamic quantities. 
Intriguingly, our microscopic expressions for the Hessian (and $\kappa_{ij}$) directly involve the pairwise forces in addition to the structure, which is encoded through $g^0_{ij}$. 
This is in contrast to the more familiar microscopic expressions for these quantities, which simply include other measures of structure.
In particular, the Hessian of the free energy can be obtained by examining the long wavelength behavior of elements of the multicomponent static structure factor tensor~\cite{Kirkwood1951, Ben1974, Cheng2022}.
In the SI~\cite{Anderson2020,Note1}, we verify the equivalency of the established expressions for the Hessian of the bulk free energy with our force-based expression [Eq.~\eqref{eq:hessian_expression}].
Additionally, we numerically verify that our (approximate) microscopic expression for $\kappa_{ij}$ is indeed symmetric upon exchange of $i$ and $j$ for a model passive system.

With the structure of the body force coefficients identified, we are now poised to demonstrate the recovery of the equilibrium coexistence criteria. 
The effective body force is nearly in gradient form [see Eq.~\eqref{eq:variational_b}] except for a factor of density outside the gradient, hence we immediately identify every element of the transformation tensor as $\mathcal{T}_{in} = - \delta_{in} (\rho_n)^{-1}$ and the species pseudopotentials as the chemical potentials with $u_i = \mu_i$.
Substituting this into our identified system of PDEs [Eq.~\eqref{eq:pseudo_sys_eq}], we indeed find that this satisfies our equations. 
We can further confirm that our body force coefficients in Eq.~\eqref{eq:body_force_coeffs_eq}, which recover the equilibrium chemical potentials, also recover the thermodynamic pressure as the global quantity.
We indeed find that $\boldsymbol{\mathcal{E}} \sim \boldsymbol{\rho}$ satisfies Eq.~\eqref{eq:E_system_eq} and our global quantity is thus entirely consistent with the thermodynamic pressure and the equilibrium Gibbs-Duhem relation, $dP=\boldsymbol{\rho} \cdot d \boldsymbol{\mu}$.
We can identify the forms of $\mathcal{G}^{\rm bulk}$ ($P^{\rm bulk}$) and $\mathcal{G}^{\rm int}$ ($P^{\rm int}$) as:
\begin{subequations}
\begin{align}
    & P^{\rm bulk} = \sum_i^{n_c} \rho_i\mu_i^{\rm bulk} - \int \mu_i^{\rm bulk} d \rho_i, \\
    & P^{\rm int} = -\sum_i^{n_c} \sum_j^{n_c}  \Bigg[ \sum_k^{n_c}\frac{1}{2}\left(\rho_i \frac{\partial\kappa_{ij}}{\partial\rho_k}-\kappa_{ij}\delta_{ik}\right) \partial_z \rho_j \partial_z \rho_k \nonumber\\
    & + \rho_i \kappa_{ij} \partial_{zz}^2 \rho_j \Bigg ].
\end{align}
\end{subequations}
As shown in the SI, the interfacial contribution to the pressure is a generalization of the one-component (and 1D) interfacial Korteweg stress~\cite{Korteweg1904, Yang1976, Note1}.
We have also confirmed in the SI that our microscopic expression for the bulk thermodynamic pressure is indeed consistent with the generalization of the one-component virial theorem~\cite{Chandler1987, Hansen2013}.

\section{Nonequilibrium Model Application}
\label{sec:Nonequilibrium_Model}
To demonstrate the effectiveness of our theory, we introduce simple phenomenological nonequilibrium field theories as a testing platform.
We can then compare the phase diagrams obtained by numerically solving for the complete spatial profiles of the species densities with those more simply obtained from our coexistence criteria.

Phenomenological continuum models for single-component density fields, such as Active Model B+ (AMB+)~\cite{Wittkowski2014, Tjhung2018, Cates2019}, and multicomponent systems, such as the Nonreciprocal Cahn-Hilliard (NRCH) model~\cite{Saha2020ScalarModel,You2020}, have been introduced to explore nonequilibrium nucleation, kinetics, and phase behavior. 
Here, we introduce Multicomponent Active Model B+ (MAMB+), which combines the more general structure of the interfacial terms in AMB+ (in comparison to the simpler Active Model B~\cite{Wittkowski2014, Cates2019}) with the multicomponent interactions of the NRCH model.
In one dimension (this can be straightforwardly generalized), MAMB+ has the following dynamics for each of the $n_c$ conserved scalar order-parameter fields, $\rho_i(z,t)$, at position $z$ and time $t$~\cite{Note1}:
\begin{subequations}
\label{eq:MAMB+_definition}
\begin{align}
    & \partial_t \rho_i  = -\partial_z J_i, \label{eq:mamb+_drhoidt}\\
    & J_i = \sum_j^{n_c} M_{i j} b^{\rm eff}_j, \label{eq:mamb+_ji}\\
    & b^{\rm eff}_j = -\rho_j\Bigg(\partial_z u_j^0  + \sum_k^{n_c} \eta_{jk}^{\rm N} \partial_z\rho_k - \sum_k^{n_c} \sum_n^{n_c} \xi_{j k n}^{{\rm A}}\partial^2_{zz}\rho_k \partial_z \rho_n \nonumber\\
    & - \sum_k^{n_c} \sum_n^{n_c} \sum_{m}^{n_c}\theta_{j k n m} \partial_z \rho_k \partial_z \rho_n \partial_z \rho_m \Bigg), \label{eq:mamb+_bj}\\
    & u_j^0 = \frac{\delta \mathcal{F}}{\delta \rho_j} + \tau_j^{\rm N} - \sum_k^{n_c} \sum_n^{n_c} \lambda_{jkn}\partial_z\rho_k \partial_z\rho_n - \sum_k^{n_c} \pi_{j k}^{{\rm A'}} \partial^2_{zz} \rho_k, \label{eq:mamb+_uj}
\end{align}
\end{subequations}
where $J_i$ is, again, the density flux of species $i$ in the $z$-direction.
Here, we have proposed that $J_i$ follows a linear transport relation in which a flux of species $i$ is directly generated by the body force on species $i$ and \textit{indirectly} driven by the body forces on every other species. 
These fluxes and forces are coupled through a positive definite mobility tensor with elements $M_{i j}$.
In addition to being Markovian, this form ensures that a force-free (all $b_i^{\rm eff} = 0$) stationary state (all $J_i = 0$) is reached.

The proposed form of the body force consists of four distinct contributions. 
The first term resembles that of passive systems (recall $b_j^{\rm eff} = -\rho_j\partial_z\mu_j$ for systems at equilibrium) with a key distinction that the defined species pseudopotentials cannot be obtained variationally.
The next three terms are ``non-gradient'' that prevent the simple identification of the species pseudopotentials. 
Each $\eta_{jk}^{\rm N}$ captures first-order contributions to $b_j^{\rm eff}$ that cannot be integrated into the bulk portion of $u_j^0$.
The next two non-gradient contributions are third-order, the first coming from each $\xi_{jkn}^{{\rm A}}$ where $\xi_{jkn}^{{\rm A}} = - \xi_{jnk}^{{\rm A}}$.
The final non-gradient contribution comes from each $\theta_{jknm}$, which is symmetric under the exchange of the $k$, $n$, and $m$ indices.

In the absence of the non-integrable body force terms, $u_j^0$ acts as the species pseudopotential which consists of four contributions. 
The first is variational, the functional derivative of a free energy $\mathcal{F}$ [see Eq.~\eqref{eq:variational_b_free_energy}], which again is simply the chemical potential [see Eq.~\eqref{eq:equil_chemical_potential_kappa}] of a passive system ($\mu_j \equiv \delta \mathcal{F} / \delta \rho_j$).
The next three contributions are non-variational.
$\tau_j^{\rm N}$ captures contributions to $u_j^0$ that cannot be integrated into $f^{\rm bulk}$ while the next two terms capture second-order non-variational effects.
For systems at equilibrium, the coefficients on the Laplacian and square-gradient terms are always related as both emerge from the functional differentiation of the free energy.
Here, each $\lambda_{jkn}$ (which is symmetric under the exchange of $k$ and $n$) breaks this relation.
Finally, each $\pi_{j k}^{{\rm A'}}$ captures broken symmetries (with respect to exchanging $j$ and $k$ and permuting the $j$, $k$, and $n$ indices of the derivative $\partial \pi_{j k}^{{\rm A'}} / \partial \rho_n$) that cannot be achieved at equilibrium.

In the SI~\cite{Note1}, we demonstrate that the proposed form of MAMB+ captures all possible first and third-order gradient contributions to $b_j^{\rm eff}$, and all possible bulk and second-order contributions to $u_j^0$, while preserving clear limits in which gradient and passive dynamics are recovered.
MAMB+ reduces to AMB+~\cite{Tjhung2018, Cates2019} when there is a single-component ($n_c=1$).
Introducing multiple chemical species provides additional avenues to achieve nonequilibrium systems in MAMB+.
In AMB+, there is one parameter (analogous to our $\xi$) that breaks the gradient structure of $b^{\rm eff}$, and one parameter (analogous to our $\lambda$) that breaks the variational structure of the would-be species pseudopotential, $u^0$.
For a single-component system in one spatial dimension and when $\xi$ is a constant, $\lambda$ can be redefined to include $\xi$ such that the effective body force is always in gradient form.
When $n_c>1$ we can break the passive structure of the effective body forces and species pseudopotentials in more ways.
Counting the number of unique, nonzero $\eta_{jk}^{\rm N}$, $\xi_{jkn}^{{\rm A}}$, and $\theta_{jknm}$, we see that we have $\frac{n_c^4}{6} + n_c^3 + \frac{5}{6} n_c^2 - n_c$ terms that can break the gradient structure of the effective body forces.
Performing the same counting for $\tau_j^{\rm N}$, $\lambda_{jkn}$, and $\pi_{jk}^{\rm A'}$, we see that there are $\frac{1}{2}n_c^3 + n_c^2 + \frac{1}{2} n_c - 1$ terms that can break the variational structure of the species pseudopotentials.
Additionally, contrasting the one-component case, MAMB+ is distinct from MAMB (defined as dynamics with every $\eta_{jk}^{\rm N}$, $\xi_{jkn}^{{\rm A}}$, and $\theta_{jknm}$ set to zero) in one spatial dimension with constant nonequilibrium coefficients if any $\xi_{jkn}^{{\rm A}}$ is nonzero.

\begin{figure*}
	\centering
	\includegraphics[width=.95\textwidth]{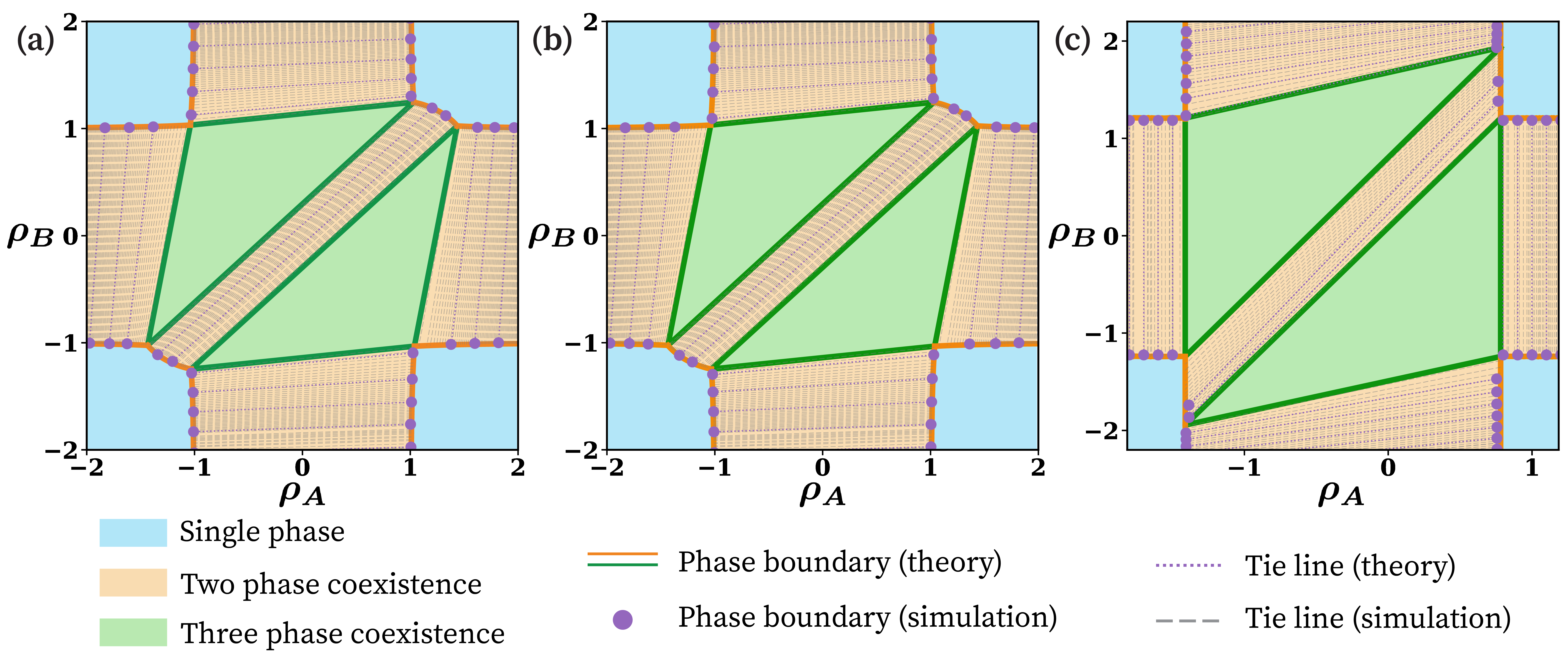}
	\caption{ \label{fig:phase_diagrams} 
    Theoretical and numerical phase diagrams of the considered MAMB models.
    NRCH models with (a) $\overline{\kappa}_{AB} (\chi - \alpha) = \overline{\kappa}_{BA} (\chi + \alpha)$, where $\boldsymbol{\mathcal{E}}^{\rm bulk}=\boldsymbol{\mathcal{E}}^{\rm int}$ and (b) $\overline{\kappa}_{AB} (\chi - \alpha) \neq \overline{\kappa}_{BA} (\chi + \alpha)$, where $\boldsymbol{\mathcal{E}}^{\rm bulk} \neq \boldsymbol{\mathcal{E}}^{\rm int}$. 
    For both (a) and (b) we set $a=-1$, $\chi = -0.5$, $\alpha = -0.2$, and $\overline{\kappa}_{AA} = \overline{\kappa}_{BB} = 1/(6\pi)$, with  $\overline{\kappa}_{AB} = \overline{\kappa}_{BA} = 0$ for (a) and $\overline{\kappa}_{AB} = \overline{\kappa}_{BA} = 0.01/(6\pi)$ for (b).
    (c) A MAMB system which results in an $\boldsymbol{\mathcal{E}}$ with nonlinear elements, achieved by introducing $\lambda_{AAA}=0.01$, $\chi = - \alpha = -1$ with $a= -1.5$, $\overline{\kappa}_{AA} = \overline{\kappa}_{BB} = 1/8\pi$, and $\overline{\kappa}_{AB} = \overline{\kappa}_{BA} = \lambda_{AAB} = \lambda_{ABB} = \lambda_{ABA} = 0$.
 }
\end{figure*}

For gradient dynamics (denoted here as MAMB), every $\eta_{jk}^{\rm N}$, $\xi_{jkn}^{{\rm A}}$, and $\theta_{jknm}$ are zero such that $\mathcal{T}_{i j} = -\rho_i^{-1} \delta_{i j}$ and $u_j=u_j^0$ solve Eq.~\eqref{eq:pseudo_sys_eq} with $J_i = -\sum_j^{n_c} M_{i j}\rho_j \partial_z u_j$.
Moreover, for passive systems, every $\tau_j^{\rm N}$, $\lambda_{j k n}$, and $\pi_{jk}^{\rm A'}$ are zero, meaning the species pseudopotential is the chemical potential, $u^0_j = \delta \mathcal{F} / \delta \rho_j$.
In this limit, MAMB becomes Multicomponent Model B (MMB).
Making any component of these nonequilibrium tensors nonzero breaks the passive structure, disallowing the definition of a global free energy.

For simplicity, we validate our theory using MAMB, where we can tune each $\tau_j^{\rm N}$, $\lambda_{j k n}$, and $\pi_{jk}^{\rm A'}$ and only need to find $\boldsymbol{\mathcal{E}}$ as the transformation tensor takes its passive form.
We adopt the following simple functional forms:
\begin{subequations}
    \begin{align}
        & f^{\rm bulk} = \sum_i^{n_c} \left( \frac{a_i}{2}\rho_i^2+\frac{b_i}{4}\rho_i^4 + \frac{1}{2} \sum_{j}^{n_c} \chi_{i j}^{\rm S} \rho_i \rho_j \right), \\
        & \tau_i^{\rm N} = \sum_{j}^{n_c} \chi^{\rm A}_{i j} \rho_j,
    \end{align}
\end{subequations}
where $f^{\rm bulk}$ appears in the free energy functional [see Eq.~\eqref{eq:variational_b_free_energy}], and $\chi_{ij}^{\rm S}$ and $\chi_{i j}^{\rm A}$ are symmetric and antisymmetric, respectively, upon exchange of $i$ and $j$.
We further assume that every $a_i$, $b_i$, $\chi_{ij}$, $\chi_{i j}^{\rm A}$, $\lambda_{jkn}$,  $\kappa_{ij}$, and $\pi_{ij}^{\rm A'}$ are constant. 
It proves convenient to define $\overline{\kappa}_{ij} \equiv \kappa_{ij} + \pi_{ij}^{\rm A'}$.
Furthermore, we now focus on two-component systems with the two species denoted as $A$ and $B$.
To further simplify, we set $a_A=a_B=a$, $b_A=b_B=1$, $\chi_{AA}^{\rm S} = \chi_{BB}^{\rm S} = 0$, $\chi_{AB}^{\rm S} = \chi_{BA}^{\rm S} = \chi$, $\chi_{AB}^{\rm A} = -\chi_{BA}^{\rm A} = \alpha$, and $\lambda_{B k n} = 0 \ \forall k, n \in \{A, B\}$.

We derive $\boldsymbol{\mathcal{E}}$ for the considered systems using the system of PDEs in Eq.~\eqref{eq:E_system_eq}, augmented with the derived compatibility conditions for a two-component system, as detailed in the SI~\cite{Note1}.
First, we examine the coexistence criteria of the NRCH model, a subclass of MAMB where every $\lambda_{jkn}$ is zero.
Coexistence criteria have been recently proposed for this model when the parameters are constrained such that $\overline{\kappa}_{AB} (\chi - \alpha) = \overline{\kappa}_{BA} (\chi + \alpha)$~\cite{Saha2024,Greve2024}. 
Interestingly, Frohoff \textit{et al.} found that this choice of parameters allows the dynamics to be mapped to ``spurious'' gradient dynamics~\cite{Frohoff2023}.
Here, we look to recover the corresponding coexistence criteria using the systematic formalism of our theory while extending the criteria to cases where $\overline{\kappa}_{AB} (\chi - \alpha) \neq \overline{\kappa}_{BA} (\chi + \alpha)$.

Using the procedure outlined in Section~\ref{sec:noneq_coex}, $\mathcal{G}$ cannot be identified for arbitrary values of the remaining parameters.
However, we can identify $\boldsymbol{\mathcal{E}}^{\rm bulk}$ and $\boldsymbol{\mathcal{E}}^{\rm int}$ (detailed in Appendix~\ref{ap:partial}):
\begin{subequations}
    \begin{align}
    & \boldsymbol{\mathcal{E}}^{\rm bulk} = \begin{bmatrix}
        \rho_A & \frac{\chi+\alpha}{\chi-\alpha}\rho_B
    \end{bmatrix}^{\rm T},\\
    & \boldsymbol{\mathcal{E}}^{\rm int} = \begin{bmatrix}
        \rho_A & \frac{\overline{\kappa}_{AB}}{\overline{\kappa}_{BA}}\rho_B
    \end{bmatrix}^{\rm T}.
\end{align}
\end{subequations}
If the coupling parameter ratio is satisfied, $\overline{\kappa}_{AB} (\chi-\alpha) = \overline{\kappa}_{BA} (\chi+\alpha)$, $\boldsymbol{\mathcal{E}}^{\rm bulk}= \boldsymbol{\mathcal{E}}^{\rm int}$ and Eq.~\eqref{eq:final_coex_criteria} is the exact coexistence criteria.
It is in this restricted case that the coexistence criteria found using our theory reduce to those of Refs.~\cite{Saha2024,Greve2024} and the system has spurious gradient dynamics~\cite{Greve2024,Frohoff2023}. 
If the coupling parameter ratio is not satisfied, $\overline{\kappa}_{AB} (\chi-\alpha) \neq \overline{\kappa}_{BA} (\chi+\alpha)$, the set of criteria corresponding to equality of $\mathcal{G}^{\rm bulk}$ [Eq.~\eqref{eq:G_coex_criteria}] can be approximated by defining $\mathcal{G}^{\rm bulk} \equiv \boldsymbol{\mathcal{E}}^{\rm bulk} \cdot \mathbf{u}^{\rm bulk} - \int \mathbf{u}^{\rm bulk} \cdot d \boldsymbol{\mathcal{E}}^{\rm bulk}$ and equating it across phases.

We now compare the phase diagrams predicted using our theory to those obtained from numerically determining the complete spatial profiles, $\rho_A^{\rm c}(z)$ and $\rho_B^{\rm c}(z)$.
To obtain these profiles, we use the numerical scheme outlined in Refs.~\cite{Thiele2019,Frohoff2021,Uecker2021,Note1}.
Figure~\ref{fig:phase_diagrams}(a) and (b) displays the numerically determined coexistence densities for the NRCH model alongside those predicted using our theory.
In Fig.~\ref{fig:phase_diagrams}(a), we find $\boldsymbol{\mathcal{E}}^{\rm bulk} = \boldsymbol{\mathcal{E}}^{\rm int}$ and hence we construct the predicted phase diagram using the exact criteria [Eq.~\eqref{eq:final_coex_criteria}].
In Fig.~\ref{fig:phase_diagrams}(b), $\boldsymbol{\mathcal{E}}^{\rm bulk} \neq \boldsymbol{\mathcal{E}}^{\rm int}$ and we must approximate the criterion in Eq.~\eqref{eq:G_coex_criteria} by equating $\mathcal{G}^{\rm bulk}$ (defined in terms of $\boldsymbol{\mathcal{E}}^{\rm bulk}$) across phases.
We find excellent agreement between our theory and simulations, reproducing the two phase and three phase coexistence boundaries quantitatively for all parameters considered, including when $\overline{\kappa}_{AB} (\chi - \alpha) \neq \overline{\kappa}_{BA} (\chi + \alpha)$. 
The theory for nonequilibrium multicomponent coexistence in Section~\ref{sec:noneq_coex} is able to precisely determine the phase diagram solely using the derived state functions and coexistence criteria.

We can understand the high accuracy of the approximate criterion used in Fig.~\ref{fig:phase_diagrams}(b) by computing $\Delta^{\alpha \beta} \mathcal{G}^{\rm bulk}$ [Eq.~\eqref{eq:deltaG}]:
\begin{multline}
    \Delta^{\alpha \beta} \mathcal{G}^{\rm bulk} \\ = \left( \frac{\chi + \alpha}{\chi - \alpha} - \frac{\overline{\kappa}_{AB}}{\overline{\kappa}_{BA}} \right) \int_{\rho_B^{\alpha}}^{\rho_B^{\beta}} \left[ u^{\rm bulk}_B (\boldsymbol{\rho}^{\rm c}) - u^{\rm coexist}_B \right] d \rho_B^{\rm c}
\end{multline}
As shown in the SI~\cite{Note1}, we find $\Delta^{\alpha \beta} \mathcal{G}^{\rm bulk} \approx 0$ and hence equating $\mathcal{G}^{\rm bulk}$ across phases works well as approximate criteria, as evidenced by the high accuracy of the predicted phase diagram in Fig.~\ref{fig:phase_diagrams}(b).

To further explore the robustness of our theory, we now explore systems in which the elements of the Maxwell construction vector contain nonlinear terms, contrasting with passive systems and systems following NRCH dynamics. 
We find that this is achieved by adjusting the model parameters to the following: $\chi = -\alpha$, $\lambda_{AAA} \neq 0$, $\overline{\kappa}_{AA} = \overline{\kappa}_{BB} \neq 0 $, and the remaining elements of  $\overline{\kappa}_{jk}$ $\lambda_{jkn}$ set to zero.
For these dynamics, we find the following generalized Maxwell construction vector:
\begin{equation}
    \boldsymbol{\mathcal{E}} = \begin{bmatrix}
        \exp\left(\frac{2\lambda_{AAA}}{\overline{\kappa}_{AA}}\rho_A\right) & 0
    \end{bmatrix}^{\rm T},
\end{equation}
which allows us to fully define $\mathcal{G}$. 
Using the coexistence criteria derived from this Maxwell construction vector, we construct the phase diagram for this system using the exact criteria [Eq.~\eqref{eq:final_coex_criteria}], shown in Fig.~\ref{fig:phase_diagrams}(c).
The agreement between the theoretically predicted phase boundaries and those obtained numerically is once again excellent.

Notably, the phase diagrams corresponding to the examined NRCH systems [Figs.~\ref{fig:phase_diagrams}(a) and (b)] have an inversion symmetry with respect to the two density fields ($\boldsymbol{\rho} \rightarrow -\boldsymbol{\rho})$. 
This symmetry is embedded in the species body forces corresponding to the NRCH model which is also invariant under $\boldsymbol{\rho} \rightarrow -\boldsymbol{\rho}$.
This inversion symmetry with respect to the density fields is broken for the phase diagram of the MAMB system considered in Fig.~\ref{fig:phase_diagrams}(c) and is indeed consistent with the considered MAMB body force which does not impose this symmetry on the species density fields.

\section{Discussion and Conclusion}
We have presented a theory for multicomponent coexistence rooted in mechanics that is applicable to both equilibrium and nonequilibrium systems. 
This theory outlines a procedure for converting the coupled species static mechanical balances into equality of bulk state functions that we term species ``pseudopotentials''. 
At equilibrium, we have verified that these state functions are simply the species chemical potentials.
More generally, these species pseudopotentials can then be used to determine additional criteria in the form of either a generalized Maxwell construction or equality of a global quantity which becomes the pressure at equilibrium. 
Equivalently, the coexistence criteria can also be expressed as a generalized common tangent construction on a suitably defined bulk free energy.
Crucially, while all coexisting phases at equilibrium satisfy equality of species chemical potentials and pressure regardless of the system details, this is not the case out of equilibrium. 
The functional forms of the species pseudopotentials and the global quantity directly follow from the details of the species mechanics, which, in general, cannot be determined variationally for nonequilibrium systems. 
To validate that our theory indeed reproduces the phase boundaries of nonequilibrium systems, we introduce the phenomenological Multicomponent Active Model B+ (MAMB+).
By explicitly determining the full spatial profiles for several model systems (including some under the umbrella of the Nonreciprocal Cahn-Hilliard model), we numerically determined the phase boundaries and found excellent quantitative agreement with our theory.

Our mechanical theory rests on the determination of mechanical equations of state in the form of the effective body force coefficients. 
In passive systems, when the coexistence criteria are \textit{always} equality of the species chemical potentials and pressure, a number of experimental and computational techniques have been developed to obtain the functional form of these equations of state.
Experimental determination of the activity coefficients that characterize the non-ideal contributions to the chemical potential has a rich and long history for multicomponent fluids~\cite{Sandler2017}. 
Moreover, several particle-based simulation techniques exist for probing these equations of state~\cite{Frenkel2001,Chipot2007}.
The success of the synergy between theory, experiment, and computation in our understanding of multicomponent \textit{equilibrium} phase behavior is evident in a number of disparate contexts.
The development of theoretical, computational, and experimental approaches for determining the mechanical body force coefficients will be needed if a similar level of success is to be found \textit{out of equilibrium}  

There are a number of theoretical tools for determining the body force coefficients required in our theory for nonequilibrium multicomponent coexistence, including the formal expansion of the exact form of the body force as determined through the Irving-Kirkwood procedure.
Whether these theories can be validated or informed through experimental or computational investigations will be important for the practical implementation of this coexistence theory.
One route to determining the coefficients may come through dynamical measurements.
To that end, an indirect method to experimentally or computationally determine these mechanical coefficients may come through measuring the \textit{flux} generated by sufficiently small values of $\mathbf{b}^{\rm eff}$. 
By keeping the force small such that the functional form of $\mathbf{b}^{\rm eff}$ is not altered, we can propose the following constitutive relation between the species fluxes and effective body forces:
\begin{subequations}
    \begin{equation}
        \mathbf{J} = \mathbf{M} \cdot \mathbf{b}^{\rm eff} + \mathbf{M}^{(3)} \vdots \mathbf{b}^{\rm eff} \mathbf{b}^{\rm eff} \mathbf{b}^{\rm eff} + \mathcal{O} \left( \left[ \mathbf{b}^{\rm eff} \right]^5 \right),
    \end{equation}
    where we again take the fluxes and effective body forces to occur along the $z$-direction.
    Here, $\mathbf{M}$ is a positive-definite rank 2 tensor coupling the species fluxes and body forces, and $\mathbf{M}^{(3)}$ is the analogous rank 4 tensor capturing the effects of nonlinear forces.
    The retention of the first nonlinear transport coefficient is necessary in order to capture all terms up to third-order in spatial gradients of the species densities upon substitution of our body force expansion:
\begin{multline}
    \label{eq:experimental_constitutive}
    \mathbf{J} = \mathbf{M} \cdot \mathbf{b}^{(1, 1)} \cdot \partial_z \boldsymbol{\rho} - \mathbf{M} \cdot \mathbf{b}^{(3,3)}\cdot \partial^3_{zzz}\boldsymbol{\rho} - \mathbf{M} \cdot \mathbf{b}^{(3,2)} \partial^2_{zz}\boldsymbol{\rho} \partial_{z}\boldsymbol{\rho} \\ - \left( \mathbf{M} \cdot \mathbf{b}^{(3,1)} - \mathbf{M}^{(3)} \vdots \mathbf{b}^{(1, 1)} \mathbf{b}^{(1, 1)} \mathbf{b}^{(1, 1)} \right) \vdots \partial_z\boldsymbol{\rho} \partial_z\boldsymbol{\rho} \partial_z\boldsymbol{\rho} \\ + \mathcal{O}\left( [\partial_z \boldsymbol{\rho}]^5 \right).
\end{multline}
\end{subequations}
Notably, the mutual diffusion coefficient tensor can be identified as $\mathbf{D} = -\mathbf{M} \cdot \mathbf{b}^{(1, 1)}$.
While the flux expression introduces additional unknowns that are not needed for our coexistence theory in the form of the mobility tensors, we nevertheless anticipate that the information encoded in the dynamical measurements will be useful in determining the required body force coefficients.
One potential avenue to determine the coefficients in Eq.~\eqref{eq:experimental_constitutive} is to perhaps leverage recent machine learning approaches for learning hydrodynamic equations of motion~\cite{Joshi2022, Supekar2023, Mandal2024}.

There remain a number of theoretical avenues to pursue. 
While our theory outlines a procedure for the determination of phase boundaries solely in terms of bulk equations of state, \textit{solutions} for the species pseudopotentials and global quantity are not guaranteed.
In the event that these functions cannot be determined analytically or numerically, the coexistence criteria cannot be cast as equality of state functions between phases.
In some limiting cases when the species pseudopotentials can be found but the global quantity cannot be identified, approximate criteria can still be found in the form of a generalized Maxwell construction.
Within our framework, there may be nonequilibrium systems that display macroscopic coexistence but with phase boundaries that cannot be determined from bulk equations of state alone, as shown in the SI~\cite{Note1}. 
Our theory does not yield bulk criteria that apply to every nonequilibrium coexistence scenario, and instead should be interpreted as a framework, applicable to every nonequilibrium system, that allows one to determine if bulk coexistence criteria exist and, if so, what they are.
Understanding the nature of the class of nonequilibrium systems (perhaps using the structure of MAMB+) that elude our framework is the subject of ongoing work. 

Finally, we note this work focused on the species densities as the order parameters characterizing coexistence, however the framework is more generally applicable to \textit{conserved} order parameters. 
One can use our approach after deriving the appropriate forces that must balance in a flux-free steady-steady for the conserved order parameters of interest, which for species densities, are simply the effective body forces that govern the species linear momentum balance.
In a separate work~\cite{Evans2023}, we develop the analogous theory for a single conserved quantity coupled to any number of \textit{nonconserved} order parameters.
Combining those results with the present works enables the determination of coexistence criteria for systems described by any number of conserved and nonconserved order parameters and allows us to describe, for example, multicomponent crystallization~\cite{Chiu2023} and nematic mixtures~\cite{Skamrahl2023,Bhattacharyya2024}.
  
\begin{acknowledgments}
We thank Eric Weiner for helpful discussions and feedback on this manuscript. 
Y.-J.C. acknowledges partial support by the UC Berkeley College of Engineering Jane Lewis Fellowship.
D.E. acknowledges partial support by the U.S. Department of Defense through the National Defense Science and Engineering Graduate Fellowship Program.
\end{acknowledgments}

\appendix

\section{Derivation of Criteria for Multicomponent Coexistence}
\subsection{Transformation Tensor}
\label{ap:transformation_tensor}
We look to derive the transformation tensor, $\boldsymbol{\mathcal{T}}$, that can convert the per-species effective body forces into per-species pseudopotentials. 
We first explicitly write out the summations over the species for the expansion of $\mathbf{b}^{\rm eff}$ [see Eq.~\eqref{eq:body_force_expansion}]:
\begin{subequations}
\label{apeq:body_force_expansion}
\begin{equation}
    b^{\rm eff}_i \left( z\right)
    \approx b^{(1)}_i + b^{(3)}_i,
\end{equation}
where:
\begin{align}
    b^{(1)}_i & = \sum_j^{n_c} b_{ij}^{(1,1)} \partial_z \rho_j, \\
    b^{(3)}_i & = - \sum_j^{n_c}\sum_k^{n_c}\sum_l^{n_c} b_{ijkl}^{(3,1)}\partial_z \rho_j \partial_z \rho_k \partial_z \rho_l \nonumber \\
    & - \sum_j^{n_c} \sum_k^{n_c} b_{ijk}^{(3,2)} \partial^2_{zz} \rho_j \partial_z \rho_k - \sum_j^{n_c} b_{ij}^{(3,3)}\partial^3_{zzz} \rho_j.
\end{align}
Notably, only the contributions to $b_{ijkl}^{(3,1)}$ that are symmetric upon exchange of the $j$, $k$, and $l$ indices will remain after triple contraction into a symmetric rank 3 tensor with components $\partial_z \rho_j \partial_z \rho_k \partial_z \rho_l$.
We define $\mathbf{b}^{(3, 1)}$ to satisfy this symmetry.
\end{subequations}
Proceeding similarly for $\mathbf{u}$ [see Eq.~\eqref{eq:pseudo_pot_expansion}]:
\begin{subequations}
\label{eq:pseudo_pot_expansion_ap}
\begin{equation}
    u_n \approx u_n^{\rm bulk} + u_n^{\rm int},
\end{equation}
where:
\begin{equation}
    u_n^{\rm int} = -\sum_j^{n_c} \sum_k^{n_c} u_{njk}^{(2,1)} \partial_z \rho_j \partial_z \rho_k 
    - \sum_j^{n_c} u_{nj}^{(2,2)} \partial^2_{zz} \rho_j.
\end{equation}
\end{subequations}
Here, we note that only the symmetric contributions to $u_{njk}^{(2,1)}$ upon exchange of $j$ and $k$ will contribute to the pseudopotential of species $n$, as these indices are contracted into a symmetric matrix with components $\partial_z\rho_j\partial_z\rho_k$. 
While the contraction with a symmetric tensor will ensure that any antisymmetric contributions to $u_{njk}^{(2,1)}$ will vanish, we nevertheless define $\mathbf{u}^{(2, 1)}$ to satisfy this symmetry in order to reduce the number of unknown coefficients. 

The transformation tensor, $\boldsymbol{\mathcal{T}}$, and species pseudopotentials, $\mathbf{u}$, are defined to satisfy the relationship: 
\begin{equation}
\label{eq:body_force_transformation}
    \sum_{i}^{n_c} \mathcal{T}_{ni} b^{\rm eff}_i= \partial_z u_n.
\end{equation}
We now aim to identify when this ansatz holds, and if so, what the components of $\boldsymbol{\mathcal{T}}$ and $\mathbf{u}$ are.

Differentiating $u_n$ in the $z$-direction we have:
\begin{multline}
    \label{eq:psi_zderive}
    \partial_z u_n = \sum_j^{n_c}\frac{\partial u_n^{\rm bulk}}{\partial \rho_j} \partial_z \rho_j
    - \sum_j^{n_c}\sum_k^{n_c}\sum_l^{n_c} \frac{\partial u_{njk}^{(2,1)}}{\partial \rho_l} \partial_z \rho_j \partial_z \rho_k \partial_z \rho_l\\
    - \sum_j^{n_c}\sum_k^{n_c} \left( \frac{\partial u_{nj}^{(2,2)}}{\partial \rho_k} + 2 u_{njk}^{(2,1)} \right) \partial^2_{zz} \rho_j \partial_z \rho_k\\
    - \sum_j^{n_c} u_{nj}^{(2,2)} \partial^3_{zzz} \rho_j,
\end{multline}
where we again note that only the contributions to $\partial u_{njk}^{(2,1)} / \partial \rho_l$ that are symmetric upon the exchange of indices $j$, $k$ and $l$ will contribute to $\partial_z u_n$.
Substituting the forms of $b^{\rm eff}_i$ and $\partial_z u_n$ into Eq.~\eqref{eq:body_force_transformation}, we identify the following relations that must hold for $\boldsymbol{\mathcal{T}}$ and $\mathbf{u}$ to satisfy Eq.~\eqref{eq:body_force_transformation}:
\begin{subequations}
\label{eq:T_system_relation}
    \begin{align}
        & \sum_{i}^{n_c} \mathcal{T}_{ni} b_{ij}^{(1,1)} = \frac{\partial u_n^{\rm bulk}}{\partial \rho_j} \ \forall n, j \in \mathcal{C}, \label{eq:t_bulk_relation}\\
        & \sum_{i}^{n_c} \mathcal{T}_{ni} b_{ijkl} ^{(3,1)} = \left[ \frac{\partial u_{njk}^{(2,1)}}{\partial \rho_l}\right]^{{\rm S}_{jkl}} \ \forall n, j, k, l \in \mathcal{C}, \label{eq:t_int_relation_1} \\
        & \sum_{i}^{n_c} \mathcal{T}_{ni} b_{ijk}^{(3,2)} = \frac{\partial u_{nj}^{(2,2)}}{\partial \rho_k} + 2 u_{njk}^{(2,1)} \ \forall n, j, k \in \mathcal{C}, \label{eq:t_int_relation_2}\\
        & \sum_{i}^{n_c} \frac{\partial}{\partial\rho_k} \left(\mathcal{T}_{ni}  b_{ij}^{(3,3)} \right)= \frac{\partial u_{nj}^{(2,2)}}{\partial\rho_k}  \ \forall n, j, k \in \mathcal{C},\label{eq:t_int_relation_3}
    \end{align}
\end{subequations}
where the superscript ${\rm S}_{jkl}$ denotes the symmetry of permuting the indices $j$, $k$, $l$. 
For example, $[A_{ijkl}]^{{\rm S}_{jkl}} \equiv \frac{1}{6} (A_{ijkl} + A_{ijlk} + A_{ikjl} + A_{iklj} + A_{iljk} + A_{ilkj})$.
From the system of partial different equations presented in Eq.~\eqref{eq:T_system_relation}, we see that there are $n_c^2+2n_c^3+\frac{1}{6}n_c^2(n_c+1)(n_c+2)$ equations for $n_c+2n_c^2+\frac{1}{2}n_c^2(n_c+1)$ variables (elements of both the transformation tensor and species pseudopotentials) and hence this system of PDEs is overdetermined.

We look to combine Eq.~\eqref{eq:T_system_relation} to find equations for the elements $\mathcal{T}_{ni}$ solely in terms of the known body force coefficients, i.e.,~eliminating the species pseudopotential coefficients from Eq.~\eqref{eq:T_system_relation}.
However, one must ensure the commutativity of partial differentiation with respect to the species densities, something that is generally not guaranteed.
Given forms for the coefficients in the effective body forces, one can generally differentiate Eq.~\eqref{eq:T_system_relation} to obtain compatibility conditions (equations that must be satisfied to ensure the commutativity of partial differentiation).
This process, known as completion to involution~\cite{Seiler2010}, depends on the form of the effective body forces and the number of species, and hence we cannot perform it generally.
Here, we only demonstrate how to eliminate the species pseudopotential coefficients from the equations in Eq.~\eqref{eq:T_system_relation}, noting that any solution to the following combined equations can be subsequently checked to see if partial differentiation is commutative.

We first examine Eq.~\eqref{eq:t_bulk_relation}, where differentiation with respect to $\rho_k$ results in $\partial^2  u_n^{\rm bulk} / \partial \rho_k\partial \rho_j$ which must be symmetric with respect to exchange of indices $j$ and $k$.
This symmetry allows us to identify the first relation that $\mathcal{T}_{ni}$ must satisfy: 
\begin{multline}
\label{eq:bulkTFrelationship}
    \sum_i^{n_c} \Bigg[ \frac{\partial \mathcal{T}_{ni}}{\partial \rho_k} b_{ij}^{(1,1)} - \frac{\partial \mathcal{T}_{ni}}{\partial \rho_j} b_{ik}^{(1,1)}\\ 
    + \mathcal{T}_{ni} \frac{\partial b_{ij}^{(1,1)}}{\partial \rho_k} - \mathcal{T}_{ni} \frac{\partial b_{ik}^{(1,1)}}{\partial \rho_j} \Bigg]
    = 0 \ \forall n, j, k \in \mathcal{C}.
\end{multline}
Next, we note $u_{njk}^{(2, 1)} = u_{nkj}^{(2, 1)} \ \forall n, j, k \in \mathcal{C}$ and hence:
\begin{multline}
\label{eq:intsymTFrelationship}
    \sum_{i}^{n_c} \bigg[ \mathcal{T}_{ni} b_{ijk}^{(3,2)} - \mathcal{T}_{ni} b_{ikj}^{(3,2)} - \frac{\partial \mathcal{T}_{ni}}{\partial \rho_k} b_{ij}^{(3, 3)} + \frac{\partial \mathcal{T}_{ni}}{\partial \rho_j} b_{ik}^{(3, 3)} \\ - \mathcal{T}_{ni} \frac{\partial b_{ij}^{(3, 3)}}{\partial \rho_k} + \mathcal{T}_{ni} \frac{\partial b_{ik}^{(3, 3)}}{\partial \rho_j} \bigg] = 0 \ \forall n, j, k \in \mathcal{C}.
\end{multline}
Lastly, we find an additional condition by differentiating Eqs.~\eqref{eq:t_int_relation_2} and \eqref{eq:t_int_relation_3} and subsequently combining the resulting relations with Eq.~\eqref{eq:t_int_relation_1} to obtain:
\begin{multline}
\label{eq:intTFrelationship}
    \sum_i^{n_c} \Bigg[ 6\mathcal{T}_{ni} b_{ijkl}^{(3,1)} - \frac{\partial}{\partial\rho_l}(\mathcal{T}_{ni} b_{ijk}^{(3,2)}) - \frac{\partial}{\partial\rho_k}(\mathcal{T}_{ni} b_{ijl}^{(3,2)}) \\
    - \frac{\partial}{\partial\rho_j}(\mathcal{T}_{ni} b_{ilk}^{(3,2)}) + 2\frac{\partial^2 }{\partial\rho_k\partial\rho_l}(\mathcal{T}_{ni} b_{ij}^{(3,3)}) \\ + \frac{\partial^2 }{\partial\rho_k\partial\rho_j}(\mathcal{T}_{ni} b_{i \ell}^{(3,3)}) \Bigg] = 0 \ \forall n, j, k, l \in \mathcal{C}.
\end{multline}
We emphasize that the system of partial differential equations in Eqs.~\eqref{eq:bulkTFrelationship}, \eqref{eq:intsymTFrelationship}, and \eqref{eq:intTFrelationship} does not contain all of the compatibility conditions necessary to ensure the commutativity of partial differentiation.
As an example of the process used to ensure this commutativity, we perform the procedure for the system of PDEs for $\boldsymbol{\mathcal{E}}$ and $\mathcal{G}$ (described in the next Section) when $n_c=2$ in the SI~\cite{Note1}.

\subsection{Maxwell Construction Vector}
\label{ap:variable_vector}
We look to derive the Maxwell construction vector $\boldsymbol{\mathcal{E}}$ that defines the generalized multicomponent Gibbs-Duhem relation. 
Explicitly writing the summations over the species densities, the expansion of the global quantity, $\mathcal{G}$, reads: 
\begin{subequations}
\label{eq:G_expansion_ap}
\begin{equation}
    \mathcal{G} \approx \mathcal{G}^{\rm bulk} + \mathcal{G}^{\rm int},
\end{equation}
where:
\begin{equation}
    \mathcal{G}^{\rm int} = -\sum_j^{n_c} \sum_k^{n_c} \mathcal{G}_{jk}^{(2,1)}\partial_z \rho_j \partial_z \rho_k
    - \sum_j^{n_c} \mathcal{G}_{j}^{(2,2)} \partial^2_{zz} \rho_j.
\end{equation}
\end{subequations}
Here, we again note that only contributions to $\mathcal{G}_{jk}^{(2,1)}$ that are symmetric upon the exchange of the $j$ and $k$ indices will remain after contraction into the symmetric matrix with components $\partial_z \rho_j \partial_z \rho_k$.

The Maxwell construction vector and the global quantity are defined to satisfy the generalized multicomponent Gibbs-Duhem relation:
\begin{equation}
\label{eq:generalized_GD_ap}
    \sum_i^{n_c} \mathcal{E}_i du_i = d\mathcal{G}.
\end{equation}
The explicit expansion of the species pseudopotentials, $\mathbf{u}$, can be found in Eq.~\eqref{eq:pseudo_pot_expansion_ap}.
We now aim to identify when the generalized Gibbs-Duhem relation holds, and when the relation does hold, what $\mathcal{G}$ and the components of $\boldsymbol{\mathcal{E}}$ are.

Differentiating $\mathcal{G}$ in the $z$-direction we have:
\begin{multline}
    \label{eq:G_zderive}
    \partial_z \mathcal{G} = \sum_j^{n_c}\frac{\partial \mathcal{G}^{\rm bulk}}{\partial \rho_j} \partial_z \rho_j
    - \sum_j^{n_c}\sum_k^{n_c}\sum_l^{n_c} \frac{\partial \mathcal{G}_{jk}^{(2,1)}}{\partial \rho_l} \partial_z \rho_j \partial_z \rho_k \partial_z \rho_l\\
    - \sum_j^{n_c}\sum_k^{n_c} \left( \frac{\partial \mathcal{G}_{j}^{(2,2)}}{\partial \rho_k} + 2 \mathcal{G}_{jk}^{(2,1)} \right) \partial^2_{zz} \rho_j \partial_z \rho_k\\
    - \sum_j^{n_c} \mathcal{G}_{j}^{(2,2)} \partial^3_{zzz} \rho_j,
\end{multline}
where we again note that only the contributions to $\partial \mathcal{G}_{jk}^{(2,1)}/\partial \rho_l$ that are symmetric upon the exchange of the indices $j$, $k$ and $l$ will remain after contraction into the symmetric matrix with components $\partial_z \rho_j \partial_z \rho_k \partial_z \rho_l$.
Substituting $\partial_z u_i$ [Eq.~\eqref{eq:psi_zderive}] and $\partial_z \mathcal{G}$ [Eq.~\eqref{eq:G_zderive}] into Eq.~\eqref{eq:generalized_GD_ap}, we identify the following relations that must be satisfied for Eq.~\eqref{eq:generalized_GD_ap} to hold:
\begin{subequations}
\label{eq:E_relations_appendix}
    \begin{align}
        &\sum_i^{n_c} \mathcal{E}_i \frac{\partial u_i^{\rm bulk}}{\partial\rho_j} = \frac{\partial \mathcal{G}^{\rm bulk}}{\partial \rho_j} \ \forall j \in \mathcal{C}, \label{eq:E_bulk_relation} \\
        &\sum_i^{n_c}\frac{\partial}{\partial\rho_k}(\mathcal{E}_i u_{ij}^{(2, 2)}) = \frac{\partial \mathcal{G}_{j}^{(2,2)}}{\partial\rho_k} \ \forall j, k \in \mathcal{C}, \label{eq:E_int_relation1} \\
        &\sum_i^{n_c} \mathcal{E}_i \left( 2 u_{ijk}^{(2, 1)} +\frac{\partial u_{ij}^{(2, 2)}}{\partial \rho_k}\right) = 2 \mathcal{G}_{jk}^{(2,1)} + \frac{\partial \mathcal{G}_{j}^{(2,2)}}{\partial\rho_k} \ \forall j, k \in \mathcal{C}, \label{eq:E_int_relation2} \\
        &\sum_i^{n_c} \mathcal{E}_i \left[\frac{\partial u_{ijk}^{(2, 1)}}{\partial\rho_l} \right]^{{\rm S}_{jkl}}= \left[\frac{\partial \mathcal{G}_{jk}^{(2,1)}}{\partial \rho_l}\right]^{{\rm S}_{jkl}} \ \forall j, k, l \in \mathcal{C}. \label{eq:E_int_relation3}
    \end{align}
\end{subequations}
From the system of PDEs in Eq.~\eqref{eq:E_relations_appendix}, we see that there are ${n_c+2n_c^2+\frac{1}{6}n_c(n_c+1)(n_c+2)}$ equations for ${1+2n_c+\frac{1}{2}n_c(n_c+1)}$ variables (elements of both the Maxwell construction vector and the global quantity) and hence this system of PDEs is overdetermined.

We look to express Eq.~\eqref{eq:E_relations_appendix} solely in terms of the known coefficients of the species pseudopotentials, i.e.~eliminating the coefficients in $\mathcal{G}$ from the equations by differentiating and combining the relations presented in Eq.~\eqref{eq:E_relations_appendix}.
As was the case for $\boldsymbol{\mathcal{T}}$, an $\boldsymbol{\mathcal{E}}$ that satisfies this system of PDEs is not guaranteed to respect the commutativity of partial differentiation.
Here, we demonstrate how to eliminate the coefficients in $\mathcal{G}$ from the equations in Eq.~\eqref{eq:E_relations_appendix} for any number of species.
Solutions to these equations can subsequently be checked to see if they respect the commutativity of partial differentiation.
In the SI~\cite{Note1}, we explicitly differentiate (prolong) the system of PDEs in Eq.~\eqref{eq:E_relations_appendix} when $n_c=2$ to obtain the compatibility conditions that ensure partial differentiation with respect to the species densities is commutative (involutivity).

We first identify the relationships required to determine the bulk contribution to the Maxwell construction vector from Eq.~\eqref{eq:E_bulk_relation}.
Notably, the Hessian of $\mathcal{G}^{\rm bulk}$ must be symmetric, $\partial^2 \mathcal{G}^{\rm bulk} / \partial \rho_j \partial \rho_k = \partial^2 \mathcal{G}^{\rm bulk} / \partial \rho_k \partial \rho_j$, yielding the relationship:
\begin{equation}
\label{eq:E_bulk_appendix}
    \sum_i^{n_c}\Bigg[E_{ij} \frac{\partial u_i^{\rm bulk}}{\partial\rho_k} - E_{ik} \frac{\partial u_i^{\rm bulk}}{\partial\rho_j} \Bigg]= 0 \ \forall j, k \in \mathcal{C},
\end{equation}
where $E_{ij} \equiv \partial \mathcal{E}_i / \partial\rho_j$.
Additionally, $\mathcal{G}_{jk}^{(2, 1)}$ must be symmetric with respect to exchanging $j$ and $k$:
\begin{equation}
    \label{eq:symmG_ap}
    \sum_i^{n_c} \left[ u_{ij}^{(2,2)}E_{ik} - u_{ik}^{(2,2)}E_{ij} \right] = 0 \ \forall j, k \in \mathcal{C}.
\end{equation}
Lastly, we find a set of equations by differentiating Eqs.~\eqref{eq:E_int_relation1} and \eqref{eq:E_int_relation2} and combining them with Eq.~\eqref{eq:E_int_relation3}:
\begin{multline}
    \label{eq:E_int_appendix3}
    \sum_i^{n_c} \bigg[ 2 E_{il} u_{ijk}^{(2,1)} + 2 E_{ik} u_{ijl}^{(2,1)} + 2 E_{ij} u_{ilk}^{(2,1)} - \frac{\partial}{\partial\rho_l}(E_{ik} u_{ij}^{(2,2)}) \\ -  \frac{\partial}{\partial\rho_k}(E_{il} u_{ij}^{(2,2)}) - \frac{\partial}{\partial\rho_j}(E_{ik} u_{il}^{(2,2)}) \bigg] = 0 \ \forall j, k, l \in \mathcal{C}.
\end{multline}

\section{Impact of Additive and Multiplicative Constants on Coexistence Criteria}
Here, we show that both $\boldsymbol{\mathcal{T}}$ and $\boldsymbol{\mathcal{E}}$ only need to be determined up to certain multiplicative/additive constants.
From Eq.~\eqref{eq:T_intro}, it is clear that introducing a global multiplicative constant $c$ into the transformation tensor, $\boldsymbol{\mathcal{T}} \rightarrow c \boldsymbol{\mathcal{T}}$, does not affect the coexistence criteria as the species pseudopotentials may be suitably redefined, $\mathbf{u} \rightarrow c \mathbf{u}$.
Moreover, contracting the transformation tensor with a diagonal matrix of constants only acts to rescale the species pseudopotentials, leaving the conditions of spatially uniform species pseudopotentials unchanged.
To see this, we express the transformation tensor as $\boldsymbol{\mathcal{T}} = \mathbf{A} \cdot \tilde{\boldsymbol{\mathcal{T}}}$, i.e.,~as the product of a diagonal matrix of constants, $\mathbf{A}$, contracted into a bare transformation tensor, $\tilde{\boldsymbol{\mathcal{T}}}$.
Substituting this into Eq.~\eqref{eq:T_intro} and multiplying each side by the inverse of $\mathbf{A}$, $\mathbf{A}^{-1}$, we find $\tilde{\boldsymbol{\mathcal{T}}} \cdot \mathbf{b}^{\rm eff} = \mathbf{A}^{-1} \cdot \partial_z \mathbf{u}$.
As $\mathbf{A}$ (and consequently its inverse) is constant and diagonal, we see that we can define rescaled pseudopotentials, $\mathbf{A}^{-1} \cdot \partial_z \mathbf{u} = \partial_z \tilde{\mathbf{u}}$, such that $\tilde{\boldsymbol{\mathcal{T}}} \cdot \mathbf{b}^{\rm eff} = \partial_z \tilde{\mathbf{u}}$.
We then see that using the bare transformation tensor, $\tilde{\boldsymbol{\mathcal{T}}}$, and the transformation tensor contracted with a diagonal matrix of constants, $\boldsymbol{\mathcal{T}}$, result in the same species pseudopotentials up to a multiplicative factor, and hence $\mathbf{u} = \mathbf{u}^{\rm coexist}$ and $\tilde{\mathbf{u}} = \tilde{\mathbf{u}}^{\rm coexist}$ are equivalent steady-state conditions.
This has the effect that one may always introduce an arbitrary multiplicative constant into the species pseudopotentials without affecting the coexistence criteria.

It is also clear from the generalized Maxwell construction [Eq.~\eqref{eq:G_maxwell_dE_form}] that additive constants to the elements of $\boldsymbol{\mathcal{E}}$ are inconsequential.
Additionally, re-scaling each element of $\boldsymbol{\mathcal{E}}$ by a constant also does not impact Eq.~\eqref{eq:G_maxwell_dE_form}.
As a result, the coexistence criteria resulting from $\boldsymbol{\mathcal{E}}$ and $a\boldsymbol{\mathcal{E}} + \mathbf{c}$ (where $a$ is a constant and $\mathbf{c}$ is a vector of constants whose elements need not be identical) will yield identical results.

\section{Partial Solutions for the Maxwell Construction Vector}
\label{ap:partial}
While a solution to the generalized Gibbs-Duhem relation [Eq.~\eqref{eq:generalized_GD} and correspondingly Eq.~\eqref{eq:E_system_eq}] ensures the system has exact coexistence criteria, in many cases this solution does not exist.
In these scenarios, one may be able to solve for bulk and interfacial Maxwell construction vectors: $\boldsymbol{\mathcal{E}}^{\rm bulk}$ (derived from $\mathcal{G}^{\rm bulk}$ and $\mathbf{u}^{\rm bulk}$) and $\boldsymbol{\mathcal{E}}^{\rm int}$ (derived from $\mathcal{G}^{\rm int}$ and $\mathbf{u}^{\rm int}$), respectively.
This decomposition is achieved by separating the terms in the generalized Gibbs-Duhem relation [Eq.~\eqref{eq:generalized_GD}]:
\begin{subequations}
\begin{align}
    & \boldsymbol{\mathcal{E}}^{\rm bulk} \cdot d\mathbf{u}^{\rm bulk} = d\mathcal{G}^{\rm bulk}, \label{eqap:E_bulk}\\
    & \boldsymbol{\mathcal{E}}^{\rm int} \cdot d\mathbf{u}^{\rm int} = d\mathcal{G}^{\rm int}. \label{eqap:E_int}
\end{align}
\end{subequations}
We now identify the equations in Eq.~\eqref{eq:E_system_eq} as bulk or interfacial.
The bulk relation for Eq.~\eqref{eqap:E_bulk} is:
\begin{equation}
    \boldsymbol{\mathcal{E}}^{\rm bulk} \cdot \partial_{\boldsymbol{\rho}}\mathbf{u}^{\rm bulk} = \partial_{\boldsymbol{\rho}}\mathcal{G}^{\rm bulk}, \label{eq:G_bulk_ap}
\end{equation}
while the interfacial relations for Eq.~\eqref{eqap:E_int} are:
\begin{subequations}
\label{eq:G_int_ap}
\begin{align}
        & \partial_{\boldsymbol{\rho}}(\boldsymbol{\mathcal{E}}^{\rm int} \cdot \mathbf{u}^{(2, 2)}) = \partial_{\boldsymbol{\rho}} \boldsymbol{\mathcal{G}}^{(2,2)},  \label{eq:G22_ap}\\
        & \boldsymbol{\mathcal{E}}^{\rm int} \cdot \left(2 \mathbf{u}^{(2, 1)}+ \partial_{\boldsymbol{\rho}} \mathbf{u}^{(2, 2)}\right) = 2 \boldsymbol{\mathcal{G}}^{(2,1)} + \partial_{\boldsymbol{\rho}} \boldsymbol{\mathcal{G}}^{(2,2)}, \label{eq:G22_G21_ap}\\
        & \boldsymbol{\mathcal{E}}^{\rm int} \cdot \left[\partial_{\boldsymbol{\rho}} \mathbf{u}^{(2, 1)}\right]^{\rm S}= \left[\partial_{\boldsymbol{\rho}} \boldsymbol{\mathcal{G}}^{(2,1)}\right]^{\rm S'}, \label{eq:G21_ap}
\end{align}
\end{subequations}
as derived in Appendix~\ref{ap:variable_vector}.

Equation~\eqref{eq:G_bulk_ap} does not guarantee that the Hessian of $\mathcal{G}^{\rm bulk}$ is symmetric, however, one can obtain the equation that ensures this for a two-component system by differentiating Eq.~\eqref{eq:G_bulk_ap} as shown in the SI~\cite{Note1}.
This is necessary for the Maxwell construction integral in Eq.~\eqref{eq:G_maxwell_drho_form} [or alternatively Eq.~\eqref{eq:G_maxwell_dE_form}] to be path-independent, and is hence a consequence of the $\boldsymbol{\mathcal{E}}$ solution satisfying the bulk relations.
We now show that Eq.~\eqref{eq:G_int_ap} ensures the weighted-area construction with $\mathbf{E}=\mathbf{E}^{\rm int} \equiv \partial_{\boldsymbol{\rho}} \boldsymbol{\mathcal{E}}^{\rm int}$ vanishes when evaluated along the path of the interfacial profile between the coexisting phases, $\boldsymbol{\rho}^{\rm c}$.
The interfacial generalized Gibbs-Duhem relation [Eq.~\eqref{eqap:E_int}] can be expressed as $\partial_z \mathcal{G}^{\rm int} = \partial_z ( \boldsymbol{\mathcal{E}}^{\rm int} \cdot \mathbf{u}^{\rm int}) - \mathbf{u}^{\rm int} \cdot \mathbf{E}^{\rm int} \cdot \partial_z \boldsymbol{\rho}$.
Clearly, $\mathbf{u}^{\rm int} \cdot \mathbf{E}^{\rm int} \cdot \partial_z \boldsymbol{\rho}$ must be the $z$-derivative of a contribution to $\mathcal{G}^{\rm int}$.
Spatially integrating this quantity across the interfacial profile is thus equal to the difference in a contribution to $\mathcal{G}^{\rm int}$ across phases, which vanishes by definition in the coexisting phases.
Therefore, the weighted-area construction [Eq.~\eqref{eq:G_maxwell_drho_form}] with $\mathbf{E}=\mathbf{E}^{\rm int}$ is guaranteed to vanish when evaluated along $\boldsymbol{\rho}^{\rm c}(z)$:
\begin{multline}
    \label{eqap:approx1}
    \int_{z^{\alpha}}^{z^{\beta}} \bigg( \mathbf{u}^{\rm bulk} (\boldsymbol{\rho}^{\rm c}) - \mathbf{u}^{\rm coexist} \bigg) \cdot \mathbf{E}^{\rm int} (\boldsymbol{\rho}^{\rm c}) \cdot \partial_z \boldsymbol{\rho}^{\rm c} dz = 0,
\end{multline}
for all $\alpha, \beta \in \mathcal{P}$.
When both the bulk and interfacial $\boldsymbol{\mathcal{E}}$ are identical [i.e., solutions satisfy all relations in Eq.~\eqref{eq:E_system_eq}], the integral vanishes along any path $\boldsymbol{\rho}(z)$, i.e.,~not just along $\boldsymbol{\rho}^{\rm c}(z)$.
In this case, Eq.~\eqref{eqap:approx1} is equal to the difference in $\mathcal{G}^{\rm bulk}$ between the coexisting phases and hence $\mathcal{G}^{\rm bulk}$ is equal in the phases.

When $\boldsymbol{\mathcal{E}}^{\rm bulk} \neq \boldsymbol{\mathcal{E}}^{\rm int}$, $\boldsymbol{\mathcal{E}}^{\rm int}$ can be used in an approximate Maxwell construction. 
The weighted Maxwell construction in Eq.~\eqref{eqap:approx1} remains universally zero when evaluated along the interfacial profile but loses path-independence, and is therefore generally nonzero along other integration paths.
Upon selecting an integration path [e.g., $(\rho_A', \rho_B'):(0,0)\rightarrow(\rho_A, 0)\rightarrow(\rho_A, \rho_B)$], the weighted-area construction (using $\boldsymbol{\mathcal{E}} = \boldsymbol{\mathcal{E}}^{\rm int}$) can be performed to \textit{approximate} the coexistence criteria.
As demonstrated in the SI~\cite{Note1}, this approximation provides reasonable estimates of the coexistence binodal for specific integration paths.

Lastly, again in scenarios where $\boldsymbol{\mathcal{E}}^{\rm bulk} \neq \boldsymbol{\mathcal{E}}^{\rm int}$, approximate criteria can be formulated using $\boldsymbol{\mathcal{E}}^{\rm bulk}$ as well by defining $\mathcal{G}^{\rm bulk} \equiv \boldsymbol{\mathcal{E}}^{\rm bulk} \cdot \mathbf{u}^{\rm bulk} - \int \mathbf{u}^{\rm bulk} \cdot d \boldsymbol{\mathcal{E}}^{\rm bulk}$.
Noting Eq.~\eqref{eq:G_maxwell_drho_form} is always zero when evaluated along $\boldsymbol{\rho}^{\rm c}$, we re-express it and use the definition of $\mathcal{G}^{\rm bulk}$ to identify $\Delta^{\alpha \beta} \mathcal{G}^{\rm bulk} \equiv \mathcal{G}^{\alpha} - \mathcal{G}^{\beta}$:
\begin{multline}
    \label{eq:deltaG2}
    \Delta^{\alpha \beta} \mathcal{G}^{\rm bulk}  = \int_{z^{\alpha}}^{z^{\beta}} \big[ \mathbf{u}^{\rm bulk}(\boldsymbol{\rho}^{\rm c}) - \mathbf{u}^{\rm coexist} \big] \cdot \big( \mathbf{E}^{\rm bulk}(\boldsymbol{\rho}^{\rm c}) \\  - \mathbf{E}^{\rm int}(\boldsymbol{\rho}^{\rm c}) \big) \cdot \partial_z \boldsymbol{\rho}^{\rm c} dz,
\end{multline}
and hence $\Delta^{\alpha \beta} \mathcal{G}^{\rm bulk}=0$ when $\boldsymbol{\mathcal{E}}^{\rm bulk} = \boldsymbol{\mathcal{E}}^{\rm int}$.
We may then set $\mathcal{G}^{\rm bulk}$ equal across phases as approximate criteria, with the criteria becoming exact when the integrals in Eq.~\eqref{eq:deltaG2} vanish.
This is the case for the NRCH model in Fig.~\ref{fig:phase_diagrams}(b), where:
\begin{multline}
    \Delta^{\alpha \beta} \mathcal{G}^{\rm bulk} = \left( \frac{\chi + \alpha}{\chi - \alpha} - \frac{\overline{\kappa}_{AB}}{\overline{\kappa}_{BA}} \right) \int_{z^{\alpha}}^{z^{\beta}} \bigg[ u^{\rm bulk}_B (\boldsymbol{\rho}^{\rm c}) \\ - u^{\rm coexist}_B \bigg] \partial_z \rho_B^{\rm c} dz \approx 0,
\end{multline}
as shown in the SI~\cite{Note1}.
We are then able to predict the coexistence densities in Fig.~\ref{fig:phase_diagrams}(b) by setting $\mathcal{G}^{\rm bulk}$ equal across phases, providing a quantitatively excellent approximation.


\begin{thebibliography}{80}%
\makeatletter
\providecommand \@ifxundefined [1]{%
 \@ifx{#1\undefined}
}%
\providecommand \@ifnum [1]{%
 \ifnum #1\expandafter \@firstoftwo
 \else \expandafter \@secondoftwo
 \fi
}%
\providecommand \@ifx [1]{%
 \ifx #1\expandafter \@firstoftwo
 \else \expandafter \@secondoftwo
 \fi
}%
\providecommand \natexlab [1]{#1}%
\providecommand \enquote  [1]{``#1''}%
\providecommand \bibnamefont  [1]{#1}%
\providecommand \bibfnamefont [1]{#1}%
\providecommand \citenamefont [1]{#1}%
\providecommand \href@noop [0]{\@secondoftwo}%
\providecommand \href [0]{\begingroup \@sanitize@url \@href}%
\providecommand \@href[1]{\@@startlink{#1}\@@href}%
\providecommand \@@href[1]{\endgroup#1\@@endlink}%
\providecommand \@sanitize@url [0]{\catcode `\\12\catcode `\$12\catcode `\&12\catcode `\#12\catcode `\^12\catcode `\_12\catcode `\%12\relax}%
\providecommand \@@startlink[1]{}%
\providecommand \@@endlink[0]{}%
\providecommand \url  [0]{\begingroup\@sanitize@url \@url }%
\providecommand \@url [1]{\endgroup\@href {#1}{\urlprefix }}%
\providecommand \urlprefix  [0]{URL }%
\providecommand \Eprint [0]{\href }%
\providecommand \doibase [0]{https://doi.org/}%
\providecommand \selectlanguage [0]{\@gobble}%
\providecommand \bibinfo  [0]{\@secondoftwo}%
\providecommand \bibfield  [0]{\@secondoftwo}%
\providecommand \translation [1]{[#1]}%
\providecommand \BibitemOpen [0]{}%
\providecommand \bibitemStop [0]{}%
\providecommand \bibitemNoStop [0]{.\EOS\space}%
\providecommand \EOS [0]{\spacefactor3000\relax}%
\providecommand \BibitemShut  [1]{\csname bibitem#1\endcsname}%
\let\auto@bib@innerbib\@empty
\bibitem [{\citenamefont {Hyman}\ \emph {et~al.}(2014)\citenamefont {Hyman}, \citenamefont {Weber},\ and\ \citenamefont {J{\"{u}}licher}}]{Hyman2014}%
  \BibitemOpen
  \bibfield  {author} {\bibinfo {author} {\bibfnamefont {A.~A.}\ \bibnamefont {Hyman}}, \bibinfo {author} {\bibfnamefont {C.~A.}\ \bibnamefont {Weber}},\ and\ \bibinfo {author} {\bibfnamefont {F.}~\bibnamefont {J{\"{u}}licher}},\ }\bibfield  {title} {\bibinfo {title} {{Liquid-liquid phase separation in biology.}},\ }\href {https://api.semanticscholar.org/CorpusID:13513251} {\bibfield  {journal} {\bibinfo  {journal} {Annu. Rev. Cell Dev. Biol.}\ }\textbf {\bibinfo {volume} {30}},\ \bibinfo {pages} {39} (\bibinfo {year} {2014})}\BibitemShut {NoStop}%
\bibitem [{\citenamefont {Banani}\ \emph {et~al.}(2017)\citenamefont {Banani}, \citenamefont {Lee}, \citenamefont {Hyman},\ and\ \citenamefont {Rosen}}]{Banani2017}%
  \BibitemOpen
  \bibfield  {author} {\bibinfo {author} {\bibfnamefont {S.~F.}\ \bibnamefont {Banani}}, \bibinfo {author} {\bibfnamefont {H.~O.}\ \bibnamefont {Lee}}, \bibinfo {author} {\bibfnamefont {A.~A.}\ \bibnamefont {Hyman}},\ and\ \bibinfo {author} {\bibfnamefont {M.~K.}\ \bibnamefont {Rosen}},\ }\bibfield  {title} {\bibinfo {title} {{Biomolecular condensates: organizers of cellular biochemistry}},\ }\href {https://doi.org/10.1038/nrm.2017.7} {\bibfield  {journal} {\bibinfo  {journal} {Nat. Rev. Mol. Cell Biol.}\ }\textbf {\bibinfo {volume} {18}},\ \bibinfo {pages} {285} (\bibinfo {year} {2017})}\BibitemShut {NoStop}%
\bibitem [{\citenamefont {Berry}\ \emph {et~al.}(2018)\citenamefont {Berry}, \citenamefont {Brangwynne},\ and\ \citenamefont {Haataja}}]{Berry2018}%
  \BibitemOpen
  \bibfield  {author} {\bibinfo {author} {\bibfnamefont {J.}~\bibnamefont {Berry}}, \bibinfo {author} {\bibfnamefont {C.~P.}\ \bibnamefont {Brangwynne}},\ and\ \bibinfo {author} {\bibfnamefont {M.}~\bibnamefont {Haataja}},\ }\bibfield  {title} {\bibinfo {title} {{Physical principles of intracellular organization via active and passive phase transitions}},\ }\href {https://doi.org/10.1088/1361-6633/aaa61e} {\bibfield  {journal} {\bibinfo  {journal} {Rep. Prog. Phys.}\ }\textbf {\bibinfo {volume} {81}},\ \bibinfo {pages} {046601} (\bibinfo {year} {2018})}\BibitemShut {NoStop}%
\bibitem [{\citenamefont {Lee}(2020)}]{Lee2020}%
  \BibitemOpen
  \bibfield  {author} {\bibinfo {author} {\bibfnamefont {C.~F.}\ \bibnamefont {Lee}},\ }\bibfield  {title} {\bibinfo {title} {{Formation of liquid-like cellular organelles depends on their composition}},\ }\href {https://doi.org/10.1038/d41586-020-01280-1} {\bibfield  {journal} {\bibinfo  {journal} {Nature}\ }\textbf {\bibinfo {volume} {581}},\ \bibinfo {pages} {144} (\bibinfo {year} {2020})}\BibitemShut {NoStop}%
\bibitem [{\citenamefont {Trinkaus}\ and\ \citenamefont {Groves}(1955)}]{Trinkaus1955}%
  \BibitemOpen
  \bibfield  {author} {\bibinfo {author} {\bibfnamefont {J.~P.}\ \bibnamefont {Trinkaus}}\ and\ \bibinfo {author} {\bibfnamefont {P.~W.}\ \bibnamefont {Groves}},\ }\bibfield  {title} {\bibinfo {title} {{Differentiation in culture of mixed aggregates of dissociated tissue cells}},\ }\href {https://doi.org/10.1073/pnas.41.10.787} {\bibfield  {journal} {\bibinfo  {journal} {Proc. Natl. Acad. Sci. U. S. A.}\ }\textbf {\bibinfo {volume} {41}},\ \bibinfo {pages} {787} (\bibinfo {year} {1955})}\BibitemShut {NoStop}%
\bibitem [{\citenamefont {Weiss}\ and\ \citenamefont {Taylor}(1960)}]{Weiss1960}%
  \BibitemOpen
  \bibfield  {author} {\bibinfo {author} {\bibfnamefont {P.}~\bibnamefont {Weiss}}\ and\ \bibinfo {author} {\bibfnamefont {A.~C.}\ \bibnamefont {Taylor}},\ }\bibfield  {title} {\bibinfo {title} {{Reconstitution of complete organs from single-cell suspensions of chick embryos in advanced stages of differentiation}},\ }\href {https://doi.org/10.1073/pnas.46.9.1177} {\bibfield  {journal} {\bibinfo  {journal} {Proc. Natl. Acad. Sci. U. S. A.}\ }\textbf {\bibinfo {volume} {46}},\ \bibinfo {pages} {1177} (\bibinfo {year} {1960})}\BibitemShut {NoStop}%
\bibitem [{\citenamefont {Steinberg}(1962)}]{Steinberg1962}%
  \BibitemOpen
  \bibfield  {author} {\bibinfo {author} {\bibfnamefont {M.~S.}\ \bibnamefont {Steinberg}},\ }\bibfield  {title} {\bibinfo {title} {{Mechanism of tissue reconstruction by dissociated cells. II. Time-course of events.}},\ }\href {https://doi.org/10.1126/science.137.3532.762} {\bibfield  {journal} {\bibinfo  {journal} {Science}\ }\textbf {\bibinfo {volume} {137}},\ \bibinfo {pages} {762} (\bibinfo {year} {1962})}\BibitemShut {NoStop}%
\bibitem [{\citenamefont {Teixeira}\ \emph {et~al.}(2024)\citenamefont {Teixeira}, \citenamefont {Beatrici}, \citenamefont {Fernandes},\ and\ \citenamefont {Brunnet}}]{Teixeira2024}%
  \BibitemOpen
  \bibfield  {author} {\bibinfo {author} {\bibfnamefont {E.~F.}\ \bibnamefont {Teixeira}}, \bibinfo {author} {\bibfnamefont {C.~P.}\ \bibnamefont {Beatrici}}, \bibinfo {author} {\bibfnamefont {H.~C.~M.}\ \bibnamefont {Fernandes}},\ and\ \bibinfo {author} {\bibfnamefont {L.~G.}\ \bibnamefont {Brunnet}},\ }\bibfield  {title} {\bibinfo {title} {{Segregation in binary mixture with differential contraction among active rings}},\ }\href {https://arxiv.org/abs/2409.02814} {\bibfield  {journal} {\bibinfo  {journal} {arXiv:2409.02814}\ } (\bibinfo {year} {2024})}\BibitemShut {NoStop}%
\bibitem [{\citenamefont {Mccarthy}\ \emph {et~al.}(2024)\citenamefont {Mccarthy}, \citenamefont {Manna}, \citenamefont {Damavandi},\ and\ \citenamefont {Manning}}]{Mccarthy2024}%
  \BibitemOpen
  \bibfield  {author} {\bibinfo {author} {\bibfnamefont {E.}~\bibnamefont {McCarthy}}, \bibinfo {author} {\bibfnamefont {R.~K.}\ \bibnamefont {Manna}}, \bibinfo {author} {\bibfnamefont {O.}~\bibnamefont {Damavandi}},\ and\ \bibinfo {author} {\bibfnamefont {M.~L.}\ \bibnamefont {Manning}},\ }\bibfield  {title} {\bibinfo {title} {{Demixing in binary mixtures with differential diffusivity at high density}},\ }\href {https://doi.org/10.1103/PhysRevLett.132.098301} {\bibfield  {journal} {\bibinfo  {journal} {Phys. Rev. Lett.}\ }\textbf {\bibinfo {volume} {132}},\ \bibinfo {pages} {098301} (\bibinfo {year} {2024})}\BibitemShut {NoStop}%
\bibitem [{\citenamefont {Harris}(1976)}]{Harris1976}%
  \BibitemOpen
  \bibfield  {author} {\bibinfo {author} {\bibfnamefont {A.~K.}\ \bibnamefont {Harris}},\ }\bibfield  {title} {\bibinfo {title} {{Is cell sorting caused by differences in the work of intercellular adhesion? A critique of the steinberg hypothesis}},\ }\href {https://doi.org/https://doi.org/10.1016/0022-5193(76)90019-9} {\bibfield  {journal} {\bibinfo  {journal} {J. Theor. Biol.}\ }\textbf {\bibinfo {volume} {61}},\ \bibinfo {pages} {267} (\bibinfo {year} {1976})}\BibitemShut {NoStop}%
\bibitem [{\citenamefont {Doan}\ \emph {et~al.}(2024)\citenamefont {Doan}, \citenamefont {Alshareedah}, \citenamefont {Singh}, \citenamefont {Banerjee},\ and\ \citenamefont {Shin}}]{Doan2024}%
  \BibitemOpen
  \bibfield  {author} {\bibinfo {author} {\bibfnamefont {V.~S.}\ \bibnamefont {Doan}}, \bibinfo {author} {\bibfnamefont {I.}~\bibnamefont {Alshareedah}}, \bibinfo {author} {\bibfnamefont {A.}~\bibnamefont {Singh}}, \bibinfo {author} {\bibfnamefont {P.~R.}\ \bibnamefont {Banerjee}},\ and\ \bibinfo {author} {\bibfnamefont {S.}~\bibnamefont {Shin}},\ }\bibfield  {title} {\bibinfo {title} {{Diffusiophoresis promotes phase separation and transport of biomolecular condensates}},\ }\href {https://doi.org/10.1038/s41467-024-51840-6} {\bibfield  {journal} {\bibinfo  {journal} {Nat. Commun}\ }\textbf {\bibinfo {volume} {15}},\ \bibinfo {pages} {1} (\bibinfo {year} {2024})}\BibitemShut {NoStop}%
\bibitem [{\citenamefont {Fily}\ and\ \citenamefont {Marchetti}(2012)}]{Fily2012}%
  \BibitemOpen
  \bibfield  {author} {\bibinfo {author} {\bibfnamefont {Y.}~\bibnamefont {Fily}}\ and\ \bibinfo {author} {\bibfnamefont {M.~C.}\ \bibnamefont {Marchetti}},\ }\bibfield  {title} {\bibinfo {title} {{Athermal phase separation of self-propelled particles with no alignment}},\ }\href {https://doi.org/https://doi.org/10.1103/PhysRevLett.108.235702} {\bibfield  {journal} {\bibinfo  {journal} {Phys. Rev. Lett.}\ }\textbf {\bibinfo {volume} {108}},\ \bibinfo {pages} {235702} (\bibinfo {year} {2012})}\BibitemShut {NoStop}%
\bibitem [{\citenamefont {Redner}\ \emph {et~al.}(2013)\citenamefont {Redner}, \citenamefont {Hagan},\ and\ \citenamefont {Baskaran}}]{Redner2013}%
  \BibitemOpen
  \bibfield  {author} {\bibinfo {author} {\bibfnamefont {G.~S.}\ \bibnamefont {Redner}}, \bibinfo {author} {\bibfnamefont {M.~F.}\ \bibnamefont {Hagan}},\ and\ \bibinfo {author} {\bibfnamefont {A.}~\bibnamefont {Baskaran}},\ }\bibfield  {title} {\bibinfo {title} {{Structure and dynamics of a phase-separating active colloidal fluid}},\ }\href {https://doi.org/10.1103/PhysRevLett.110.055701} {\bibfield  {journal} {\bibinfo  {journal} {Phys. Rev. Lett.}\ }\textbf {\bibinfo {volume} {110}},\ \bibinfo {pages} {55701} (\bibinfo {year} {2013})}\BibitemShut {NoStop}%
\bibitem [{\citenamefont {Wittkowski}\ \emph {et~al.}(2014)\citenamefont {Wittkowski}, \citenamefont {Tiribocchi}, \citenamefont {Stenhammar}, \citenamefont {Allen}, \citenamefont {Marenduzzo},\ and\ \citenamefont {Cates}}]{Wittkowski2014}%
  \BibitemOpen
  \bibfield  {author} {\bibinfo {author} {\bibfnamefont {R.}~\bibnamefont {Wittkowski}}, \bibinfo {author} {\bibfnamefont {A.}~\bibnamefont {Tiribocchi}}, \bibinfo {author} {\bibfnamefont {J.}~\bibnamefont {Stenhammar}}, \bibinfo {author} {\bibfnamefont {R.~J.}\ \bibnamefont {Allen}}, \bibinfo {author} {\bibfnamefont {D.}~\bibnamefont {Marenduzzo}},\ and\ \bibinfo {author} {\bibfnamefont {M.~E.}\ \bibnamefont {Cates}},\ }\bibfield  {title} {\bibinfo {title} {{Scalar {$\phi$}4 field theory for active-particle phase separation}},\ }\href {https://www.nature.com/articles/ncomms5351} {\bibfield  {journal} {\bibinfo  {journal} {Nat. Commun}\ } (\bibinfo {year} {2014})}\BibitemShut {NoStop}%
\bibitem [{\citenamefont {Takatori}\ and\ \citenamefont {Brady}(2015)}]{Takatori2015}%
  \BibitemOpen
  \bibfield  {author} {\bibinfo {author} {\bibfnamefont {S.~C.}\ \bibnamefont {Takatori}}\ and\ \bibinfo {author} {\bibfnamefont {J.~F.}\ \bibnamefont {Brady}},\ }\bibfield  {title} {\bibinfo {title} {{Towards a thermodynamics of active matter}},\ }\href {https://doi.org/10.1103/PhysRevE.91.032117} {\bibfield  {journal} {\bibinfo  {journal} {Phys. Rev. E}\ }\textbf {\bibinfo {volume} {91}},\ \bibinfo {pages} {32117} (\bibinfo {year} {2015})}\BibitemShut {NoStop}%
\bibitem [{\citenamefont {Speck}(2016)}]{Speck2016}%
  \BibitemOpen
  \bibfield  {author} {\bibinfo {author} {\bibfnamefont {T.}~\bibnamefont {Speck}},\ }\bibfield  {title} {\bibinfo {title} {{Stochastic thermodynamics for active matter}},\ }\href {https://doi.org/10.1209/0295-5075/114/30006} {\bibfield  {journal} {\bibinfo  {journal} {Europhys. Lett.}\ }\textbf {\bibinfo {volume} {114}},\ \bibinfo {pages} {30006} (\bibinfo {year} {2016})}\BibitemShut {NoStop}%
\bibitem [{\citenamefont {Solon}\ \emph {et~al.}(2018)\citenamefont {Solon}, \citenamefont {Stenhammar}, \citenamefont {Cates}, \citenamefont {Kafri},\ and\ \citenamefont {Tailleur}}]{Solon2018}%
  \BibitemOpen
  \bibfield  {author} {\bibinfo {author} {\bibfnamefont {A.~P.}\ \bibnamefont {Solon}}, \bibinfo {author} {\bibfnamefont {J.}~\bibnamefont {Stenhammar}}, \bibinfo {author} {\bibfnamefont {M.~E.}\ \bibnamefont {Cates}}, \bibinfo {author} {\bibfnamefont {Y.}~\bibnamefont {Kafri}},\ and\ \bibinfo {author} {\bibfnamefont {J.}~\bibnamefont {Tailleur}},\ }\bibfield  {title} {\bibinfo {title} {{Generalized thermodynamics of phase equilibria in scalar active matter}},\ }\href {https://doi.org/10.1103/PhysRevE.97.020602} {\bibfield  {journal} {\bibinfo  {journal} {Phys. Rev. E}\ }\textbf {\bibinfo {volume} {97}},\ \bibinfo {pages} {20602} (\bibinfo {year} {2018})}\BibitemShut {NoStop}%
\bibitem [{\citenamefont {Hermann}\ \emph {et~al.}(2019)\citenamefont {Hermann}, \citenamefont {Krinninger}, \citenamefont {de~las Heras},\ and\ \citenamefont {Schmidt}}]{Hermann2019}%
  \BibitemOpen
  \bibfield  {author} {\bibinfo {author} {\bibfnamefont {S.}~\bibnamefont {Hermann}}, \bibinfo {author} {\bibfnamefont {P.}~\bibnamefont {Krinninger}}, \bibinfo {author} {\bibfnamefont {D.}~\bibnamefont {de~las Heras}},\ and\ \bibinfo {author} {\bibfnamefont {M.}~\bibnamefont {Schmidt}},\ }\bibfield  {title} {\bibinfo {title} {{Phase coexistence of active Brownian particles}},\ }\href {https://doi.org/10.1103/PhysRevE.100.052604} {\bibfield  {journal} {\bibinfo  {journal} {Phys. Rev. E.}\ }\textbf {\bibinfo {volume} {100}},\ \bibinfo {pages} {52604} (\bibinfo {year} {2019})}\BibitemShut {NoStop}%
\bibitem [{\citenamefont {Agudo-Canalejo}\ and\ \citenamefont {Golestanian}(2019)}]{Agudo2019}%
  \BibitemOpen
  \bibfield  {author} {\bibinfo {author} {\bibfnamefont {J.}~\bibnamefont {Agudo-Canalejo}}\ and\ \bibinfo {author} {\bibfnamefont {R.}~\bibnamefont {Golestanian}},\ }\bibfield  {title} {\bibinfo {title} {{Active phase separation in mixtures of chemically interacting particles}},\ }\href {https://journals.aps.org/prl/abstract/10.1103/PhysRevLett.123.018101} {\bibfield  {journal} {\bibinfo  {journal} {Phys. Rev. Lett.}\ }\textbf {\bibinfo {volume} {123}} (\bibinfo {year} {2019})}\BibitemShut {NoStop}%
\bibitem [{\citenamefont {Fruchart}\ \emph {et~al.}(2020)\citenamefont {Fruchart}, \citenamefont {Hanai}, \citenamefont {Littlewood},\ and\ \citenamefont {Vitelli}}]{Fruchart2020}%
  \BibitemOpen
  \bibfield  {author} {\bibinfo {author} {\bibfnamefont {M.}~\bibnamefont {Fruchart}}, \bibinfo {author} {\bibfnamefont {R.}~\bibnamefont {Hanai}}, \bibinfo {author} {\bibfnamefont {P.~B.}\ \bibnamefont {Littlewood}},\ and\ \bibinfo {author} {\bibfnamefont {V.}~\bibnamefont {Vitelli}},\ }\bibfield  {title} {\bibinfo {title} {{Non-reciprocal phase transitions}},\ }\href {https://doi.org/10.1038/s41586-021-03375-9} {\bibfield  {journal} {\bibinfo  {journal} {Nature}\ }\textbf {\bibinfo {volume} {592}},\ \bibinfo {pages} {363} (\bibinfo {year} {2020})}\BibitemShut {NoStop}%
\bibitem [{\citenamefont {Hermann}\ \emph {et~al.}(2021)\citenamefont {Hermann}, \citenamefont {de~las Heras},\ and\ \citenamefont {Schmidt}}]{Hermann2021}%
  \BibitemOpen
  \bibfield  {author} {\bibinfo {author} {\bibfnamefont {S.}~\bibnamefont {Hermann}}, \bibinfo {author} {\bibfnamefont {D.}~\bibnamefont {de~las Heras}},\ and\ \bibinfo {author} {\bibfnamefont {M.}~\bibnamefont {Schmidt}},\ }\bibfield  {title} {\bibinfo {title} {{Phase separation of active Brownian particles in two dimensions: anything for a quiet life}},\ }\href {https://doi.org/10.1080/00268976.2021.1902585} {\bibfield  {journal} {\bibinfo  {journal} {Mol. Phys.}\ }\textbf {\bibinfo {volume} {119}},\ \bibinfo {pages} {15} (\bibinfo {year} {2021})}\BibitemShut {NoStop}%
\bibitem [{\citenamefont {Omar}\ \emph {et~al.}(2023)\citenamefont {Omar}, \citenamefont {Row}, \citenamefont {Mallory},\ and\ \citenamefont {Brady}}]{Omar2023}%
  \BibitemOpen
  \bibfield  {author} {\bibinfo {author} {\bibfnamefont {A.~K.}\ \bibnamefont {Omar}}, \bibinfo {author} {\bibfnamefont {H.}~\bibnamefont {Row}}, \bibinfo {author} {\bibfnamefont {S.~A.}\ \bibnamefont {Mallory}},\ and\ \bibinfo {author} {\bibfnamefont {J.~F.}\ \bibnamefont {Brady}},\ }\bibfield  {title} {\bibinfo {title} {{Mechanical theory of nonequilibrium coexistence and motility-induced phase separation}},\ }\href {https://doi.org/10.1073/pnas.2219900120} {\bibfield  {journal} {\bibinfo  {journal} {Proc. Natl. Acad. Sci. U. S. A.}\ }\textbf {\bibinfo {volume} {120}},\ \bibinfo {pages} {e2219900120} (\bibinfo {year} {2023})}\BibitemShut {NoStop}%
\bibitem [{\citenamefont {Evans}\ and\ \citenamefont {Omar}(2023)}]{Evans2023}%
  \BibitemOpen
  \bibfield  {author} {\bibinfo {author} {\bibfnamefont {D.}~\bibnamefont {Evans}}\ and\ \bibinfo {author} {\bibfnamefont {A.~K.}\ \bibnamefont {Omar}},\ }\bibfield  {title} {\bibinfo {title} {{Theory of nonequilibrium symmetry-breaking coexistence and active crystallization}},\ }\href {https://arxiv.org/abs/2309.10341} {\bibfield  {journal} {\bibinfo  {journal} {arXiv:2309.10341}\ } (\bibinfo {year} {2023})}\BibitemShut {NoStop}%
\bibitem [{\citenamefont {Petroff}\ \emph {et~al.}(2015)\citenamefont {Petroff}, \citenamefont {Wu},\ and\ \citenamefont {Libchaber}}]{Petroff2015}%
  \BibitemOpen
  \bibfield  {author} {\bibinfo {author} {\bibfnamefont {A.~P.}\ \bibnamefont {Petroff}}, \bibinfo {author} {\bibfnamefont {X.-L.}\ \bibnamefont {Wu}},\ and\ \bibinfo {author} {\bibfnamefont {A.}~\bibnamefont {Libchaber}},\ }\bibfield  {title} {\bibinfo {title} {{Fast-moving bacteria self-organize into active two-dimensional crystals of rotating cells}},\ }\href {https://doi.org/10.1103/PhysRevLett.114.158102} {\bibfield  {journal} {\bibinfo  {journal} {Phys. Rev. Lett.}\ }\textbf {\bibinfo {volume} {114}},\ \bibinfo {pages} {158102} (\bibinfo {year} {2015})}\BibitemShut {NoStop}%
\bibitem [{\citenamefont {Natan}\ \emph {et~al.}(2022)\citenamefont {Natan}, \citenamefont {Worlitzer}, \citenamefont {Ariel},\ and\ \citenamefont {Be’er}}]{Natan2022}%
  \BibitemOpen
  \bibfield  {author} {\bibinfo {author} {\bibfnamefont {G.}~\bibnamefont {Natan}}, \bibinfo {author} {\bibfnamefont {V.~M.}\ \bibnamefont {Worlitzer}}, \bibinfo {author} {\bibfnamefont {G.}~\bibnamefont {Ariel}},\ and\ \bibinfo {author} {\bibfnamefont {A.}~\bibnamefont {Be’er}},\ }\bibfield  {title} {\bibinfo {title} {{Mixed-species bacterial swarms show an interplay of mixing and segregation across scales}},\ }\href {https://doi.org/10.1038/s41598-022-20644-3} {\bibfield  {journal} {\bibinfo  {journal} {Sci. Rep.}\ }\textbf {\bibinfo {volume} {12}},\ \bibinfo {pages} {1} (\bibinfo {year} {2022})}\BibitemShut {NoStop}%
\bibitem [{\citenamefont {Tan}\ \emph {et~al.}(2022)\citenamefont {Tan}, \citenamefont {Mietke}, \citenamefont {Li}, \citenamefont {Chen}, \citenamefont {Higinbotham}, \citenamefont {Foster}, \citenamefont {Gokhale}, \citenamefont {Dunkel},\ and\ \citenamefont {Fakhri}}]{Tan2022}%
  \BibitemOpen
  \bibfield  {author} {\bibinfo {author} {\bibfnamefont {T.~H.}\ \bibnamefont {Tan}}, \bibinfo {author} {\bibfnamefont {A.}~\bibnamefont {Mietke}}, \bibinfo {author} {\bibfnamefont {J.}~\bibnamefont {Li}}, \bibinfo {author} {\bibfnamefont {Y.}~\bibnamefont {Chen}}, \bibinfo {author} {\bibfnamefont {H.}~\bibnamefont {Higinbotham}}, \bibinfo {author} {\bibfnamefont {P.~J.}\ \bibnamefont {Foster}}, \bibinfo {author} {\bibfnamefont {S.}~\bibnamefont {Gokhale}}, \bibinfo {author} {\bibfnamefont {J.}~\bibnamefont {Dunkel}},\ and\ \bibinfo {author} {\bibfnamefont {N.}~\bibnamefont {Fakhri}},\ }\bibfield  {title} {\bibinfo {title} {{Odd dynamics of living chiral crystals}},\ }\href {https://doi.org/10.1038/s41586-022-04889-6} {\bibfield  {journal} {\bibinfo  {journal} {Nature}\ }\textbf {\bibinfo {volume} {607}},\ \bibinfo {pages} {287} (\bibinfo {year} {2022})}\BibitemShut {NoStop}%
\bibitem [{\citenamefont {Kang}\ \emph {et~al.}(2024)\citenamefont {Kang}, \citenamefont {Ma}, \citenamefont {Liu}, \citenamefont {Wang}, \citenamefont {Zhou}, \citenamefont {Xue}, \citenamefont {Xu},\ and\ \citenamefont {Li}}]{Kang2024}%
  \BibitemOpen
  \bibfield  {author} {\bibinfo {author} {\bibfnamefont {W.}~\bibnamefont {Kang}}, \bibinfo {author} {\bibfnamefont {X.}~\bibnamefont {Ma}}, \bibinfo {author} {\bibfnamefont {C.}~\bibnamefont {Liu}}, \bibinfo {author} {\bibfnamefont {S.}~\bibnamefont {Wang}}, \bibinfo {author} {\bibfnamefont {Y.}~\bibnamefont {Zhou}}, \bibinfo {author} {\bibfnamefont {C.}~\bibnamefont {Xue}}, \bibinfo {author} {\bibfnamefont {Y.}~\bibnamefont {Xu}},\ and\ \bibinfo {author} {\bibfnamefont {B.}~\bibnamefont {Li}},\ }\bibfield  {title} {\bibinfo {title} {{Liquid-liquid phase separation (LLPS) in synthetic biosystems}},\ }\href {https://doi.org/https://doi.org/10.1016/j.mser.2023.100762} {\bibfield  {journal} {\bibinfo  {journal} {Mat. Sci. Eng. R.}\ }\textbf {\bibinfo {volume} {157}},\ \bibinfo {pages} {100762} (\bibinfo {year} {2024})}\BibitemShut {NoStop}%
\bibitem [{\citenamefont {Clerk-Maxwell}(1875)}]{Maxwell1875}%
  \BibitemOpen
  \bibfield  {author} {\bibinfo {author} {\bibfnamefont {J.}~\bibnamefont {Clerk-Maxwell}},\ }\bibfield  {title} {\bibinfo {title} {{On the dynamical evidence of the molecular constitution of bodies}},\ }\href@noop {} {\bibfield  {journal} {\bibinfo  {journal} {Nature}\ }\textbf {\bibinfo {volume} {11}},\ \bibinfo {pages} {357} (\bibinfo {year} {1875})}\BibitemShut {NoStop}%
\bibitem [{\citenamefont {Aifantis}\ and\ \citenamefont {Serrin}(1983)}]{Aifantis1983}%
  \BibitemOpen
  \bibfield  {author} {\bibinfo {author} {\bibfnamefont {E.~C.}\ \bibnamefont {Aifantis}}\ and\ \bibinfo {author} {\bibfnamefont {J.~B.}\ \bibnamefont {Serrin}},\ }\bibfield  {title} {\bibinfo {title} {{The mechanical theory of fluid interfaces and Maxwell's rule}},\ }\href {https://doi.org/10.1016/0021-9797(83)90053-X} {\bibfield  {journal} {\bibinfo  {journal} {J. Colloid Interface Sci.}\ }\textbf {\bibinfo {volume} {96}},\ \bibinfo {pages} {517} (\bibinfo {year} {1983})}\BibitemShut {NoStop}%
\bibitem [{\citenamefont {Dinelli}\ \emph {et~al.}(2023)\citenamefont {Dinelli}, \citenamefont {O’Byrne}, \citenamefont {Curatolo}, \citenamefont {Zhao}, \citenamefont {Sollich},\ and\ \citenamefont {Tailleur}}]{Dinelli2023}%
  \BibitemOpen
  \bibfield  {author} {\bibinfo {author} {\bibfnamefont {A.}~\bibnamefont {Dinelli}}, \bibinfo {author} {\bibfnamefont {J.}~\bibnamefont {O’Byrne}}, \bibinfo {author} {\bibfnamefont {A.}~\bibnamefont {Curatolo}}, \bibinfo {author} {\bibfnamefont {Y.}~\bibnamefont {Zhao}}, \bibinfo {author} {\bibfnamefont {P.}~\bibnamefont {Sollich}},\ and\ \bibinfo {author} {\bibfnamefont {J.}~\bibnamefont {Tailleur}},\ }\bibfield  {title} {\bibinfo {title} {{Non-reciprocity across scales in active mixtures}},\ }\href {https://doi.org/10.1038/s41467-023-42713-5} {\bibfield  {journal} {\bibinfo  {journal} {Nat. Commun}\ }\textbf {\bibinfo {volume} {14}},\ \bibinfo {pages} {1} (\bibinfo {year} {2023})}\BibitemShut {NoStop}%
\bibitem [{\citenamefont {Saha}(2024)}]{Saha2024}%
  \BibitemOpen
  \bibfield  {author} {\bibinfo {author} {\bibfnamefont {S.}~\bibnamefont {Saha}},\ }\bibfield  {title} {\bibinfo {title} {{Phase coexistence in the Non-reciprocal Cahn-Hilliard model}},\ }\href {https://doi.org/10.48550/arXiv.2402.10057} {\bibfield  {journal} {\bibinfo  {journal} {arXiv:2402.10057}\ } (\bibinfo {year} {2024})}\BibitemShut {NoStop}%
\bibitem [{\citenamefont {Greve}\ \emph {et~al.}(2024)\citenamefont {Greve}, \citenamefont {Lovato}, \citenamefont {Frohoff-H{\"{u}}lsmann},\ and\ \citenamefont {Thiele}}]{Greve2024}%
  \BibitemOpen
  \bibfield  {author} {\bibinfo {author} {\bibfnamefont {D.}~\bibnamefont {Greve}}, \bibinfo {author} {\bibfnamefont {G.}~\bibnamefont {Lovato}}, \bibinfo {author} {\bibfnamefont {T.}~\bibnamefont {Frohoff-H{\"{u}}lsmann}},\ and\ \bibinfo {author} {\bibfnamefont {U.}~\bibnamefont {Thiele}},\ }\bibfield  {title} {\bibinfo {title} {{Coexistence of uniform and oscillatory states resulting from nonreciprocity and conservation laws}},\ }\href {https://arxiv.org/abs/2402.08634} {\bibfield  {journal} {\bibinfo  {journal} {arXiv:2402.08634}\ } (\bibinfo {year} {2024})}\BibitemShut {NoStop}%
\bibitem [{\citenamefont {Grosberg}\ and\ \citenamefont {Joanny}(2015)}]{Grosberg2015}%
  \BibitemOpen
  \bibfield  {author} {\bibinfo {author} {\bibfnamefont {A.~Y.}\ \bibnamefont {Grosberg}}\ and\ \bibinfo {author} {\bibfnamefont {J.-F.}\ \bibnamefont {Joanny}},\ }\bibfield  {title} {\bibinfo {title} {{Nonequilibrium statistical mechanics of mixtures of particles in contact with different thermostats}},\ }\href {https://doi.org/10.1103/PhysRevE.92.032118} {\bibfield  {journal} {\bibinfo  {journal} {Phys. Rev. E.}\ }\textbf {\bibinfo {volume} {92}},\ \bibinfo {pages} {32118} (\bibinfo {year} {2015})}\BibitemShut {NoStop}%
\bibitem [{\citenamefont {Weber}\ \emph {et~al.}(2016)\citenamefont {Weber}, \citenamefont {Weber},\ and\ \citenamefont {Frey}}]{Weber2016}%
  \BibitemOpen
  \bibfield  {author} {\bibinfo {author} {\bibfnamefont {S.~N.}\ \bibnamefont {Weber}}, \bibinfo {author} {\bibfnamefont {C.~A.}\ \bibnamefont {Weber}},\ and\ \bibinfo {author} {\bibfnamefont {E.}~\bibnamefont {Frey}},\ }\bibfield  {title} {\bibinfo {title} {{Binary mixtures of particles with different diffusivities demix}},\ }\href {https://doi.org/10.1103/PhysRevLett.116.058301} {\bibfield  {journal} {\bibinfo  {journal} {Phys. Rev. Lett.}\ }\textbf {\bibinfo {volume} {116}},\ \bibinfo {pages} {58301} (\bibinfo {year} {2016})}\BibitemShut {NoStop}%
\bibitem [{\citenamefont {Han}\ \emph {et~al.}(2017)\citenamefont {Han}, \citenamefont {Yan}, \citenamefont {Granick},\ and\ \citenamefont {Luijten}}]{Han2017}%
  \BibitemOpen
  \bibfield  {author} {\bibinfo {author} {\bibfnamefont {M.}~\bibnamefont {Han}}, \bibinfo {author} {\bibfnamefont {J.}~\bibnamefont {Yan}}, \bibinfo {author} {\bibfnamefont {S.}~\bibnamefont {Granick}},\ and\ \bibinfo {author} {\bibfnamefont {E.}~\bibnamefont {Luijten}},\ }\bibfield  {title} {\bibinfo {title} {{Effective temperature concept evaluated in an active colloid mixture}},\ }\href {https://www.pnas.org/doi/abs/10.1073/pnas.1706702114} {\bibfield  {journal} {\bibinfo  {journal} {Proc. Natl. Acad. Sci. U. S. A.}\ }\textbf {\bibinfo {volume} {114}},\ \bibinfo {pages} {7513} (\bibinfo {year} {2017})}\BibitemShut {NoStop}%
\bibitem [{\citenamefont {Angelani}\ \emph {et~al.}(2011)\citenamefont {Angelani}, \citenamefont {Maggi}, \citenamefont {Bernardini}, \citenamefont {Rizzo},\ and\ \citenamefont {Di~Leonardo}}]{Angelani2011}%
  \BibitemOpen
  \bibfield  {author} {\bibinfo {author} {\bibfnamefont {L.}~\bibnamefont {Angelani}}, \bibinfo {author} {\bibfnamefont {C.}~\bibnamefont {Maggi}}, \bibinfo {author} {\bibfnamefont {M.~L.}\ \bibnamefont {Bernardini}}, \bibinfo {author} {\bibfnamefont {A.}~\bibnamefont {Rizzo}},\ and\ \bibinfo {author} {\bibfnamefont {R.}~\bibnamefont {Di~Leonardo}},\ }\bibfield  {title} {\bibinfo {title} {{Effective interactions between colloidal particles suspended in a bath of swimming cells}},\ }\href {https://doi.org/10.1103/PhysRevLett.107.138302} {\bibfield  {journal} {\bibinfo  {journal} {Phys. Rev. Lett.}\ }\textbf {\bibinfo {volume} {107}},\ \bibinfo {pages} {138302} (\bibinfo {year} {2011})}\BibitemShut {NoStop}%
\bibitem [{\citenamefont {Stenhammar}\ \emph {et~al.}(2015)\citenamefont {Stenhammar}, \citenamefont {Wittkowski}, \citenamefont {Marenduzzo},\ and\ \citenamefont {Cates}}]{Stenhammar2015}%
  \BibitemOpen
  \bibfield  {author} {\bibinfo {author} {\bibfnamefont {J.}~\bibnamefont {Stenhammar}}, \bibinfo {author} {\bibfnamefont {R.}~\bibnamefont {Wittkowski}}, \bibinfo {author} {\bibfnamefont {D.}~\bibnamefont {Marenduzzo}},\ and\ \bibinfo {author} {\bibfnamefont {M.~E.}\ \bibnamefont {Cates}},\ }\bibfield  {title} {\bibinfo {title} {{Activity-induced phase separation and self-assembly in mixtures of active and passive particles}},\ }\href {https://doi.org/https://doi.org/10.1103/PhysRevLett.114.018301} {\bibfield  {journal} {\bibinfo  {journal} {Phys. Rev. Lett.}\ }\textbf {\bibinfo {volume} {114}},\ \bibinfo {pages} {018301} (\bibinfo {year} {2015})}\BibitemShut {NoStop}%
\bibitem [{\citenamefont {Wittkowski}\ \emph {et~al.}(2017)\citenamefont {Wittkowski}, \citenamefont {Stenhammar},\ and\ \citenamefont {Cates}}]{Wittkowski2017}%
  \BibitemOpen
  \bibfield  {author} {\bibinfo {author} {\bibfnamefont {R.}~\bibnamefont {Wittkowski}}, \bibinfo {author} {\bibfnamefont {J.}~\bibnamefont {Stenhammar}},\ and\ \bibinfo {author} {\bibfnamefont {M.~E.}\ \bibnamefont {Cates}},\ }\bibfield  {title} {\bibinfo {title} {{Nonequilibrium dynamics of mixtures of active and passive colloidal particles}},\ }\href {https://doi.org/10.1088/1367-2630/aa8195} {\bibfield  {journal} {\bibinfo  {journal} {New J. Phys.}\ }\textbf {\bibinfo {volume} {19}},\ \bibinfo {pages} {105003} (\bibinfo {year} {2017})}\BibitemShut {NoStop}%
\bibitem [{\citenamefont {Omar}\ \emph {et~al.}(2019)\citenamefont {Omar}, \citenamefont {Wu}, \citenamefont {Wang},\ and\ \citenamefont {Brady}}]{Omar2019}%
  \BibitemOpen
  \bibfield  {author} {\bibinfo {author} {\bibfnamefont {A.~K.}\ \bibnamefont {Omar}}, \bibinfo {author} {\bibfnamefont {Y.}~\bibnamefont {Wu}}, \bibinfo {author} {\bibfnamefont {Z.~G.}\ \bibnamefont {Wang}},\ and\ \bibinfo {author} {\bibfnamefont {J.~F.}\ \bibnamefont {Brady}},\ }\bibfield  {title} {\bibinfo {title} {{Swimming to stability: Structural and dynamical control via active doping}},\ }\href {https://doi.org/10.1021/acsnano.8b07421} {\bibfield  {journal} {\bibinfo  {journal} {ACS Nano}\ }\textbf {\bibinfo {volume} {13}},\ \bibinfo {pages} {560} (\bibinfo {year} {2019})}\BibitemShut {NoStop}%
\bibitem [{\citenamefont {Batton}\ and\ \citenamefont {Rotskoff}(2024)}]{Batton2024}%
  \BibitemOpen
  \bibfield  {author} {\bibinfo {author} {\bibfnamefont {C.~H.}\ \bibnamefont {Batton}}\ and\ \bibinfo {author} {\bibfnamefont {G.~M.}\ \bibnamefont {Rotskoff}},\ }\bibfield  {title} {\bibinfo {title} {{Microscopic origin of tunable assembly forces in chiral active environments}},\ }\href {https://pubs.rsc.org/en/content/articlelanding/2024/sm/d4sm00247d} {\bibfield  {journal} {\bibinfo  {journal} {Soft Matter}\ } (\bibinfo {year} {2024})}\BibitemShut {NoStop}%
\bibitem [{\citenamefont {Mason}\ \emph {et~al.}(2024)\citenamefont {Mason}, \citenamefont {Jack},\ and\ \citenamefont {Bruna}}]{Manson2024}%
  \BibitemOpen
  \bibfield  {author} {\bibinfo {author} {\bibfnamefont {J.}~\bibnamefont {Mason}}, \bibinfo {author} {\bibfnamefont {R.~L.}\ \bibnamefont {Jack}},\ and\ \bibinfo {author} {\bibfnamefont {M.}~\bibnamefont {Bruna}},\ }\bibfield  {title} {\bibinfo {title} {{Dynamical patterns in active-passive particle mixtures with non-reciprocal interactions: Exact hydrodynamic analysis}},\ }\href {https://arxiv.org/abs/2408.03932v1} {\bibfield  {journal} {\bibinfo  {journal} {arXiv:2408.03932}\ } (\bibinfo {year} {2024})}\BibitemShut {NoStop}%
\bibitem [{\citenamefont {Kreienkamp}\ and\ \citenamefont {Klapp}(2024)}]{Kreienkamp2024}%
  \BibitemOpen
  \bibfield  {author} {\bibinfo {author} {\bibfnamefont {K.~L.}\ \bibnamefont {Kreienkamp}}\ and\ \bibinfo {author} {\bibfnamefont {S.~H.~L.}\ \bibnamefont {Klapp}},\ }\bibfield  {title} {\bibinfo {title} {{Dynamical structures in phase-separating non-reciprocal polar active mixtures}},\ }\href {https://arxiv.org/abs/2404.06305v2} {\bibfield  {journal} {\bibinfo  {journal} {arXiv:2404.06305}\ } (\bibinfo {year} {2024})}\BibitemShut {NoStop}%
\bibitem [{\citenamefont {You}\ \emph {et~al.}(2020)\citenamefont {You}, \citenamefont {Baskaran},\ and\ \citenamefont {Marchetti}}]{You2020}%
  \BibitemOpen
  \bibfield  {author} {\bibinfo {author} {\bibfnamefont {Z.}~\bibnamefont {You}}, \bibinfo {author} {\bibfnamefont {A.}~\bibnamefont {Baskaran}},\ and\ \bibinfo {author} {\bibfnamefont {M.~C.}\ \bibnamefont {Marchetti}},\ }\bibfield  {title} {\bibinfo {title} {{Nonreciprocity as a generic route to traveling states}},\ }\href {https://doi.org/10.1073/pnas.2010318117} {\bibfield  {journal} {\bibinfo  {journal} {Proc. Natl. Acad. Sci. USA}\ }\textbf {\bibinfo {volume} {117}},\ \bibinfo {pages} {19767} (\bibinfo {year} {2020})}\BibitemShut {NoStop}%
\bibitem [{\citenamefont {Fruchart}\ \emph {et~al.}(2021)\citenamefont {Fruchart}, \citenamefont {Hanai}, \citenamefont {Littlewood},\ and\ \citenamefont {Vitelli}}]{Fruchart2021}%
  \BibitemOpen
  \bibfield  {author} {\bibinfo {author} {\bibfnamefont {M.}~\bibnamefont {Fruchart}}, \bibinfo {author} {\bibfnamefont {R.}~\bibnamefont {Hanai}}, \bibinfo {author} {\bibfnamefont {P.~B.}\ \bibnamefont {Littlewood}},\ and\ \bibinfo {author} {\bibfnamefont {V.}~\bibnamefont {Vitelli}},\ }\bibfield  {title} {\bibinfo {title} {{Non-reciprocal phase transitions}},\ }\href {https://doi.org/10.1038/s41586-021-03375-9} {\bibfield  {journal} {\bibinfo  {journal} {Nature}\ }\textbf {\bibinfo {volume} {592}},\ \bibinfo {pages} {363} (\bibinfo {year} {2021})}\BibitemShut {NoStop}%
\bibitem [{\citenamefont {Chiu}\ and\ \citenamefont {Omar}(2023)}]{Chiu2023}%
  \BibitemOpen
  \bibfield  {author} {\bibinfo {author} {\bibfnamefont {Y.-J.}\ \bibnamefont {Chiu}}\ and\ \bibinfo {author} {\bibfnamefont {A.~K.}\ \bibnamefont {Omar}},\ }\bibfield  {title} {\bibinfo {title} {{Phase coexistence implications of violating Newton’s third law}},\ }\href {https://doi.org/10.1063/5.0146822} {\bibfield  {journal} {\bibinfo  {journal} {J. Chem. Phys.}\ }\textbf {\bibinfo {volume} {158}},\ \bibinfo {pages} {164903} (\bibinfo {year} {2023})}\BibitemShut {NoStop}%
\bibitem [{\citenamefont {Gibbs}(1878)}]{Gibbs1878}%
  \BibitemOpen
  \bibfield  {author} {\bibinfo {author} {\bibfnamefont {J.~W.}\ \bibnamefont {Gibbs}},\ }\bibfield  {title} {\bibinfo {title} {{On the equilibrium of heterogeneous substances}},\ }\href {https://doi.org/10.2475/ajs.s3-16.96.441} {\bibfield  {journal} {\bibinfo  {journal} {Am. J. Sci.}\ }\textbf {\bibinfo {volume} {s3-16}},\ \bibinfo {pages} {441} (\bibinfo {year} {1878})}\BibitemShut {NoStop}%
\bibitem [{\citenamefont {Plischke}\ and\ \citenamefont {Bergersen}(1994)}]{Plischke1994}%
  \BibitemOpen
  \bibfield  {author} {\bibinfo {author} {\bibfnamefont {M.}~\bibnamefont {Plischke}}\ and\ \bibinfo {author} {\bibfnamefont {B.}~\bibnamefont {Bergersen}},\ }\href@noop {} {\emph {\bibinfo {title} {{Equilibrium Statistical Physics}}}},\ \bibinfo {edition} {3rd}\ ed.\ (\bibinfo  {publisher} {World scientific},\ \bibinfo {address} {Singapore},\ \bibinfo {year} {1994})\BibitemShut {NoStop}%
\bibitem [{\citenamefont {Omar}\ \emph {et~al.}(2020)\citenamefont {Omar}, \citenamefont {Wang},\ and\ \citenamefont {Brady}}]{Omar2020}%
  \BibitemOpen
  \bibfield  {author} {\bibinfo {author} {\bibfnamefont {A.~K.}\ \bibnamefont {Omar}}, \bibinfo {author} {\bibfnamefont {Z.-G.}\ \bibnamefont {Wang}},\ and\ \bibinfo {author} {\bibfnamefont {J.~F.}\ \bibnamefont {Brady}},\ }\bibfield  {title} {\bibinfo {title} {{Microscopic origins of the swim pressure and the anomalous surface tension of active matter}},\ }\href {https://doi.org/10.1103/PhysRevE.101.012604} {\bibfield  {journal} {\bibinfo  {journal} {Phys. Rev. E.}\ }\textbf {\bibinfo {volume} {101}},\ \bibinfo {pages} {12604} (\bibinfo {year} {2020})}\BibitemShut {NoStop}%
\bibitem [{\citenamefont {Irving}\ and\ \citenamefont {Kirkwood}(1950)}]{Irving1950}%
  \BibitemOpen
  \bibfield  {author} {\bibinfo {author} {\bibfnamefont {J.~H.}\ \bibnamefont {Irving}}\ and\ \bibinfo {author} {\bibfnamefont {J.~G.}\ \bibnamefont {Kirkwood}},\ }\bibfield  {title} {\bibinfo {title} {{The statistical mechanical theory of transport processes. IV. The equations of hydrodynamics}},\ }\href {https://doi.org/10.1063/1.1747782} {\bibfield  {journal} {\bibinfo  {journal} {J. Chem. Phys.}\ }\textbf {\bibinfo {volume} {18}},\ \bibinfo {pages} {817} (\bibinfo {year} {1950})}\BibitemShut {NoStop}%
\bibitem [{\citenamefont {Hansen}\ and\ \citenamefont {McDonald}(2013)}]{Hansen2013}%
  \BibitemOpen
  \bibfield  {author} {\bibinfo {author} {\bibfnamefont {J.-P.}\ \bibnamefont {Hansen}}\ and\ \bibinfo {author} {\bibfnamefont {I.~R.}\ \bibnamefont {McDonald}},\ }\href@noop {} {\emph {\bibinfo {title} {{Theory of Simple Liquids}}}},\ \bibinfo {edition} {4th}\ ed.\ (\bibinfo  {publisher} {Elsevier},\ \bibinfo {address} {New York},\ \bibinfo {year} {2013})\BibitemShut {NoStop}%
\bibitem [{\citenamefont {Seiler}(2010)}]{Seiler2010}%
  \BibitemOpen
  \bibfield  {author} {\bibinfo {author} {\bibfnamefont {W.~M.}\ \bibnamefont {Seiler}},\ }\href@noop {} {\emph {\bibinfo {title} {{Involution}}}},\ Vol.~\bibinfo {volume} {24}\ (\bibinfo  {publisher} {Springer Berlin Heidelberg},\ \bibinfo {address} {Berlin},\ \bibinfo {year} {2010})\BibitemShut {NoStop}%
\bibitem [{\citenamefont {Saha}\ \emph {et~al.}(2020)\citenamefont {Saha}, \citenamefont {Agudo-Canalejo},\ and\ \citenamefont {Golestanian}}]{Saha2020ScalarModel}%
  \BibitemOpen
  \bibfield  {author} {\bibinfo {author} {\bibfnamefont {S.}~\bibnamefont {Saha}}, \bibinfo {author} {\bibfnamefont {J.}~\bibnamefont {Agudo-Canalejo}},\ and\ \bibinfo {author} {\bibfnamefont {R.}~\bibnamefont {Golestanian}},\ }\bibfield  {title} {\bibinfo {title} {{Scalar active mixtures: The nonreciprocal Cahn-Hilliard model}},\ }\href {https://doi.org/10.1103/PhysRevX.10.041009} {\bibfield  {journal} {\bibinfo  {journal} {Phys. Rev. X.}\ }\textbf {\bibinfo {volume} {10}},\ \bibinfo {pages} {041009} (\bibinfo {year} {2020})}\BibitemShut {NoStop}%
\bibitem [{Note1()}]{Note1}%
  \BibitemOpen
  \bibinfo {note} {Supporting information [URL].}\BibitemShut {Stop}%
\bibitem [{Note2()}]{Note2}%
  \BibitemOpen
  \bibinfo {note} {Our definition differs from the often presented, ``textbook'' definition of degrees of freedom, which includes an additional degree of freedom corresponding to the system temperature. In principle, an arbitrary number of microscopic or external parameters can impact $\protect \mathbf {u}$ and $\protect \mathcal {G}$ and subsequently the phase behavior. Here, we avoid a subjective accounting of the dependencies of our state functions.}\BibitemShut {Stop}%
\bibitem [{\citenamefont {Langford}\ and\ \citenamefont {Omar}(2024)}]{Langford2024}%
  \BibitemOpen
  \bibfield  {author} {\bibinfo {author} {\bibfnamefont {L.}~\bibnamefont {Langford}}\ and\ \bibinfo {author} {\bibfnamefont {A.~K.}\ \bibnamefont {Omar}},\ }\bibfield  {title} {\bibinfo {title} {{Phase separation, capillarity, and odd surface flows in chiral active matter}},\ }\href {https://arxiv.org/abs/2408.14686} {\bibfield  {journal} {\bibinfo  {journal} {arXiv:2408.14686}\ } (\bibinfo {year} {2024})}\BibitemShut {NoStop}%
\bibitem [{\citenamefont {Zwanzig}(2001)}]{Zwanzig2001}%
  \BibitemOpen
  \bibfield  {author} {\bibinfo {author} {\bibfnamefont {R.}~\bibnamefont {Zwanzig}},\ }\href@noop {} {\emph {\bibinfo {title} {{Nonequilibrium Statistical Mechanics}}}}\ (\bibinfo  {publisher} {Oxford University Press},\ \bibinfo {address} {Oxford},\ \bibinfo {year} {2001})\BibitemShut {NoStop}%
\bibitem [{\citenamefont {Lindell}(1993)}]{Lindell1993}%
  \BibitemOpen
  \bibfield  {author} {\bibinfo {author} {\bibfnamefont {I.~V.}\ \bibnamefont {Lindell}},\ }\bibfield  {title} {\bibinfo {title} {{Delta function expansions, complex delta functions and the steepest descent method}},\ }\href {https://doi.org/10.1119/1.17238} {\bibfield  {journal} {\bibinfo  {journal} {Am. J. Phys.}\ }\textbf {\bibinfo {volume} {61}},\ \bibinfo {pages} {438} (\bibinfo {year} {1993})}\BibitemShut {NoStop}%
\bibitem [{\citenamefont {Lurie-Gregg}\ \emph {et~al.}(2014)\citenamefont {Lurie-Gregg}, \citenamefont {Schulte},\ and\ \citenamefont {Roundy}}]{Lurie2014}%
  \BibitemOpen
  \bibfield  {author} {\bibinfo {author} {\bibfnamefont {P.}~\bibnamefont {Lurie-Gregg}}, \bibinfo {author} {\bibfnamefont {J.~B.}\ \bibnamefont {Schulte}},\ and\ \bibinfo {author} {\bibfnamefont {D.}~\bibnamefont {Roundy}},\ }\bibfield  {title} {\bibinfo {title} {{Approach to approximating the pair distribution function of inhomogeneous hard-sphere fluids}},\ }\href {https://doi.org/https://doi.org/10.1103/PhysRevE.90.042130} {\bibfield  {journal} {\bibinfo  {journal} {Phys. Rev. E}\ }\textbf {\bibinfo {volume} {90}},\ \bibinfo {pages} {042130} (\bibinfo {year} {2014})}\BibitemShut {NoStop}%
\bibitem [{\citenamefont {De~Groot}\ and\ \citenamefont {Mazur}(2013)}]{DeGroot2013}%
  \BibitemOpen
  \bibfield  {author} {\bibinfo {author} {\bibfnamefont {S.~R.}\ \bibnamefont {De~Groot}}\ and\ \bibinfo {author} {\bibfnamefont {P.}~\bibnamefont {Mazur}},\ }\href@noop {} {\emph {\bibinfo {title} {{Non-equilibrium Thermodynamics}}}}\ (\bibinfo  {publisher} {Dover Publication Inc},\ \bibinfo {address} {New York},\ \bibinfo {year} {2013})\BibitemShut {NoStop}%
\bibitem [{\citenamefont {Kirkwood}\ and\ \citenamefont {Buff}(1951)}]{Kirkwood1951}%
  \BibitemOpen
  \bibfield  {author} {\bibinfo {author} {\bibfnamefont {J.~G.}\ \bibnamefont {Kirkwood}}\ and\ \bibinfo {author} {\bibfnamefont {F.~P.}\ \bibnamefont {Buff}},\ }\bibfield  {title} {\bibinfo {title} {{The statistical mechanical theory of solutions. I}},\ }\href {https://doi.org/10.1063/1.1748352} {\bibfield  {journal} {\bibinfo  {journal} {J. Chem. Phys}\ }\textbf {\bibinfo {volume} {19}},\ \bibinfo {pages} {774} (\bibinfo {year} {1951})}\BibitemShut {NoStop}%
\bibitem [{\citenamefont {Ben-Naim}(1974)}]{Ben1974}%
  \BibitemOpen
  \bibfield  {author} {\bibinfo {author} {\bibfnamefont {A.}~\bibnamefont {Ben-Naim}},\ }\bibfield  {title} {\bibinfo {title} {{Theory of Solutions}},\ }in\ \href@noop {} {\emph {\bibinfo {booktitle} {Water and Aqueous Solutions}}}\ (\bibinfo  {publisher} {Plenum Press},\ \bibinfo {address} {New York},\ \bibinfo {year} {1974})\ pp.\ \bibinfo {pages} {123--176}\BibitemShut {NoStop}%
\bibitem [{\citenamefont {Cheng}(2022)}]{Cheng2022}%
  \BibitemOpen
  \bibfield  {author} {\bibinfo {author} {\bibfnamefont {B.}~\bibnamefont {Cheng}},\ }\bibfield  {title} {\bibinfo {title} {{Computing chemical potentials of solutions from structure factors}},\ }\href {https://doi.org/10.1063/5.0107059} {\bibfield  {journal} {\bibinfo  {journal} {J. Chem. Phys.}\ }\textbf {\bibinfo {volume} {157}},\ \bibinfo {pages} {121101} (\bibinfo {year} {2022})}\BibitemShut {NoStop}%
\bibitem [{\citenamefont {Anderson}\ \emph {et~al.}(2020)\citenamefont {Anderson}, \citenamefont {Glaser},\ and\ \citenamefont {Glotzer}}]{Anderson2020}%
  \BibitemOpen
  \bibfield  {author} {\bibinfo {author} {\bibfnamefont {J.~A.}\ \bibnamefont {Anderson}}, \bibinfo {author} {\bibfnamefont {J.}~\bibnamefont {Glaser}},\ and\ \bibinfo {author} {\bibfnamefont {S.~C.}\ \bibnamefont {Glotzer}},\ }\bibfield  {title} {\bibinfo {title} {{HOOMD-blue: A Python package for high-performance molecular dynamics and hard particle Monte Carlo simulations}},\ }\href {https://doi.org/10.1016/j.commatsci.2019.109363} {\bibfield  {journal} {\bibinfo  {journal} {Comput. Mater. Sci.}\ }\textbf {\bibinfo {volume} {173}},\ \bibinfo {pages} {109363} (\bibinfo {year} {2020})}\BibitemShut {NoStop}%
\bibitem [{\citenamefont {Korteweg}(1904)}]{Korteweg1904}%
  \BibitemOpen
  \bibfield  {author} {\bibinfo {author} {\bibfnamefont {D.~J.}\ \bibnamefont {Korteweg}},\ }\bibfield  {title} {\bibinfo {title} {{Archives neerl}},\ }\href@noop {} {\bibfield  {journal} {\bibinfo  {journal} {Sci. Exacts. Nat}\ }\textbf {\bibinfo {volume} {6}} (\bibinfo {year} {1904})}\BibitemShut {NoStop}%
\bibitem [{\citenamefont {Yang}\ \emph {et~al.}(1976)\citenamefont {Yang}, \citenamefont {Fleming},\ and\ \citenamefont {Gibbs}}]{Yang1976}%
  \BibitemOpen
  \bibfield  {author} {\bibinfo {author} {\bibfnamefont {A.~J.}\ \bibnamefont {Yang}}, \bibinfo {author} {\bibfnamefont {P.~D.}\ \bibnamefont {Fleming}},\ and\ \bibinfo {author} {\bibfnamefont {J.~H.}\ \bibnamefont {Gibbs}},\ }\bibfield  {title} {\bibinfo {title} {{Molecular theory of surface tension}},\ }\href {https://doi.org/10.1063/1.432687} {\bibfield  {journal} {\bibinfo  {journal} {J. Chem. Phys.}\ }\textbf {\bibinfo {volume} {64}},\ \bibinfo {pages} {3732} (\bibinfo {year} {1976})}\BibitemShut {NoStop}%
\bibitem [{\citenamefont {{David Chandler}}(1987)}]{Chandler1987}%
  \BibitemOpen
  \bibfield  {author} {\bibinfo {author} \bibfnamefont {D.}~{\bibnamefont {{Chandler}}},\ }\href@noop {} {\emph {\bibinfo {title} {{Introduction to Modern Statistical Mechanics}}}}\ (\bibinfo  {publisher} {Oxford University Press},\ \bibinfo {address} {New York},\ \bibinfo {year} {1987})\BibitemShut {NoStop}%
\bibitem [{\citenamefont {Tjhung}\ \emph {et~al.}(2018)\citenamefont {Tjhung}, \citenamefont {Nardini},\ and\ \citenamefont {Cates}}]{Tjhung2018}%
  \BibitemOpen
  \bibfield  {author} {\bibinfo {author} {\bibfnamefont {E.}~\bibnamefont {Tjhung}}, \bibinfo {author} {\bibfnamefont {C.}~\bibnamefont {Nardini}},\ and\ \bibinfo {author} {\bibfnamefont {M.~E.}\ \bibnamefont {Cates}},\ }\bibfield  {title} {\bibinfo {title} {{Cluster phases and bubbly phase separation in active fluids: Reversal of the Ostwald process}},\ }\href {https://doi.org/https://doi.org/10.1103/PhysRevX.8.031080} {\bibfield  {journal} {\bibinfo  {journal} {Phys. Rev. X.}\ }\textbf {\bibinfo {volume} {8}},\ \bibinfo {pages} {031080} (\bibinfo {year} {2018})}\BibitemShut {NoStop}%
\bibitem [{\citenamefont {Cates}(2019)}]{Cates2019}%
  \BibitemOpen
  \bibfield  {author} {\bibinfo {author} {\bibfnamefont {M.~E.}\ \bibnamefont {Cates}},\ }\bibfield  {title} {\bibinfo {title} {{Active field theories}},\ }\href {https://arxiv.org/abs/1904.01330v1} {\bibfield  {journal} {\bibinfo  {journal} {arXiv:1904.01330}\ } (\bibinfo {year} {2019})}\BibitemShut {NoStop}%
\bibitem [{\citenamefont {Frohoff-H{\"{u}}lsmann}\ \emph {et~al.}(2023)\citenamefont {Frohoff-H{\"{u}}lsmann}, \citenamefont {Holl}, \citenamefont {Knobloch}, \citenamefont {Gurevich},\ and\ \citenamefont {Thiele}}]{Frohoff2023}%
  \BibitemOpen
  \bibfield  {author} {\bibinfo {author} {\bibfnamefont {T.}~\bibnamefont {Frohoff-H{\"{u}}lsmann}}, \bibinfo {author} {\bibfnamefont {M.~P.}\ \bibnamefont {Holl}}, \bibinfo {author} {\bibfnamefont {E.}~\bibnamefont {Knobloch}}, \bibinfo {author} {\bibfnamefont {S.~V.}\ \bibnamefont {Gurevich}},\ and\ \bibinfo {author} {\bibfnamefont {U.}~\bibnamefont {Thiele}},\ }\bibfield  {title} {\bibinfo {title} {{Stationary broken parity states in active matter models}},\ }\href {https://doi.org/10.1103/PhysRevE.107.064210} {\bibfield  {journal} {\bibinfo  {journal} {Phys. Rev. E.}\ }\textbf {\bibinfo {volume} {107}},\ \bibinfo {pages} {64210} (\bibinfo {year} {2023})}\BibitemShut {NoStop}%
\bibitem [{\citenamefont {Thiele}\ \emph {et~al.}(2019)\citenamefont {Thiele}, \citenamefont {Frohoff-H{\"{u}}lsmann}, \citenamefont {Engelnkemper}, \citenamefont {Knobloch},\ and\ \citenamefont {Archer}}]{Thiele2019}%
  \BibitemOpen
  \bibfield  {author} {\bibinfo {author} {\bibfnamefont {U.}~\bibnamefont {Thiele}}, \bibinfo {author} {\bibfnamefont {T.}~\bibnamefont {Frohoff-H{\"{u}}lsmann}}, \bibinfo {author} {\bibfnamefont {S.}~\bibnamefont {Engelnkemper}}, \bibinfo {author} {\bibfnamefont {E.}~\bibnamefont {Knobloch}},\ and\ \bibinfo {author} {\bibfnamefont {A.~J.}\ \bibnamefont {Archer}},\ }\bibfield  {title} {\bibinfo {title} {{First order phase transitions and the thermodynamic limit}},\ }\href {https://doi.org/10.1088/1367-2630/AB5CAF} {\bibfield  {journal} {\bibinfo  {journal} {New J. Phys.}\ }\textbf {\bibinfo {volume} {21}},\ \bibinfo {pages} {123021} (\bibinfo {year} {2019})}\BibitemShut {NoStop}%
\bibitem [{\citenamefont {Frohoff-H{\"{u}}lsmann}\ \emph {et~al.}(2021)\citenamefont {Frohoff-H{\"{u}}lsmann}, \citenamefont {Wrembel},\ and\ \citenamefont {Thiele}}]{Frohoff2021}%
  \BibitemOpen
  \bibfield  {author} {\bibinfo {author} {\bibfnamefont {T.}~\bibnamefont {Frohoff-H{\"{u}}lsmann}}, \bibinfo {author} {\bibfnamefont {J.}~\bibnamefont {Wrembel}},\ and\ \bibinfo {author} {\bibfnamefont {U.}~\bibnamefont {Thiele}},\ }\bibfield  {title} {\bibinfo {title} {{Suppression of coarsening and emergence of oscillatory behavior in a Cahn-Hilliard model with nonvariational coupling}},\ }\href {https://doi.org/10.1103/PhysRevE.103.042602} {\bibfield  {journal} {\bibinfo  {journal} {Phys. Rev. E.}\ }\textbf {\bibinfo {volume} {103}},\ \bibinfo {pages} {42602} (\bibinfo {year} {2021})}\BibitemShut {NoStop}%
\bibitem [{\citenamefont {Uecker}(2021)}]{Uecker2021}%
  \BibitemOpen
  \bibfield  {author} {\bibinfo {author} {\bibfnamefont {H.}~\bibnamefont {Uecker}},\ }\href@noop {} {\emph {\bibinfo {title} {{Numerical Continuation and Bifurcation in Nonlinear PDEs}}}}\ (\bibinfo  {publisher} {SIAM},\ \bibinfo {address} {Philadelphia},\ \bibinfo {year} {2021})\BibitemShut {NoStop}%
\bibitem [{\citenamefont {Sandler}(2017)}]{Sandler2017}%
  \BibitemOpen
  \bibfield  {author} {\bibinfo {author} {\bibfnamefont {S.~I.}\ \bibnamefont {Sandler}},\ }\href@noop {} {\emph {\bibinfo {title} {{Chemical, Biochemical, and Engineering Thermodynamics}}}},\ \bibinfo {edition} {5th}\ ed.\ (\bibinfo  {publisher} {John Wiley {\&} Sons},\ \bibinfo {address} {New York},\ \bibinfo {year} {2017})\BibitemShut {NoStop}%
\bibitem [{\citenamefont {Frenkel}\ and\ \citenamefont {Smit}(2001)}]{Frenkel2001}%
  \BibitemOpen
  \bibfield  {author} {\bibinfo {author} {\bibfnamefont {D.}~\bibnamefont {Frenkel}}\ and\ \bibinfo {author} {\bibfnamefont {B.}~\bibnamefont {Smit}},\ }\href@noop {} {\emph {\bibinfo {title} {{Understanding Molecular Simulation: From Algorithms to Applications}}}},\ \bibinfo {edition} {2nd}\ ed.\ (\bibinfo  {publisher} {Academic Press},\ \bibinfo {address} {San Diego},\ \bibinfo {year} {2001})\BibitemShut {NoStop}%
\bibitem [{\citenamefont {Chipot}\ and\ \citenamefont {Pohorille}(2007)}]{Chipot2007}%
  \BibitemOpen
  \bibfield  {author} {\bibinfo {author} {\bibfnamefont {C.}~\bibnamefont {Chipot}}\ and\ \bibinfo {author} {\bibfnamefont {A.}~\bibnamefont {Pohorille}},\ }\href@noop {} {\emph {\bibinfo {title} {{Free Energy Calculations: Theory and Applications in Chemistry and Biology}}}}\ (\bibinfo  {publisher} {Springer},\ \bibinfo {address} {Berlin, Heidelberg},\ \bibinfo {year} {2007})\BibitemShut {NoStop}%
\bibitem [{\citenamefont {Joshi}\ \emph {et~al.}(2022)\citenamefont {Joshi}, \citenamefont {Ray}, \citenamefont {Lemma}, \citenamefont {Varghese}, \citenamefont {Sharp}, \citenamefont {Dogic}, \citenamefont {Baskaran},\ and\ \citenamefont {Hagan}}]{Joshi2022}%
  \BibitemOpen
  \bibfield  {author} {\bibinfo {author} {\bibfnamefont {C.}~\bibnamefont {Joshi}}, \bibinfo {author} {\bibfnamefont {S.}~\bibnamefont {Ray}}, \bibinfo {author} {\bibfnamefont {L.~M.}\ \bibnamefont {Lemma}}, \bibinfo {author} {\bibfnamefont {M.}~\bibnamefont {Varghese}}, \bibinfo {author} {\bibfnamefont {G.}~\bibnamefont {Sharp}}, \bibinfo {author} {\bibfnamefont {Z.}~\bibnamefont {Dogic}}, \bibinfo {author} {\bibfnamefont {A.}~\bibnamefont {Baskaran}},\ and\ \bibinfo {author} {\bibfnamefont {M.~F.}\ \bibnamefont {Hagan}},\ }\bibfield  {title} {\bibinfo {title} {{Data-driven discovery of active nematic hydrodynamics}},\ }\href {https://doi.org/10.1103/PhysRevLett.129.258001} {\bibfield  {journal} {\bibinfo  {journal} {Phys. Rev. Lett.}\ }\textbf {\bibinfo {volume} {129}},\ \bibinfo {pages} {258001} (\bibinfo {year} {2022})}\BibitemShut {NoStop}%
\bibitem [{\citenamefont {Supekar}\ \emph {et~al.}(2023)\citenamefont {Supekar}, \citenamefont {Song}, \citenamefont {Hastewell}, \citenamefont {Choi}, \citenamefont {Mietke},\ and\ \citenamefont {Dunkel}}]{Supekar2023}%
  \BibitemOpen
  \bibfield  {author} {\bibinfo {author} {\bibfnamefont {R.}~\bibnamefont {Supekar}}, \bibinfo {author} {\bibfnamefont {B.}~\bibnamefont {Song}}, \bibinfo {author} {\bibfnamefont {A.}~\bibnamefont {Hastewell}}, \bibinfo {author} {\bibfnamefont {G.~P.}\ \bibnamefont {Choi}}, \bibinfo {author} {\bibfnamefont {A.}~\bibnamefont {Mietke}},\ and\ \bibinfo {author} {\bibfnamefont {J.}~\bibnamefont {Dunkel}},\ }\bibfield  {title} {\bibinfo {title} {{Learning hydrodynamic equations for active matter from particle simulations and experiments}},\ }\href {https://doi.org/10.1073/pnas.2206994120} {\bibfield  {journal} {\bibinfo  {journal} {Proc. Natl. Acad. Sci. U. S. A.}\ }\textbf {\bibinfo {volume} {120}},\ \bibinfo {pages} {e2206994120} (\bibinfo {year} {2023})}\BibitemShut {NoStop}%
\bibitem [{\citenamefont {Mandal}\ \emph {et~al.}(2024)\citenamefont {Mandal}, \citenamefont {Huang}, \citenamefont {Fruchart}, \citenamefont {Moerman}, \citenamefont {Vaikuntanathan}, \citenamefont {Murugan},\ and\ \citenamefont {Vitelli}}]{Mandal2024}%
  \BibitemOpen
  \bibfield  {author} {\bibinfo {author} {\bibfnamefont {R.}~\bibnamefont {Mandal}}, \bibinfo {author} {\bibfnamefont {R.}~\bibnamefont {Huang}}, \bibinfo {author} {\bibfnamefont {M.}~\bibnamefont {Fruchart}}, \bibinfo {author} {\bibfnamefont {P.~G.}\ \bibnamefont {Moerman}}, \bibinfo {author} {\bibfnamefont {S.}~\bibnamefont {Vaikuntanathan}}, \bibinfo {author} {\bibfnamefont {A.}~\bibnamefont {Murugan}},\ and\ \bibinfo {author} {\bibfnamefont {V.}~\bibnamefont {Vitelli}},\ }\bibfield  {title} {\bibinfo {title} {{Learning dynamical behaviors in physical systems}},\ }\href {https://arxiv.org/abs/2406.07856v1} {\bibfield  {journal} {\bibinfo  {journal} {arXiv:2406.07856}\ } (\bibinfo {year} {2024})}\BibitemShut {NoStop}%
\bibitem [{\citenamefont {Skamrahl}\ \emph {et~al.}(2023)\citenamefont {Skamrahl}, \citenamefont {Sch{\"{u}}nemann}, \citenamefont {Mukenhirn}, \citenamefont {Pang}, \citenamefont {Gottwald}, \citenamefont {Jipp}, \citenamefont {Ferle}, \citenamefont {R{\"{u}}beling}, \citenamefont {Oswald}, \citenamefont {Honigmann},\ and\ \citenamefont {Janshoff}}]{Skamrahl2023}%
  \BibitemOpen
  \bibfield  {author} {\bibinfo {author} {\bibfnamefont {M.}~\bibnamefont {Skamrahl}}, \bibinfo {author} {\bibfnamefont {J.}~\bibnamefont {Sch{\"{u}}nemann}}, \bibinfo {author} {\bibfnamefont {M.}~\bibnamefont {Mukenhirn}}, \bibinfo {author} {\bibfnamefont {H.}~\bibnamefont {Pang}}, \bibinfo {author} {\bibfnamefont {J.}~\bibnamefont {Gottwald}}, \bibinfo {author} {\bibfnamefont {M.}~\bibnamefont {Jipp}}, \bibinfo {author} {\bibfnamefont {M.}~\bibnamefont {Ferle}}, \bibinfo {author} {\bibfnamefont {A.}~\bibnamefont {R{\"{u}}beling}}, \bibinfo {author} {\bibfnamefont {T.~A.}\ \bibnamefont {Oswald}}, \bibinfo {author} {\bibfnamefont {A.}~\bibnamefont {Honigmann}},\ and\ \bibinfo {author} {\bibfnamefont {A.}~\bibnamefont {Janshoff}},\ }\bibfield  {title} {\bibinfo {title} {{Cellular segregation in cocultures is driven by differential adhesion and contractility on distinct timescales}},\ }\href {https://doi.org/10.1073/pnas.2213186120} {\bibfield  {journal} {\bibinfo  {journal} {Proc. Natl. Acad. Sci. U. S. A.}\
  }\textbf {\bibinfo {volume} {120}},\ \bibinfo {pages} {e2213186120} (\bibinfo {year} {2023})}\BibitemShut {NoStop}%
\bibitem [{\citenamefont {Bhattacharyya}\ and\ \citenamefont {Yeomans}(2024)}]{Bhattacharyya2024}%
  \BibitemOpen
  \bibfield  {author} {\bibinfo {author} {\bibfnamefont {S.}~\bibnamefont {Bhattacharyya}}\ and\ \bibinfo {author} {\bibfnamefont {J.~M.}\ \bibnamefont {Yeomans}},\ }\bibfield  {title} {\bibinfo {title} {{Phase ordering in binary mixtures of active nematic fluids}},\ }\href {https://arxiv.org/abs/2404.19061} {\bibfield  {journal} {\bibinfo  {journal} {arXiv:2404.19061}\ } (\bibinfo {year} {2024})}\BibitemShut {NoStop}%
\end{thebibliography}
\end{document}


\title{Supporting Information -- Theory of Nonequilibrium Multicomponent Coexistence}

\author{Yu-Jen Chiu}
\thanks{These authors contributed equally to this work}
\affiliation{Department of Materials Science and Engineering, University of California, Berkeley, California 94720, USA}

\author{Daniel Evans}
\thanks{These authors contributed equally to this work}
\affiliation{Department of Materials Science and Engineering, University of California, Berkeley, California 94720, USA}
\affiliation{Materials Sciences Division, Lawrence Berkeley National Laboratory, Berkeley, California 94720, USA}

\author{Ahmad K. Omar}
\email{aomar@berkeley.edu}
\affiliation{Department of Materials Science and Engineering, University of California, Berkeley, California 94720, USA}
\affiliation{Materials Sciences Division, Lawrence Berkeley National Laboratory, Berkeley, California 94720, USA}

\maketitle

\tableofcontents

\section{Recovery of Equilibrium Coexistence Criteria for Passive Particles with Pairwise Interactions}
\label{sec:equilibrium_recovery}
In this section, we derive the exact form of the effective body forces for a multicomponent system of passive particles with isotropic pairwise interactions. 
We then formally expand this expression with respect to species density gradients. 
By noting the relation between the effective body forces and chemical potential gradients, we obtain a microscopic form of the species chemical potentials.
Comparing this form to that obtained through functional differentiation of a multicomponent ``square-gradient'' free energy functional, we are able to obtain a microscopic force-based expressions for the Hessian of the bulk free energy and the phenomenological square-gradient coefficients.
The consistency of the two forms of the chemical potentials requires that certain microscopic relations hold. 
We explore the validity of these relations numerically using particle-based Brownian dynamics simulations. 
Additionally, we corroborate the force-based expressions we derive for the Hessian matrix of the bulk free energy by comparing it with the Kirkwood-Buff Hessian matrix and the virial pressure expression.

\subsection{Species Mechanics: Exact Form}
\label{sec:species_mechanics_exact}

Our theory begins with the static mechanical balance for each species. 
Exact expressions for each term obtained in the mechanical balance can be derived beginning from the microscopic equations of motion and performing an Irving-Kirkwood procedure~\cite{Irving1950}. 
We consider a system of $n_c$ species with $N_i$ particles of species $i$ such that the total number of particles is $N = \sum_{i}^{n_c}N_i$. 
Particles interact through isotropic conservative pairwise forces and are maintained at a constant temperature through contact with a thermal bath. 
Microscopically, while the dynamical details of how the system is maintained at a constant temperature can vary, these details do not impact the equilibrium properties of the system.
We can thus thermostat our system by choosing the most convenient equation of motion that will result in an equilibrium distribution of microstates. 
We therefore introduce a fictitious local bath that exerts nonconservative drag forces and stochastic Brownian forces on our particles. 
We further neglect particle inertia to arrive at the following overdamped Langevin equation:
\begin{equation}
    \label{eq:langevin}
    \dot{\mathbf{r}}^\alpha_{i} = \frac{1}{\zeta_i} \left[\sum_j^{n_c} \sum_{\beta\neq \alpha}^{N_j}\mathbf{F}_{ij}(\mathbf{r}^\alpha_i - \mathbf{r}^\beta_j;t) + \mathbf{F}_{s}^\alpha(t) \right],
\end{equation}
where $\mathbf{\dot{r}}^\alpha_i$ ($\mathbf{r}^\alpha_i$) is the velocity (position) of the $\alpha$th particle of species $i$, $\zeta_i$ is the drag coefficient of species $i$, $\mathbf{F}_{ij}(\mathbf{r}^\alpha_i - \mathbf{r}^\beta_j;t) \equiv - \partial U_{ij} / \partial (\mathbf{r}^\alpha_i - \mathbf{r}^\beta_j)$ is the pairwise interaction force between particle $\alpha$ (at $\mathbf{r}^\alpha_i$) and $\beta$ (at $\mathbf{r}^\beta_j$) of species $i$ and $j$, respectively. 
The stochastic Brownian force satisfies the fluctuation-dissipation theorem with a mean of $\langle\mathbf{F}_{s}^\alpha(t) \rangle = \mathbf{0}$ and variance of $\langle \mathbf{F}_{s}^\alpha(t)\mathbf{F}_{s}^\beta(t')\rangle = 2k_BT \zeta_i \delta^{\alpha\beta}\delta(t-t')\mathbf{I}$, where $\delta (t-t')$ is the Dirac delta function, $\delta^{\alpha\beta}$ is the Kronecker delta function, and $\mathbf{I}$ is the identity tensor.

$f_N(\boldsymbol{\Gamma}; t)$ is the $N$-body probability density of finding the system in microstate $\boldsymbol{\Gamma}$ at time $t$.
From Eq.~\eqref{eq:langevin} we see that our microstate is simply the set of all $N$ particle positions with $\boldsymbol{\Gamma} = \mathbf{r}^N$.  
For our stochastic system, the distribution satisfies a Fokker-Planck equation with $\partial f_N / \partial t = \mathcal{L} f_N$ where $\mathcal{L}$ is the Fokker-Planck operator consistent with Eq.~\eqref{eq:langevin}~\cite{Zwanzig2001}:
\begin{equation}
    \mathcal{L} = \sum_i^{n_c}\sum_\alpha^{N_j} \left[ \frac{\partial}{\partial \mathbf{r}_i^\alpha}  \cdot \left(-\frac{1}{\zeta_i} \sum_j^{n_c}\sum_{\beta\neq\alpha}^{N_j} \mathbf{F}_{ij}(\mathbf{r}^\alpha_i - \mathbf{r}^\beta_j) + \frac{k_BT}{\zeta_i} \frac{\partial}{\partial \mathbf{r}_i^\alpha} \right)\right].
\end{equation}
We can now obtain the exact evolution equation for the statistics of any microscopic quantity, $\hat{\mathcal{O}}(\boldsymbol{\Gamma})$. 
To obtain the evolution equation for the ensemble average, $\mathcal{O}(t) \equiv \langle \hat{\mathcal{O}}\rangle \equiv \int_\gamma d \boldsymbol{\Gamma} f_N(\boldsymbol{\Gamma}; t)\hat{\mathcal{O}(\boldsymbol{\Gamma}})$ (where here the phase space volume is $\gamma = V^N$ and $V$ is the system volume), we can make use of the adjoint $\mathcal{L}^{\dagger}$ to our Fokker-Planck operator with $\partial \mathcal{O} / \partial t = \int_\gamma d\boldsymbol{\Gamma}f_N \mathcal{L}^{\dagger}\hat{\mathcal{O}}$~\cite{Zwanzig2001}.
The exact form of the evolution equation for the average species density field $\rho_i(\mathbf{x}; t) = \langle \sum_{\alpha = 1}^{N_i} \delta(\mathbf{x} - \mathbf{r}_i^\alpha)\rangle$ can now be obtained with: 
\begin{equation}
\label{eq:IKcontinuity}
\frac{\partial}{\partial t}\rho_i = - \boldsymbol{\nabla}\cdot \mathbf{J}_i,
 \end{equation}   
where $\boldsymbol{\nabla}\equiv \partial / \partial \mathbf{x}$ and we have made use of the following property of Dirac delta functions,~$\partial \delta (x - y)/ \partial y = - \partial \delta (x - y)/ \partial x$.
The species flux takes the following exact form:
\begin{equation}
\label{eq:IKflux}
    \mathbf{J}_i = \frac{1}{\zeta_i}\int_\gamma d\boldsymbol{\Gamma}f_N\left[ \sum_\alpha^{N_i} \left(\sum_j^{n_c}\sum_{\beta\neq\alpha}^{N_j}\mathbf{F}_{ij}(\mathbf{r}^\alpha_i - \mathbf{r}^\beta_j) + k_BT\frac{\partial}{\partial\mathbf{r}_{i}^\alpha}\right) \delta(\mathbf{x} - \mathbf{r}_i^\alpha) \right],
\end{equation}
We can again exploit the properties of the delta function to convert derivatives of particle positions to derivatives with respect to the spatial field position. 
Doing so and rearranging Eq.~\eqref{eq:IKflux} results in:
\begin{equation}
\label{eq:IKoverdampedmech}
    \mathbf{0} = - k_BT\boldsymbol{\nabla}\rho_i + \sum_j^{n_c} \int_\gamma d\boldsymbol{\Gamma}f_N \sum_\alpha^{N_i} \sum_{\beta\neq\alpha}^{N_j} \mathbf{F}_{ij} (\mathbf{r}^\alpha_i - \mathbf{r}^\beta_j) \delta(\mathbf{x} - \mathbf{r}_i^\alpha) - \zeta_i \mathbf{J}_i.
\end{equation}

As discussed in the main text and Ref.~\cite{Omar2023}, solving for stationary states requires solving the static mechanical balance. 
In the overdamped limit, the momentum density is ill-defined, but we can still see the emergence of the overdamped mechanical balance (which should take the form of $\mathbf{0} = \boldsymbol{\nabla} \cdot \boldsymbol{\sigma}_i + \mathbf{b}_i$) in Eq.~\eqref{eq:IKoverdampedmech}. 
The divergence of the ideal gas contribution should appear in the mechanical balance as $\boldsymbol{\nabla} \cdot (-k_BT \rho_i \mathbf{I})$ which we have indeed recovered and a finite flux would result in a drag force density $-\zeta_i \mathbf{J}_i$ which we also have obtained. 
The term arising from the particle interactions, in general, will result in terms that can be expressed as a stress and others that can only be expressed as body forces. 
Here, we will not provide the distinct forms of the stress as we are ultimately interested in combining these contributions into an effective body force. 
We can reexpress the interaction term appearing in our mechanical balance by introducing the two-body correlation function, defined as~\cite{Chandler1987}:
\begin{equation}
f_2^{ij} (\mathbf{r}_1, \mathbf{r}_2) = 
\Bigg\{
\begin{array}{cc}
     &  N_i N_j \int (d\mathbf{r}_i)^{N_i -1} \int (d\mathbf{r}_j)^{N_j -1} \prod_{k \neq i,j}^{n_c} \int d(\mathbf{r}_k)^{N_k} f_N, \; \text{if} \; i \neq j\\
     & N_i (N_i -1) \int (d\mathbf{r}_i)^{N_i -2} \prod_{k \neq i}^{n_c} \int d(\mathbf{r}_k)^{N_k} f_N, \; \text{if} \; i = j
\end{array},
\end{equation}
where $f_2^{ij}(\mathbf{r}_1, \mathbf{r}_2)$ is the joint distribution function of finding a particle of species $i$ at position $\mathbf{r}_1$ and a particle of species $j$ at $\mathbf{r}_2$.
Here, we have adopted the notation of $\int d(\mathbf{r}_k)^{N_k} \equiv \int_V d \mathbf{r}_k^{1} \int_V d \mathbf{r}_k^{2} \cdots \int_V d \mathbf{r}_k^{N_k} $ and $\int d(\mathbf{r}_k)^{N_k - 1} \equiv \int_V d \mathbf{r}_k^{1} \int_V d \mathbf{r}_k^{2} \cdots \int_V d \mathbf{r}_k^{N_k -1}$ as products of $N_k$ and $N_k - 1$ integrals of species $k$, respectively.
In defining this distribution, we use the fact that all particle pairs consistent with the desired species are equivalent, and we can thus integrate out the positions of all particles except for a single pair. 
We can now express the average of any pairwise property $a_{ij}^{\alpha\beta}$ using either distribution with $\int_\gamma d \boldsymbol{\Gamma} f_N \sum_\alpha^{N_i} \sum_{\beta\neq\alpha}^{N_j} a^{\alpha\beta}_{ij} \equiv \int d\mathbf{r}_1 d\mathbf{r}_2 f_2^{ij} a_{ij}^{12}$.
We now invoke the definition of the dimensionless pair-correlation function, $g_{ij}(\mathbf{r}_1, \mathbf{r}_2)$, with~\cite{Chandler1987}:
\begin{equation}
\label{eq:pair_distribution}
    f_2^{ij} (\mathbf{r}_1,\mathbf{r}_2) = \rho_i(\mathbf{r}_1) \rho_j( \mathbf{r}_2) g_{ij}(\mathbf{r}_1, \mathbf{r}_2),
\end{equation}
where $\rho_i(\mathbf{r}_1)$ [$\rho_j(\mathbf{r}_2)$] is the density of species $i$ [$j$] evaluated at position $\mathbf{r}_1$ [$\mathbf{r}_2$].
Substituting these definitions into our mechanical balance, we arrive at the following:  
\begin{equation}
    \label{eq:beff}
    \mathbf{0} = \boldsymbol{\nabla} \cdot \boldsymbol{\sigma}_i + \mathbf{b}_i
    \equiv - k_BT \boldsymbol{\nabla}\rho_i(\mathbf{x}) + \rho_i(\mathbf{x}) \sum_{j}^{n_c} \int d \mathbf{x}' \rho_j(\mathbf{x}') g_{ij}(\mathbf{x},\mathbf{x}') \mathbf{F}_{ij}(\mathbf{x}-\mathbf{x}') - \zeta_i\mathbf{J}_i,
\end{equation}
where the integral over $\mathbf{r}_1$ has been eliminated by the Dirac delta function in Eq.~\eqref{eq:IKoverdampedmech} and, for notional convenience, we use $\mathbf{x}'$ in place of $\mathbf{r}_2$.
For flux-free states, we finally arrive at the static momentum balance and the exact form of the effective body force: 
\begin{equation}
    \mathbf{0} = \boldsymbol{\nabla} \cdot \boldsymbol{\sigma}_i + \mathbf{b}_i \equiv \mathbf{b}_i^{\rm eff} = - k_BT \boldsymbol{\nabla}\rho_i(\mathbf{x}) + \rho_i(\mathbf{x}) \sum_{j}^{n_c} \int d \mathbf{x}' \rho_j(\mathbf{x}') g_{ij}(\mathbf{x},\mathbf{x}') \mathbf{F}_{ij}(\mathbf{x}-\mathbf{x}').
\end{equation}

\subsection{Species Mechanics: Gradient Expansion}
\label{sec:species_mechanics_expansion}

With an exact form for the effective body force acting on each species, $\mathbf{b}_i^{\rm eff}(\mathbf{x})$, we can now formally perform the gradient expansion required for our theory.
In alignment with our coexistence theory, we consider a quasi-1D system where variations occur only along the $z$-direction with translational invariance in the remaining two dimensions. 
Consequently, the z-component of the exact effective body force now simplifies to:
\begin{equation}
\label{eq:beff_z}
    \frac{d}{dz}\sigma_i +b_i = - k_BT \frac{d}{dz} \rho_i (z)-\rho_i(z) \sum_{j}^{n_c}\int d x' \int d y' \int d z' \rho_j(z') g_{ij}(z,z', \Delta x, \Delta y) F_{ij}(\Delta x, \Delta y, \Delta z),
\end{equation}
where $\Delta z = z'-z$, $\Delta x = x' - x$, and $\Delta y = y'-y$ represent the relative distances in three dimensions with ($x$, $y$, $z$) as the position at which we are evaluating the mechanical balance.
Here, $F_{ij}(\Delta x, \Delta y, \Delta z)$ is the $z$ component of the interaction force.
The translational invariance in the $x$ and $y$ directions results in the pair distribution function only depending on $\Delta x$ and $\Delta y$, while variations in the $z$ dimension result in an explicit dependence of $g_{ij}$ on $z$ and $z'$.
For simplicity, we only denote the $z$ dependence of the species densities as they are invariant in $x$ and $y$ [making their choice arbitrary in evaluating Eq.~\eqref{eq:beff_z}].

We now look to express the integral arising in our effective body force as a gradient expansion of the species densities at the local position $z$.
We can justify such an expansion in the event that the kernel appearing in Eq.~\eqref{eq:beff_z}, the product of the pairwise force and pair distribution function, peaks at a location near $z$ and quickly vanishes as $|\Delta z|$ approaches some microscopic distance~\cite{Lindell1993}.
These requirements of a highly localized and quickly vanishing kernel are both met when interactions are short-ranged.
While the kernel peak will be at some finite $\Delta z$ (such as the mean diameter of the interacting particles), we nevertheless perform all expansions about $|\Delta z| = 0$. 
We can then express the density of species $j$ at position $z'$ solely using features of the density at position $z$ with: 
\begin{equation}
\label{eq:density_expansion}
    \rho_j(z') = \sum_{a=0}^{\infty}\frac{\Delta z^a}{a!}\frac{d^a\rho_j}{dz^a},
\end{equation}
where $\rho_j \equiv \rho_j (z)$.

\sloppy Similarly, we can expand $g_{ij}(z,z', \Delta x, \Delta y)$ with respect to species density gradients at position $z$.
Before proceeding, it should be noted that a pair distribution function (which is simply a marginalization of the $N$-body distribution as discussed in Sec.~\ref{sec:species_mechanics_exact}) does not have species densities as natural variables. 
However, we can associate the pair distribution functions of \textit{uniform systems} with species densities $\{\rho_m\}$ as $g_{ij}^0(\Delta z, \Delta x,\Delta y;\{\rho_m\})\equiv g_{ij}^0 (\{\rho_m\})$.
For a uniform system with translational invariance, the pair distribution function solely depends on the relative distances. 
We now construct the pair distribution function of \textit{inhomogeneous systems}, $g_{ij}(z,z',\Delta x,\Delta y)$, using the spatially uniform pair distribution function but evaluated at the density of interest.
The question arises: when approximating $g_{ij}$ for an inhomogeneous system with $g_{ij}^0$, should one evaluate $g_{ij}^0$ at the densities at $z$ or $z'$ or some combination of the two? 
There are three commonly employed approximations~\cite{Lurie2014}: the midpoint approximation ${g_{ij} (z,z',\Delta x,\Delta y) = g_{ij}^0(\{\rho_m([z'+z]/2)\})}$; the mean density approximation ${g_{ij} (z,z',\Delta x,\Delta y) = g_{ij}^0( (\{\rho_m (z)\}+\{\rho_m (z')\})/2)}$; and the mean function approximation ${g_{ij} (z,z',\Delta x,\Delta y) = \left[g_{ij}^0(\{\rho_m(z)\}) + g_{ij}^0(\{\rho_m(z')\})\right]/2}$. 
Exploring all three approaches, we find that both the mean density and mean function approximations result in the same expansion and eventually recover the correct form of the chemical potentials, in contrast to the midpoint approximation.
The origins of this intriguing discrepancy between these approximations are not immediately obvious and warrant further investigation.

For convenience, we will use the mean function approach to approximate our inhomogeneous pair distribution function. 
To eliminate the dependence on the nonlocal species densities,  $\{\rho_m(z')\}$, we first expand $g_{ij}^0(\{\rho_m(z')\})$ to first-order in species densities resulting in:
\begin{subequations}
\label{eq:g_expansion}
\begin{align}
    g_{ij} (z,z',\Delta x,\Delta y) & \equiv \frac{1}{2}\left[g_{ij}^0(\{\rho_m(z)\}) + g_{ij}^0(\{\rho_m(z')\})\right] \\
    & \approx  \frac{1}{2} g_{ij}^0(\{\rho_m(z)\}) + \frac{1}{2}\left(g_{ij}^0(\{\rho_m(z)\}) + \sum_k^{n_c} (\rho_k(z')-\rho_k(z))\frac{d g_{ij}^0}{d \rho_k} \Bigg|_{\{\rho_m(z)\}}\right)\\
    & = g_{ij}^0 + \frac{1}{2} \sum_k^{n_c} (\rho_k(z')-\rho_k(z))\frac{d g_{ij}^0}{d \rho_k}\Bigg|_{\{\rho_m(z)\}},
\end{align}
\end{subequations}
where $g_{ij}^0 \equiv g_{ij}^0(\{\rho_m\} = \{\rho_m(z)\})$. 
We can again invoke our expansion for $\rho_k(z')$ [Eq.~\eqref{eq:density_expansion}] to arrive at our expanded form of $\rho_j(z') g_{ij}(z,z')$:
\begin{widetext}
\begin{equation}
\label{eq:full_rho_g_expansion}
    \rho_j (z') g_{ij}(z,z',\Delta x,\Delta y) \approx \left(g_{ij}^0 + \frac{1}{2}\sum_{k}^{n_c}\left[ \frac{d g_{ij}^0}{d \rho_k}\left(\sum_{b=1}^{\infty}\frac{(\Delta z)^b}{b!}\frac{d^b \rho_k}{dz^b}\right)\right]\right)\left(\sum_{a=0}^{\infty}\frac{(\Delta z)^a}{a!}\frac{d^a\rho_j}{dz^a}\right).
\end{equation}
\end{widetext}
Substituting Eq.~\eqref{eq:full_rho_g_expansion} into our exact expression for the body force [Eq.~\eqref{eq:beff_z}] and truncating terms beyond third-order in gradients of the species densities, we arrive at the desired form (see main text) for the expanded effective body force with: 
\begin{subequations}
\label{eq:microscopic_beff}
    \begin{align}
        & b^{\rm eff} \approx \sum_j^{n_c} b^{(1,1)}_{ij}  \partial_z \rho_j - \sum_j^{n_c}\sum_k^{n_c}\sum_l^{n_c}b^{(3,1)}_{ijkl} \partial_z \rho_j \partial_z \rho_k \partial_z \rho_l - \sum_j^{n_c}\sum_k^{n_c}b^{(3,2)}_{ijk} \partial^2_{zz} \rho_j \partial_z \rho_k - \sum_j^{n_c} b^{(3,3)}_{ij} \partial^3_{zzz} \rho_j ,\label{eq:microscopic_beff_a}\\
        & b_{ij}^{(1,1)} = -k_BT\delta_{ij} - \rho_i \sum_k^{n_c}\int dx' \int dy' \int dz' \Delta z \left(F_{ij}g^{0}_{ij} + \frac{1}{2} F_{ik}\rho_k \frac{\partial g_{ik}^0}{\partial\rho_j}\right), \\
        & b_{ijkl}^{(3,1)} = 0, \\
        & b_{ijk}^{(3,2)} = \rho_i \int dx' \int dy' \int dz' \frac{1}{4} (\Delta z)^3 \left(F_{ij} \frac{\partial g_{ij}^0}{\partial \rho_k} + F_{ik} \frac{\partial g_{ik}^0}{\partial \rho_j}\right), \\
        & b_{ij}^{(3,3)} = \rho_i \int dx' \int dy' \int dz' \frac{1}{6} (\Delta z)^3 \left(F_{ij} g_{ij}^0 +\frac{1}{2} \sum_k^{n_c} F_{ik}\rho_k \frac{\partial g_{ik}^0}{\partial\rho_j}\right),
    \end{align}
\end{subequations}
where we have defined $\partial_z \equiv \partial / \partial z$, $\partial^2_{zz} \equiv \partial^2 / \partial z^2$, and $\partial^3_{zzz} \equiv \partial^3 / \partial z^3$.
We note that the even gradient terms were omitted in Eq.~\eqref{eq:microscopic_beff_a} on the basis that the effective body force should be odd with respect to spatial inversion. 
Including the second-order terms and deriving the microscopic expression for their coefficients, we found, as anticipated, that the coefficients are identically zero. 

\subsection{Microscopic Form of Chemical Potential and Symmetry Requirements}
\label{sisec:symmetry_derivation}
For passive systems in equilibrium, we can use thermodynamics as an alternative route to obtaining effective body forces. 
We can then compare this thermodynamic approach with the statistical mechanical approach of the previous section, allowing us to both check the microscopic expressions presented in Eq.~\eqref{eq:microscopic_beff} and relate them to thermodynamic quantities. 
The thermodynamic route is rooted in the fact that the static effective body force experienced by a species is simply:
\begin{equation}
\label{eq:eq_body_force}
    \mathbf{b}_i^{\rm eff} = -\rho_i \boldsymbol{\nabla}\mu_i,
\end{equation}
where $-\boldsymbol{\nabla}\mu_i$ represents the local forces experienced by particles of species $i$ (this is precisely the force that is found in irreversible thermodynamics~\cite{DeGroot2013}) and the species density prefactor ensures that $-\rho_i \boldsymbol{\nabla}\mu_i$ is indeed a force density. 
We recognize that the sum of the species mechanical balances will result in the total system mechanical balance, $\boldsymbol{\nabla}\cdot \boldsymbol{\sigma} = \sum_i^{n_c}\mathbf{b}_i^{\rm eff}$, where $\boldsymbol{\sigma}$ is the total system stress.
In the absence of any external or active forces, the static mechanical balance of the system contains no body forces.
If we consider isotropic systems and neglect corrections to the stress due to spatial gradients, we have $\boldsymbol{\sigma} = -P^{\rm bulk}\mathbf{I}$ where $P^{\rm bulk}$ is the thermodynamic pressure. 
We can recognize that our proposed thermodynamic species body force expression [Eq.~\eqref{eq:eq_body_force}] is consistent with this mechanical expectation by invoking the Gibbs-Duhem relation, $dP^{\rm bulk} = \sum_i^{n_c}\rho_i d\mu^{\rm bulk}_i$. 

We now proceed to derive the thermodynamic form of the expanded body force.
The chemical potential of an inhomogeneous system can be found by taking a functional derivative of an appropriate free energy functional.  
To obtain an expanded body force that is entirely local and contains terms up to third-order in spatial gradients of the species densities, we consider a multicomponent ``square-gradient'' free energy functional: 
\begin{equation}
    \label{eq:eq_free_energy_functional}
    F[\{\rho_m\}] = \int_V d\mathbf{x} \left( f^{\rm bulk}(\{\rho_m\})+\frac{1}{2}\sum_i^{n_c}\sum_j^{n_c}\kappa_{ij}(\{\rho_m\})\boldsymbol{\nabla}\rho_i\cdot\boldsymbol{\nabla}\rho_j \right),
\end{equation}
where $f^{\rm bulk}$ is the free energy density of a uniform system and $\kappa_{ij}$ is the $ij$ component of a (positive) symmetric tensor of state functions that penalizes the formation of spatial density gradients.
The equilibrium species chemical potential is simply $\mu_i = \delta F / \delta \rho_i$, which in our case results in: 
\begin{equation}
\label{eq:equil_chemical_potential_kappa}
    \mu_i = \mu_i^{\rm bulk} - \sum_{j}^{n_c} \sum_{k}^{n_c}\frac{1}{2} \frac{\partial \kappa_{ij}}{\partial \rho_{k}}\boldsymbol{\nabla}\rho_j\boldsymbol{\nabla}\rho_k - \sum_{j}^{n_c}  \kappa_{ij} \boldsymbol{\nabla}^2\rho_j,
\end{equation}
where $\mu_i^{\rm bulk} = \frac{\partial f^{\rm bulk}}{\partial\rho_i}$ and we have used the symmetry of $\partial \kappa_{ij} / \partial \rho_k$ upon exchange of indices $i$, $j$, and $k$ from its definition in terms of the direct correlation function (See Chapter 6.2 in~\cite{Hansen2013}).
We note that the chemical potential naturally arises when extremizing the free energy functional with respect to the density profiles subject to the conservation of particle number. 
This results in the condition of uniform chemical potential, consistent with equilibrium expectations and our expectation that the species body forces are identically zero. 

By substituting Eq.~\eqref{eq:equil_chemical_potential_kappa} into Eq.~\eqref{eq:eq_body_force} and focusing on the $z$-dimension, we arrive at the following form for the equilibrium expanded force:
\begin{multline}
\label{eq:thermo_beff}
    b^{\rm eff}_i = - \rho_i \Bigg[\sum_j^{n_c} \frac{\partial\mu_i^{\rm bulk}}{\partial\rho_j}{\nabla}\rho_j - \sum_{j}^{n_c} \sum_{k}^{n_c} \sum_{l}^{n_c} \frac{1}{2} \frac{\partial^2 \kappa_{ij}}{\partial\rho_k\partial\rho_l} {\nabla}\rho_j {\nabla}\rho_k {\nabla}\rho_l\\
    - \sum_j^{n_c} \sum_k^{n_c} 2 \frac{\partial\kappa_{ij}}{\partial\rho_k} {\nabla}^2\rho_j {\nabla}\rho_k - \sum_j^{n_c} \kappa_{ij} {\nabla}^3\rho_j\Bigg].
\end{multline}
Equating the microscopic body force found from our Irving-Kirkwood procedure [Eq.~\eqref{eq:microscopic_beff}] and that obtained from thermodynamics [Eq.~\eqref{eq:thermo_beff}] allows us to identify:
\begin{subequations}
\label{eq:kappa_relation}
    \begin{align}
        & \frac{\partial\mu_i^{\rm bulk}}{\partial\rho_j} \equiv H_{ij}^{\rm bulk} = k_BT/\rho_i \delta_{ij} + \sum_{k}^{n_c} \int dx' \int dy' \int d z' (\Delta z) \left( F_{ij}g_{ij}^0 +\frac{1}{2} F_{ik}\rho_k\frac{\partial g_{ik}^0}{\partial \rho_j} \right), \label{eq:hessian_expression} \\
        & \kappa_{ij} = - \int dx' \int dy' \int d z' (\Delta z)^3 \frac{1}{6}  \left( F_{ij} g_{ij}^0 +\frac{1}{2}\sum_{k}^{n_c} F_{ik}\rho_k\frac{\partial g_{ik}^0}{\partial \rho_j} \right), \label{eq:kappa_relation_a}\\ 
        & \frac{\partial\kappa_{ij}}{\partial\rho_k} = - \int dx' \int dy' \int d z'(\Delta z)^3  \frac{1}{8}\left( F_{ij}\frac{\partial g_{ij}^0}{\partial \rho_k}+ F_{ik}\frac{\partial g_{ik}^0}{\partial \rho_j} \right)\label{eq:kappa_relation_b}, 
    \end{align}
\end{subequations}
where $H_{ij}^{\rm bulk}$ is the Hessian matrix of the bulk free energy density (i.e.,~$H_{ij}^{\rm bulk} = \partial^2 f^{\rm bulk}/\partial \rho_i \partial \rho_j$).
We emphasize that the expressions provided above appear to retain a dependence on $z$ but this spatial dependence is entirely the result of any spatial variation of the species density. 
It is entirely appropriate to consider these expressions as state functions that solely depend on the species densities.

To the best of our knowledge, the proposed microscopic expressions for both $H_{ij}^{\rm bulk}$ and $\kappa_{ij}$ in Eq.~\eqref{eq:kappa_relation} are not directly found in the literature. 
We therefore require independent checks that our expressions are indeed consistent with thermodynamic expectations.
As a first check, we expect that both the bulk Hessian matrix and the interfacial coefficient tensor are symmetric.
The symmetry of the Hessian is ensured if:
\begin{equation}
\label{sieq:hessian_symmetry_int}
    \int dx' \int dy' \int d z' (\Delta z) \sum_{k}^{n_c}\left(  F_{ik}\rho_k\frac{\partial g_{ik}^0}{\partial \rho_j} \right)
    = \int dx' \int dy' \int d z' (\Delta z) \sum_{k}^{n_c} \left(  F_{jk}\rho_k\frac{\partial g_{jk}^0}{\partial \rho_i} \right),
\end{equation}
while the symmetry of $\kappa_{ij}$ is ensured if: 
\begin{equation}
\label{sieq:kappa_symmetry_int}
   \int dx' \int dy' \int d z' (\Delta z)^3 \sum_{k}^{n_c}\left(  F_{ik}\rho_k\frac{\partial g_{ik}^0}{\partial \rho_j} \right)
    = \int dx' \int dy' \int d z' (\Delta z)^3 \sum_{k}^{n_c}\left(  F_{jk}\rho_k\frac{\partial g_{jk}^0}{\partial \rho_i} \right).
\end{equation}
We further note that for Eq.~\eqref{eq:kappa_relation_b} to be consistent with Eq.~\eqref{eq:kappa_relation_a}, the following relation must hold:
\begin{equation}
    \label{eq:sym_requirement1}
\int dx' \int dy' \int dz' (\Delta z)^3 F_{ij}\frac{\partial g_{ij}}{\partial\rho_k} = \int dx' \int dy' \int dz' (\Delta z)^3 F_{ik}\frac{\partial g_{ik}}{\partial\rho_j}.
\end{equation}
This relation [Eq.~\eqref{eq:sym_requirement1}] is guaranteed if our expression for $\kappa_{ij}$ is indeed symmetric and we drop all contributions from second derivatives of $g_{ij}^0$ with respect to species density, in accordance with the approximation invoked in Eq.~\eqref{eq:g_expansion}. 

In the following section, we numerically demonstrate for a passive two-component system that the distinct portions of the \textit{integrands} appearing on either side of Eqs.~\eqref{sieq:hessian_symmetry_int} and~\eqref{sieq:kappa_symmetry_int} are themselves identical, with ${\sum_{k}^{n_c}\left(  F_{ik}\rho_k\frac{\partial g_{ik}^0}{\partial \rho_j} \right) = \sum_{k}^{n_c}\left(  F_{jk}\rho_k\frac{\partial g_{jk}^0}{\partial \rho_i} \right)}$ and thus all symmetry conditions are satisfied.

\subsection{Numerical Validation of Symmetry Relations}
\label{sisec:symmetry}
We conduct particle-based Brownian dynamics simulations to validate the symmetry relations required for our microscopic expression for the chemical potential to be thermodynamically consistent. 
We consider a binary mixture of particles of species $A$ and $B$ with pairwise interaction forces resulting from Lennard-Jones potentials.
The interaction energy between two particles of species $A$ and two particles of species $B$ are set to $\varepsilon_{AA}/k_BT = 0.5$ and $\varepsilon_{BB}/k_BT = 0.6$, respectively.
The interaction energy between dissimilar species is set to $\varepsilon_{AB}/k_BT = 0.4$.
For all interactions, the Lennard-Jones diameter $\sigma$ is identical and the interaction range is cut off at $2.5\sigma$.
We select these interaction energies to prevent phase separation while ensuring that species $A$ and $B$ are in fact distinguishable. 
We define the overall volume fraction of the system to be $\phi \equiv N\pi(2^{1/6}\sigma)^3/6V$ where $V$ is the system volume and we take $2^{1/6}\sigma$ to be the physically relevant particle diameter.
The volume fraction of an individual species $i$ is simply $\phi_i \equiv (N_i/N)\phi$ as the species are of identical size.
The overdamped Langevin equation shown in Eq.~\eqref{eq:langevin} was used with $\zeta_A = \zeta_B = \zeta$.
All simulations were performed using the \texttt{HOOMD-Blue} simulation software and consisted of at least 49999 particles~\cite{Anderson2020} and run for a minimum duration of $500~\zeta \sigma^2/k_BT$. 

\begin{figure}
    \centering
    \includegraphics[width=1\linewidth]{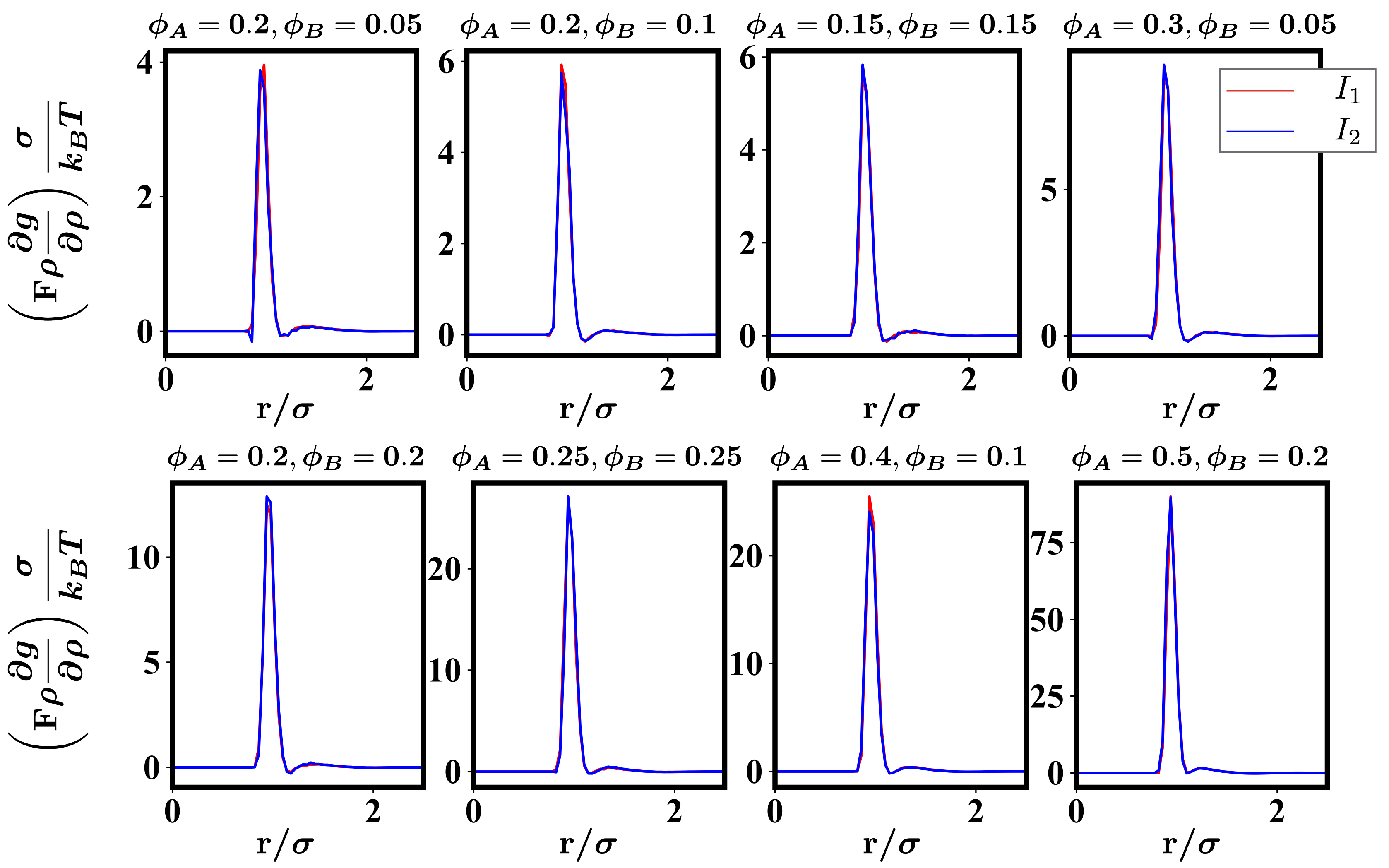}
    \caption{Comparison between the integrands $I_1$ and $I_2$ from Eq.~\eqref{sieq:twocomp_H_symmetry}. 
    The symmetry of the integrands validates the symmetry of the Hessian matrix and $\kappa_{ij}$, a requirement of equilibrium systems.
    }
    \label{sifig:H_symmetry_total}
\end{figure}

\begin{figure}
    \centering
    \includegraphics[width=0.7\linewidth]{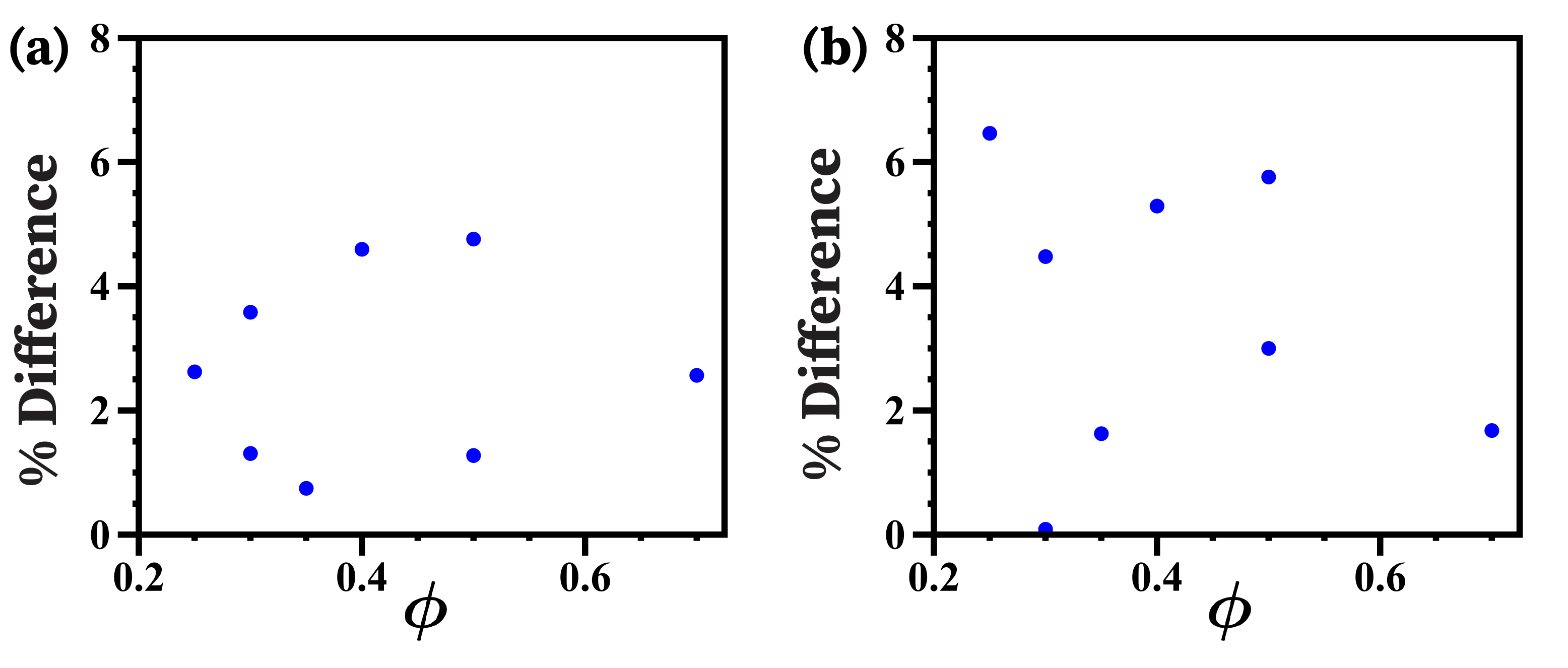}
    \caption{Percentage difference between the integrals of the (a) first and (b) third moment of the $I_1$ and $I_2$ integrands from Eq.~\eqref{sieq:twocomp_H_symmetry}.}
    \label{sifig:H_symmetry_error}
\end{figure}

We computed the steady-state pair distribution functions for uniform systems, $g_{ij}^0$, from the simulations at various volume fractions. 
The partial derivatives of $g_{ij}^0$ with respect to the species density were subsequently computed by performing simulations at various (but closely separated) volume fractions and performing a finite difference derivative.
For a two-component system ($i \in \{A,B\}$), the relevant portions of the integrands appearing in both Eqs.~\eqref{sieq:hessian_symmetry_int} and~\eqref{sieq:kappa_symmetry_int} are: 
\begin{subequations}
\label{sieq:twocomp_H_symmetry}
\begin{align}
    & I_1 = F_{AA}\rho_A \frac{\partial g^0_{AA}}{\partial \rho_B} + F_{AB}\rho_B \frac{\partial g^0_{AB}}{\partial \rho_B}, \\
    & I_2 = F_{BA}\rho_A \frac{\partial g^0_{BA}}{\partial \rho_A} + F_{BB}\rho_B \frac{\partial g^0_{BB}}{\partial \rho_A}.
\end{align}
\end{subequations}
The numerically obtained integrands are shown in Fig.~\ref{sifig:H_symmetry_total} for various volume fractions and ratios of volume fractions: the integrands themselves are nearly identical. 
To quantify the difference, we compute the relative difference in the first and third moments of $I_1$ and $I_2$ (corresponding to the symmetry relation integrals of Eqs.~\eqref{sieq:hessian_symmetry_int} and~\eqref{sieq:kappa_symmetry_int}, respectively), and find small relative errors, as shown in Fig.~\ref{sifig:H_symmetry_error}.
This numerical demonstration validates that our derived microscopic expressions for $H_{ij}^{\rm bulk}$ and $\kappa_{ij}$ coefficients are indeed symmetric with respect to exchange of $i$ and $j$ indices [see Eq.~\eqref{eq:kappa_relation}].

\subsection{Numerical Validation of Hessian Matrix Expressions}
\begin{figure}
    \centering
    \includegraphics[width=0.7\linewidth]{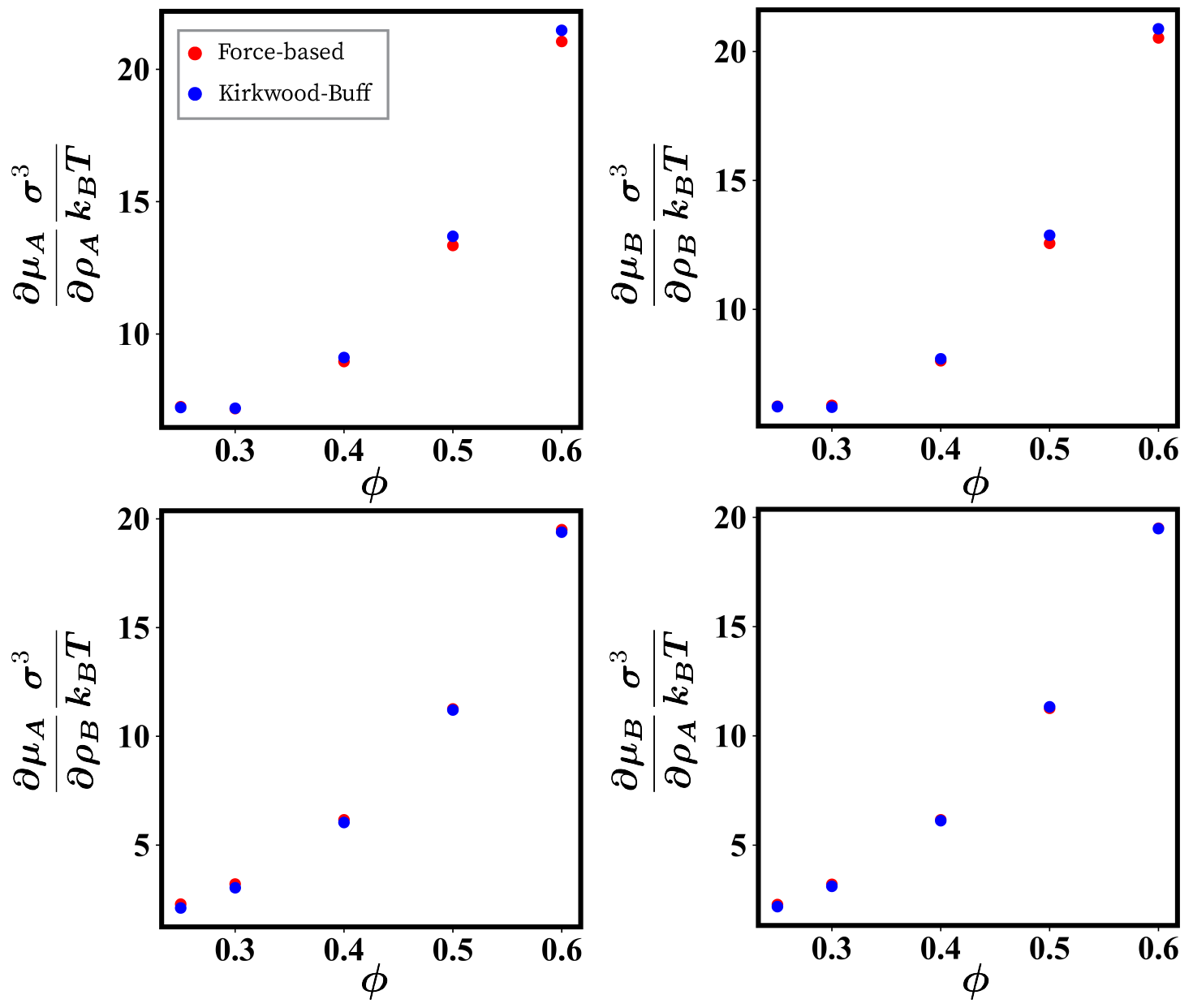}
    \caption{Components of the free energy Hessian matrix of passive systems calculated using the force-based expression [Eq.~\eqref{eq:hessian_expression}] and Kirkwood-Buff expression [Eq.~\eqref{eq:hessian_structure_factor}].
    The total system volume fraction ranges from $0.25$ to $0.6$.
    }
    \label{sifig:Hessian_matrix_compare}
\end{figure}

\begin{figure}
    \centering
    \includegraphics[width=0.5\linewidth]{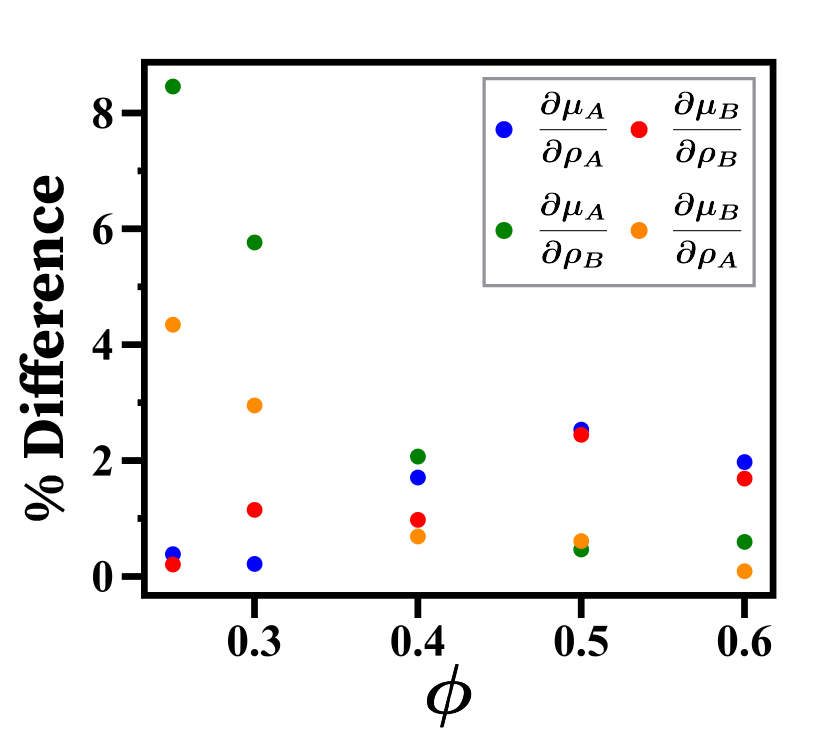}
    \caption{Relative difference between the components of the Hessian matrix calculated using the force-based expression presented here [Eq.~\eqref{eq:hessian_expression}] and the Kirkwood-Buff expression [Eq.~\eqref{eq:hessian_structure_factor}].}
    \label{sifig:Hessian_matrix_difference}
\end{figure}

We look to compare the microscopic form of the free energy Hessian matrix found in Section~\ref{sisec:symmetry_derivation} to a known form to validate the expression. 
From the Kirkwood-Buff theory of solutions, we have~\cite{Kirkwood1951,Ben1974}:
\begin{equation}
\label{eq:H_inv}
    H_{ij}^{-1} \equiv \left(\frac{\partial\rho_i}{\partial\mu_j}\right)_{T,\mu_{k\neq j}} = \frac{1}{k_BT} (\rho_i\rho_j G_{ij} + \rho_i \delta_{ij}),
\end{equation}
where $G_{ij}$ is the Kirkwood-Buff integral~\cite{Cheng2022}:
\begin{equation}
\label{eq:K_B_integral}
    G_{ij} = \frac{1}{\sqrt{\rho_i\rho_j}}\left(\lim_{\mathbf{k}\rightarrow \mathbf{0}}S_{ij}(\mathbf{k}) - \delta_{ij}\right),
\end{equation}
where the static structure factor is defined as:
\begin{equation}
\label{eq:structure_factor}
    S_{ij}(\mathbf{k}) = \frac{1}{\sqrt{\rho_i\rho_j}} \langle \tilde{\rho}_i(\mathbf{k}) \tilde{\rho}_j(-\mathbf{k}) \rangle =  \frac{1}{\sqrt{N_i N_j}}\left\langle\left(\sum_{\alpha=1}^{N_i} \exp(-{\rm i}\mathbf{k}\cdot\mathbf{r}_\alpha)\right) \left(\sum_{\beta=1}^{N_j} \exp(-{\rm i}\mathbf{k}\cdot\mathbf{r}_\beta)\right)\right\rangle,
\end{equation}
which can be calculated directly from the distribution of particle positions.
By combining Eqs.~\eqref{eq:H_inv} and~\eqref{eq:K_B_integral} above, we arrive at:
\begin{equation}
    H_{ij}^{-1} = \frac{1}{k_BT}\sqrt{\rho_i\rho_j} \lim_{\mathbf{k}\rightarrow \mathbf{0}}S_{ij}(\mathbf{k}).
\end{equation}
The components of the bulk Hessian matrix are simply:
\begin{equation}
\label{eq:hessian_structure_factor}
    H_{ij}^{\rm bulk} \equiv \left(\frac{\partial\mu_i}{\partial\rho_j}\right)_{T,\rho_{k\neq j}} = (H_{ij}^{-1})^{-1},
\end{equation}
where $\sum_j^{n_c} H_{ij}^{\rm bulk}H_{jk}^{-1} = \delta_{ik}$~\cite{Ben1974}.
We now have an alternative microscopic expression (via the structure factor) for the equilibrium bulk free energy Hessian matrix that we can compare to our mechanical expression [Eq.~\eqref{eq:hessian_expression}].  
While the formal equivalence between our force-based expression for the Hessian and the established expression in terms of static factor can be demonstrated in the dilute limit (when $g_{ij}^0 \equiv \exp(-\beta U_{ij})$, where $U_{ij}$ is the pairwise interaction potential) we seek to demonstrate their equivalence for arbitrary species densities numerically.
We again conduct particle-based Brownian dynamics simulations to validate the derived microscopic expression for the bulk free energy Hessian matrix. 
The simulation details are identical to those described in Section~\ref{sisec:symmetry}.
All simulations in this section are conducted with equal amounts of $A$ and $B$ particles with $\phi_A = \phi_B = \phi/2$.
The partial derivatives of the homogeneous pairwise distribution functions, needed for our force-based expression, are obtained following the same procedure as in Section~\ref{sisec:symmetry}.
We determine the components of $H_{ij}^{-1}$ using the definition of the static structure factor provided in Eq.~\eqref{eq:structure_factor} and subsequently invert $H_{ij}^{-1}$ to obtain $H_{ij}^{\rm bulk}$ [Eq.~\eqref{eq:hessian_structure_factor}].
All elements of the Hessian matrix, computed both using our force-based expression [Eq.~\eqref{eq:hessian_expression}] and using the Kirkwood-Buff expression [Eq.~\eqref{eq:hessian_structure_factor}], are presented in Fig.~\ref{sifig:Hessian_matrix_compare} with the relative difference between the two expressions of each component of the Hessian matrix shown in Fig.~\ref{sifig:Hessian_matrix_difference}.
We find excellent agreement between the two expressions and note that the increase in accuracy of the force-based expression with \textit{increasing density} is a consequence of better numerical statistics for the radial distribution function at higher densities.
A formal proof of the equivalence of these expressions will be explored in future work, but the numerical equivalence offered here appears to be robust. 


\subsection{Consistency of Microscopic Bulk Chemical Potentials with Virial Pressure}
The total system pressure of uniform systems is often computed in simulation from the well-known virial expression which can be found from the Irving-Kirkwood procedure or, for equilibrium systems, directly from equilibrium statistical mechanics (the single-component expressions can be found in, for example, Ref.~\cite{Chandler1987}).
For an isotropic, uniform, passive systems, the generalization of the single component pressure expression to multicomponent systems is:
\begin{equation}
\label{eq:P_virial}
    P^{\rm bulk} = \sum_i^{n_c} \sum_k^{n_c} \left(k_BT \rho_i\delta_{ik} + \frac{\rho_i\rho_k}{6}\int d\Delta \mathbf{x} (\Delta \mathbf{x})\cdot\mathbf{F}_{ik} g_{ik}^0  \right),
\end{equation}
where $\Delta \mathbf{x}$ is the particle separation vector, the argument of the pairwise forces and uniform pair distribution function.

In Section~\ref{sisec:symmetry_derivation}, we motivated identifying $\mathbf{b}^{\rm eff}_i = -\rho_i \boldsymbol{\nabla}\mu_i$ by noting that the total system pressure of isotropic systems would then satisfy $\boldsymbol{\nabla} P^{\rm bulk} = \sum_i^{n_c} \rho_i \boldsymbol{\nabla} \mu^{\rm bulk}_i$ (where gradient terms have been discarded), consistent with the equilibrium Gibbs-Duhem relation.
To confirm that our derived bulk form of the chemical potentials is indeed consistent with the established form of the total pressure [Eq.~\eqref{eq:P_virial}], we compute $\boldsymbol{\nabla}P^{\rm bulk}$ as:
\begin{subequations}
\begin{equation}
    \boldsymbol{\nabla}P^{\rm bulk} = \sum_j^{n_c} \frac{\partial P^{\rm bulk}}{\partial \rho_j} \boldsymbol{\nabla}\rho_j,
\end{equation}
\begin{equation}
\label{eq:P_virial_deriv}
    \frac{\partial P^{\rm bulk} }{\partial \rho_j} = k_BT 
    + \int d\Delta\mathbf{x} (\Delta \mathbf{x})\cdot \sum_i^{n_c}\sum_{k}^{n_c} \left( \frac{1}{3}\rho_i\mathbf{F}_{ij} g_{ij}^0 + \frac{1}{6}\rho_i\rho_k \mathbf{F}_{ik} \frac{\partial g^0_{ik}}{\partial\rho_j} \right).
\end{equation}
We can use the system isotropy to solely express Eq.~\eqref{eq:P_virial_deriv} with a single component of the force (e.g.,~the $z$ component, $F_{ij}$):
\begin{equation}
\label{eq:presssure_1d_deriv}
    \frac{\partial P^{\rm bulk} }{\partial \rho_j} = k_BT 
    + \int d\Delta\mathbf{x} (\Delta z)\sum_i^{n_c}\sum_{k}^{n_c} \left(\rho_iF_{ij} g_{ij}^0 + \frac{1}{2}\rho_i\rho_k F_{ik} \frac{\partial g^0_{ik}}{\partial\rho_j} \right).
\end{equation}
\end{subequations}
We can express $\sum_i^{n_c} \rho_i \boldsymbol{\nabla} \mu^{\rm bulk}_i$ as:
\begin{subequations}
\begin{equation}
\label{eq:chemical_deriv}
    \sum_i^{n_c} \rho_i \boldsymbol{\nabla} \mu^{\rm bulk}_i = \sum_i^{n_c} \sum_j^{n_c} \rho_i \frac{\partial \mu^{\rm bulk}_i}{\partial\rho_j} \boldsymbol{\nabla}\rho_j.
\end{equation}
Inserting our microscopic expression for $\partial \mu^{\rm bulk}_i / \partial\rho_j$ [see Eq.~\eqref{eq:hessian_expression}], we have:
\begin{equation}
\label{eq:hessian_pressure_form}
    \sum_i^{n_c} \rho_i\frac{\partial \mu^{\rm bulk}_i}{\partial\rho_j} =  k_BT + \int d\Delta\mathbf{x} (\Delta z) \sum_i^{n_c}\sum_{k}^{n_c} \left( \rho_i F_{ij}g_{ij}^0 +\frac{1}{2} \rho_i \rho_k F_{ik}\frac{\partial g_{ik}^0}{\partial \rho_j} \right).
\end{equation}
\end{subequations}
Comparing Eqs.~\eqref{eq:presssure_1d_deriv} and~\eqref{eq:hessian_pressure_form}, we confirm that our microscopic form of the bulk species chemical potential is indeed consistent with the established form of the total system pressure. 
While this only provides a check for the bulk expression for the chemical potential, our gradient terms result in the correct coexistence criteria, providing an additional, albeit indirect, validation of our expressions. 

\subsection{Derivation of Multicomponent Korteweg Stress}
Here, we derive the multicomponent version of the one-dimensional Korteweg stress~\cite{Korteweg1904, Yang1976} by using the equilibrium Gibbs-Duhem relation:
\begin{equation}
    \partial_zP = \partial_z P^{\rm bulk} + \partial_z P^{\rm int} = \sum_i^{n_c} \rho_i \partial_z \mu_i = \sum_i^{n_c} \left( \partial_z \left(\rho_i \mu_i \right) - \mu_i \partial_z \rho_i \right),
\end{equation}
where we have separated the pressure ($P$) into its bulk ($P^{\rm bulk}$) and interfacial ($P^{\rm int}$) contributions with $P = P^{\rm bulk} + P^{\rm int}$.
Substituting the equilibrium form of the species chemical potentials [Eq.~\eqref{eq:equil_chemical_potential_kappa}], we identify $\partial_z P^{\rm bulk}$ as:
\begin{subequations}
    \begin{equation}
        \partial_z P^{\rm bulk} = \sum_i^{n_c} \partial_z \left( \rho_i \mu_i^{\rm bulk} \right) - \partial_z f^{\rm bulk},
    \end{equation}
    where $\partial_z f^{\rm bulk} = \sum_i^{n_c} \mu_i^{\rm bulk} \partial_z \rho_i$ is the differential of the bulk free energy density.
    We identify $\partial_z P^{\rm int}$ as:
    \begin{multline}
        \partial_z P^{\rm int} = -\sum_i^{n_c} \partial_z \left( \rho_i \sum_j^{n_c} \left[ \kappa_{ij} \partial^2_{zz} \rho_j + \frac{1}{2}\sum_k^{n_c} \frac{\partial \kappa_{ij}}{\partial \rho_k} \partial_{z} \rho_j \partial_{z} \rho_k \right] \right) \\ + \sum_i^{n_c} \sum_j^{n_c} \left( \kappa_{ij} \partial^2_{zz} \rho_j + \frac{1}{2}\sum_k^{n_c} \frac{\partial \kappa_{ij}}{\partial \rho_k} \partial_{z} \rho_j \partial_{z} \rho_k \right) \partial_z \rho_i.
    \end{multline}
\end{subequations}
We now recognize:
\begin{equation}
    \sum_i^{n_c} \sum_j^{n_c} \left( \kappa_{ij} \partial^2_{zz} \rho_j \partial_z \rho_i + \frac{1}{2}\sum_k^{n_c} \frac{\partial \kappa_{ij}}{\partial \rho_k} \partial_{z} \rho_j \partial_{z} \rho_k \partial_z \rho_i \right) = \sum_i^{n_c} \sum_j^{n_c} \partial_z \left(\frac{\kappa_{i j}}{2} \partial_z \rho_i \partial_z \rho_j \right),
\end{equation}
and subsequently find:
\begin{subequations}
    \begin{align}
        & P^{\rm bulk} = \sum_i^{n_c} \mu_i^{\rm bulk} \rho_i - f^{\rm bulk}, \\
        & P^{\rm int} = -\sum_j^{n_c} \left( \sum_i^{n_c} \rho_i \kappa_{ij} \partial^2_{zz} \rho_j + \frac{1}{2} \sum_k^{n_c} \left[ \sum_i^{n_c}  \rho_i \frac{\partial \kappa_{ij}}{\partial \rho_k} - \kappa_{j k}  \right] \partial_z \rho_j \partial_z \rho_k \right).
    \end{align}
\end{subequations}
The above form is the anticipated extension of the 1D Korteweg stress~\cite{Korteweg1904, Yang1976} to multicomponent systems.

\section{Equilibrium Transformation Tensor and Maxwell Construction Vector}
Here, we demonstrate that our expression for the effective body forces for passive systems derived in Sec.~\ref{sec:equilibrium_recovery} is indeed consistent with a transformation tensor that has components $\mathcal{T}_{ni}= -\delta_{in}(\rho_n)^{-1}$ and a generalized Maxwell construction vector with components $\mathcal{E}_{i}^{\rm bulk} =\mathcal{E}_{i}^{\rm int} = \rho_i$.
From the main text, $\boldsymbol{\mathcal{T}}$ must satisfy:
\begin{subequations}
\begin{equation}
\label{sieq:bulk_T}
        \sum_i^{n_c} \Bigg[ \frac{\partial \mathcal{T}_{ni}}{\partial \rho_k} b_{ij}^{(1,1)} - \frac{\partial \mathcal{T}_{ni}}{\partial \rho_j} b_{ik}^{(1,1)} 
    + \mathcal{T}_{ni} \frac{\partial b_{ij}^{(1,1)}}{\partial \rho_k} - \mathcal{T}_{ni} \frac{\partial b_{ik}^{(1,1)}}{\partial \rho_j} \Bigg]
    = 0 \  \forall n, j, k \in \mathcal{C},
\end{equation}
\begin{equation}
\label{sieq:intsymTFrelationship}
    \sum_{i}^{n_c} \bigg[ \mathcal{T}_{ni} b_{ijk}^{(3,2)} - \mathcal{T}_{ni} b_{ikj}^{(3,2)} - \frac{\partial \mathcal{T}_{ni}}{\partial \rho_k} b_{ij}^{(3, 3)} + \frac{\partial \mathcal{T}_{ni}}{\partial \rho_j} b_{ik}^{(3, 3)} - \mathcal{T}_{ni} \frac{\partial b_{ij}^{(3, 3)}}{\partial \rho_k} + \mathcal{T}_{ni} \frac{\partial b_{ik}^{(3, 3)}}{\partial \rho_j} \bigg] = 0 \ \forall n, j, k \in \mathcal{C}.
\end{equation}
\begin{multline}
\label{sieq:intTFrelationship}
    \sum_i^{n_c} \Bigg[ 6\mathcal{T}_{ni}b_{ijkl}^{(3,1)} - \frac{\partial}{\partial\rho_l}(\mathcal{T}_{ni} b_{ijk}^{(3,2)}) 
    - \frac{\partial}{\partial\rho_k}(\mathcal{T}_{ni} b_{ijl}^{(3,2)}) - \frac{\partial}{\partial\rho_j}(\mathcal{T}_{ni} b_{ilk}^{(3,2)}) \\ + 2\frac{\partial^2 }{\partial\rho_k\partial\rho_l}(\mathcal{T}_{ni} b_{ij}^{(3,3)}) + \frac{\partial^2 }{\partial\rho_k\partial\rho_j}(\mathcal{T}_{ni} b_{il}^{(3,3)}) \Bigg] = 0 \ \forall n, j, k, l \in \mathcal{C}.
\end{multline}
\end{subequations}
As noted in Sec.~\ref{sec:equilibrium_recovery}, $b_{ij}^{(1,1)} = -\rho_i H_{ij}^{\rm bulk}$, where $H_{ij}^{\rm bulk}$ is the $ij$ component of the Hessian matrix associated with the bulk free energy density and is, of course, symmetric with respect to the exchange of indices $i$ and $j$. 
Additionally, the derivative of the Hessian matrix with respect to the species density is also symmetric with $\partial H^{\rm bulk}_{ij} / \partial\rho_k = \partial H^{\rm bulk}_{ik} / \partial\rho_j$.
Substitution of $b_{ij}^{(1,1)} = -\rho_i H_{ij}^{\rm bulk}$ and $\mathcal{T}_{ni}= -\delta_{in}(\rho_n)^{-1}$ into Eq.~\eqref{sieq:bulk_T} results in:
\begin{multline}
    \sum_{i}^{n_c}\bigg[\delta_{in}\delta_{nk}(\rho_{n})^{-1}\rho_{i}H_{ij}^{\rm bulk} - \delta_{in}\delta_{nj}(\rho_{n})^{-1}\rho_{i}H_{ik}^{\rm bulk} \\ + (\rho_i)^{-1} \left(\delta_{ik}H_{ij}^{\rm bulk} + \rho_i\frac{\partial H_{ij}^{\rm bulk}}{\partial\rho_k} - \delta_{ij}H_{ik}^{\rm bulk} - \rho_i\frac{\partial H_{ik}^{\rm bulk}}{\partial\rho_j} \right)\bigg]\\
     = \left[ H_{kj}^{\rm bulk} - H_{jk}^{\rm bulk} + (\rho_{n})^{-1} \left( H_{kj}^{\rm bulk} - H_{jk}^{\rm bulk} + \rho_i \left(\frac{\partial H_{ij}^{\rm bulk}}{\partial\rho_k} - \frac{\partial H_{ik}^{\rm bulk}}{\partial\rho_j} \right) \right)\right] = 0 \ \forall n, j, k \in \mathcal{C},
\end{multline}
and hence we confirm that the equilibrium transformation tensor indeed satisfies Eq.~\eqref{sieq:bulk_T}.

Section~\ref{sec:equilibrium_recovery} demonstrated that the remaining body force coefficients can be compactly expressed using the symmetric coefficient $\kappa_{ij}$, with: 
\begin{align}
    & b_{ijkl}^{(3,1)} = -\frac{1}{2}\rho_i\frac{\partial^2 \kappa_{ij}}{\partial\rho_k \partial\rho_l},\\
    & b_{ijk}^{(3,2)} = -2\rho_i\frac{\partial\kappa_{ij}}{\partial\rho_k},\\
    & b_{ij}^{(3,3)} = -\rho_i\kappa_{ij}.
\end{align}
Substituting these relations and $\mathcal{T}_{ni}= -\delta_{in}(\rho_n)^{-1}$ into Eq.~\eqref{sieq:intsymTFrelationship}, we find:
\begin{multline}
    \sum_{i}^{n_c} \Bigg[ 2 \delta_{in} (\rho_n)^{-1}  \rho_i \frac{\partial \kappa_{ij}}{\partial \rho_k} - 2 \delta_{in} (\rho_n)^{-1}  \rho_i \frac{\partial \kappa_{ik}}{\partial \rho_j} + \delta_{i n} (\rho_n)^{-2} \delta_{n k} \rho_i \kappa_{i j} - \delta_{i n} (\rho_n)^{-2} \delta_{n j} \rho_i \kappa_{i k} \\ - \delta_{in} (\rho_n)^{-1} \left( \kappa_{i j} \delta_{i k} + \rho_i \frac{\partial \kappa_{i j}}{\partial \rho_k} \right) + \delta_{in} (\rho_n)^{-1} \left( \kappa_{i k} \delta_{i j} + \rho_i \frac{\partial \kappa_{i k}}{\partial \rho_j} \right)  \Bigg] = \frac{3}{2} \frac{\partial \kappa_{nj}}{\partial \rho_k} - \frac{3}{2} \frac{\partial \kappa_{nk}}{\partial \rho_j}  = 0 \ \forall n, j, k \in \mathcal{C},
\end{multline}
where we have used the symmetry of $\partial \kappa_{ij} / \partial \rho_k$ upon exchange of indices $i$, $j$, and $k$ from its definition in terms of the direct correlation function (See Chapter 6.2 in~\cite{Hansen2013}).
Next, we substitute the expressions for the body force coefficients and $\mathcal{T}_{ni}$ into Eq.~\eqref{sieq:intTFrelationship}:
\begin{align}
    & \sum_i^{n_c} \Bigg[ 3 \delta_{in} (\rho_n)^{-1} \rho_i \frac{\partial^2 \kappa_{i j}}{\partial \rho_k \rho_l} - 2 \frac{\partial}{\partial\rho_l}\left(\delta_{in} (\rho_n)^{-1} \rho_i \frac{\partial \kappa_{ij}}{\partial \rho_k}\right) 
    - 2 \frac{\partial}{\partial\rho_k}\left(\delta_{in} (\rho_n)^{-1} \rho_i \frac{\partial \kappa_{ij}}{\partial \rho_l}\right) \nonumber \\ & - 2 \frac{\partial}{\partial\rho_j}\left(\delta_{in} (\rho_n)^{-1} \rho_i \frac{\partial \kappa_{il}}{\partial \rho_k}\right) + 2\frac{\partial^2 }{\partial\rho_k\partial\rho_l}\left(\delta_{in} (\rho_n)^{-1} \rho_i \kappa_{ij}\right) + \frac{\partial^2 }{\partial\rho_k\partial\rho_j}\left(\delta_{in} (\rho_n)^{-1} \rho_i \kappa_{il}\right) \Bigg] \nonumber \\ & = 3 \frac{\partial^2 \kappa_{n j}}{\partial \rho_k \rho_l} - 2 \frac{\partial \kappa_{nj}}{\partial \rho_l \partial \rho_k}
    - 2 \frac{\partial^2 \kappa_{nj}}{\partial \rho_k \partial \rho_l} - 2 \frac{\partial^2 \kappa_{nl}}{\partial \rho_j \partial \rho_k} + 2\frac{\partial^2 \kappa_{nj}}{\partial\rho_k\partial\rho_l} + \frac{\partial^2 \kappa_{nl}}{\partial\rho_k\partial\rho_j} = \frac{\partial^2 \kappa_{n j}}{\partial \rho_k \rho_l} - \frac{\partial^2 \kappa_{n l}}{\partial \rho_k \rho_j} = 0 \ \forall n, j, k, l \in \mathcal{C}.
\end{align}
We have now shown that the equilibrium solution for the transformation matrix indeed satisfies our derived relations.

Following the main text, we define the interfacial and bulk $\boldsymbol{\mathcal{E}}$ with $d\mathcal{G}^{\rm bulk} = \boldsymbol{\mathcal{E}}^{\rm bulk}\cdot\mathbf{u}^{\rm bulk}$ and $d\mathcal{G}^{\rm int} = \boldsymbol{\mathcal{E}}^{\rm int}\cdot\mathbf{u}^{\rm int}$, respectively. 
Each Maxwell construction vector must satisfy:
\begin{subequations}
\label{eq:E_relations}
\begin{align}
    \label{sieq:E_bulk_appendix}
    & \sum_i^{n_c}\Bigg[E_{ij}^{\rm bulk} \frac{\partial u_i^{\rm bulk}}{\partial\rho_k} - E_{ik}^{\rm bulk} \frac{\partial u_i^{\rm bulk}}{\partial\rho_j} \Bigg]= 0 \ \forall j, k \in \mathcal{C}, \\
    \label{sieq:E_int_appendix1}
    & \sum_i^{n_c} \left[u_{ij}^{(2,2)}E^{\rm int}_{ik} - u_{ik}^{(2,2)}E^{\rm int}_{ij} \right] = 0 \ \forall j, k \in \mathcal{C}, \\
    \label{sieq:E_int_appendix3}
    & \sum_i^{n_c} \bigg[ 2E_{il}^{\rm int} u_{ijk}^{(2,1)} + 2E_{ik}^{\rm int} u_{ijl}^{(2,1)} + 2E_{ij}^{\rm int} u_{ilk}^{(2,1)} - \frac{\partial}{\partial\rho_l}\left( E_{ik}^{\rm int}u_{ij}^{(2,2)} \right) \nonumber \\ & - \frac{\partial}{\partial\rho_k}\left(E_{il}^{\rm int}u_{ij}^{(2,2)} \right) - \frac{\partial}{\partial\rho_j}\left( E_{ik}^{\rm int}u_{il}^{(2,2)} \right) \bigg] = 0 \ \forall j, k, l \in \mathcal{C},
\end{align}
\end{subequations}
where $E_{ij}^{\rm bulk} \equiv \partial \mathcal{E}_i^{\rm bulk} / \partial\rho_j$ and $E_{ij}^{\rm int} \equiv \partial \mathcal{E}_i^{\rm int} / \partial\rho_j$. 

In the equilibrium limit, the species pseudopotentials are simply the chemical potentials, e.g.,~$u_i = \mu_i$.
Thus, we can relate the derivative of the species pseudopotentials to the Hessian matrix: $\partial u_i^{\rm bulk} / \partial\rho_j = \partial\mu_i^{\rm bulk} / \partial\rho_j = H_{ij}^{\rm bulk}$.
As a result, $\partial u_i^{\rm bulk} / \partial\rho_j$ is symmetric with respect to exchanging $i$ and $j$.
Substituting the equilibrium solution ($\mathcal{E}_{i}^{\rm bulk} =\mathcal{E}_{i}^{\rm int} = \rho_i$) of the generalized Maxwell construction vector into Eq.~\eqref{sieq:E_bulk_appendix}, we arrive at:
\begin{equation}
    \frac{\partial \mu_j^{\rm bulk}}{\partial\rho_k} - \frac{\partial \mu_k^{\rm bulk}}{\partial\rho_j}= 0 \ \forall j, k \in \mathcal{C},
\end{equation}
and hence confirm that the equilibrium Maxwell construction vector indeed satisfies Eq.~\eqref{sieq:E_bulk_appendix}.

Section~\ref{sec:equilibrium_recovery} demonstrated that the interfacial coefficients in the species pseudopotentials can be compactly expressed with $\kappa_{ij}$:
\begin{subequations}
\label{eq:mu_kappa}
\begin{align}
    & \mu_{ijk}^{(2,1)} = \frac{1}{2}\frac{\partial\kappa_{ij}}{\partial\rho_k}, \label{eq:mu_kappa_1}\\
    & \mu_{ij}^{(2,2)} = \kappa_{ij}. \label{eq:mu_kappa_2}
\end{align}
\end{subequations}
Substituting Eq.~\eqref{eq:mu_kappa} and the equilibrium solution ($\mathcal{E}_{i}^{\rm bulk} =\mathcal{E}_{i}^{\rm int} = \rho_i$) into our remaining relations [Eqs.~\eqref{sieq:E_int_appendix1} and~\eqref{sieq:E_int_appendix3}] results in the following relations:
\begin{subequations}
\begin{align}
    & \kappa_{kj} - \kappa_{jk} = 0 \ \forall j, k \in \mathcal{C}, \\
    & \frac{\partial \kappa_{lj}}{\partial \rho_k}+ \frac{\partial \kappa_{kj}}{\partial \rho_l} + \frac{\partial \kappa_{jl}}{\partial \rho_k} - \frac{\partial\kappa_{kj}}{\partial\rho_l} - \frac{\partial \kappa_{lj}}{\partial\rho_k} - \frac{\partial \kappa_{kl}}{\partial\rho_j} = 0 \ \forall j, k, l \in \mathcal{C},
\end{align}
\end{subequations}
which are all satisfied.
We have now shown that the equilibrium solution for the Maxwell construction vector satisfies our derived relations.

\section{One Component Coexistence Criteria}
\label{sec:one-component}

The multicomponent coexistence theory developed in our work represents a generalization of the mechanical theory of single-component nonequilibrium phase separation presented in Refs.~\cite{Aifantis1983, Solon2018, Omar2023}.
In contrast to a general multicomponent system, one of the two coexistence criteria required for single-component two phase coexistence can be immediately determined (i.e.,~without needing to find $\mathcal{T}$ or $\mathcal{E}$ which are now both scalars for a single-component system). 
This follows from the static momentum balance which takes the form:
\begin{equation}
    \mathbf{0} = \boldsymbol{\nabla}\cdot\boldsymbol{\sigma} + \mathbf{b},
\end{equation}
where $\boldsymbol{\sigma}$ and $\mathbf{b}$ represent all stresses and body forces, respectively. 
If we again consider the scenario of quasi-1D phase coexistence, we now have:
\begin{equation}
\label{eq:1dmechanics}
    0 = \frac{d}{dz}\sigma_{zz} + b_z.
\end{equation}
Up until now, Eq.~\eqref{eq:1dmechanics} looks identical to the species mechanical balance presented in the main text. 
The difference, however, is that for a single-component system in one dimension, we can \textit{always} express the body force as $b_z = d\sigma^b/dz$ where $\sigma^b$ is an effective stress arising from the body force~\cite{Omar2020, Omar2023}.
This stress can be defined so long as $b_z$ solely depends on a single field (e.g.,~the density, $\rho$).
For a multicomponent system, $b_z$ generally depends on all $n_c$ species densities and thus cannot be readily expressed as an effective stress.
For a one-component system, we are thus able to express our one-dimensional momentum balance as:
\begin{equation}
\label{eq:1d_dynamic_stress}
    0 = \frac{d}{dz}\Sigma,
\end{equation}
where $\Sigma \equiv \sigma_{zz} + \sigma^b$ is the total \textit{effective or dynamic stress}.

Integration of this mechanic balance [Eq.~\eqref{eq:1d_dynamic_stress}] results in the condition of uniform dynamic stress.
A second-order density gradient expansion of the stress takes the form:
\begin{equation}
\label{eq:1component_stress}
    -\Sigma(\rho) \equiv P(\rho) = P^{\rm bulk}(\rho) - a(\rho)\frac{d^2\rho}{dz^2} - b(\rho)\left(\frac{d\rho}{dz}\right)^2,
\end{equation}
where we have defined a dynamic pressure with a bulk contribution $P^{\rm bulk}$, and $a$ and $b$ represent interfacial coefficients. 
We emphasize that this effective or dynamic pressure reduces to the static pressure in the absence of body forces.
All equations of state ($P^{\rm bulk}$, $a$, and $b$) can be formally obtained from microscopic considerations through an Irving-Kirkwood procedure. 
Expressing the stress in this form allows us to readily identify the first of two coexistence criteria: equality of pressure with $P^{\rm bulk}(\rho^{\alpha}) = P^{\rm bulk}(\rho^{\beta}) = P^{\rm coexist}$ \textit{without needing to perform a transformation on the mechanical balance}.

The second criterion takes the form of a generalized equal-area Maxwell construction~\cite{Aifantis1983, Solon2018, Omar2023}.
This criterion is obtained by defining a variable\footnote{Here, we depart from the nomenclature adopted in Ref.~\cite{Omar2023} which defined this variable as $\mathcal{E}$.}, $\mathcal{R} (\rho)$, such that an integration of the stress with respect to $\mathcal{R}$ eliminates the gradient terms and thus the criterion \textit{does not} require solving form the complete density profile.
By using the uniform stress condition, rearranging Eq.~\eqref{eq:1component_stress}, and invoking the definition of $\mathcal{R}$, we have:
\begin{equation}
\label{eq:1component_E}
\int_{\mathcal{R}^{\beta}}^{\mathcal{R}^{\alpha}} \left[ P^{\rm bulk}(\mathcal{R})  - P^{\rm coexist} \right] \ d \mathcal{R} = \int_{\mathcal{R}^{\beta}}^{\mathcal{R}^{\alpha}} \left[ a\frac{d^2\rho}{dz^2} + b\left(\frac{d\rho}{dz}\right)^2\right] \ d \mathcal{R} = 0.
\end{equation}
The form of $\mathcal{R}$ that allows this equation to hold was found to be~\cite{Aifantis1983, Solon2018, Omar2023}: 
\begin{equation}
\label{eq:equalareavariable}
   \frac{\partial\mathcal{R}}{\partial \rho} = \frac{1}{a}   \exp\left( 2 \int \frac{b} {a} \ d\rho \right).
\end{equation}
Note that the generalized Maxwell construction variable is itself an equation-of-state that depends on the precise forms of the interfacial coefficients but not on the shape of the interface itself.
The nonequilibrium coexistence criteria of single-component two-phase coexistence thus take the form:
\begin{subequations}
\label{eq:Mech_Theory}
\begin{equation}
\label{eq:Mech_Theory_equal_P_a}
    P^{\rm bulk}(\mathcal{R}^{\alpha}) = P^{\rm bulk}(\mathcal{R}^{\beta}) = P^{\rm coexist},
\end{equation}
\begin{equation}
\label{eq:Mech_Theory_gen_eqal_area_b}   \int_{\mathcal{R}^{\beta}}^{\mathcal{R}^{\alpha}} \left[ P^{\rm bulk}(\mathcal{R})  - P^{\rm coexist} \right] \ d \mathcal{R} = 0.
\end{equation}
\end{subequations}

The criterion of the generalized Maxwell construction [Eq.~\eqref{eq:Mech_Theory_gen_eqal_area_b}] can be alternatively expressed as equality of a species ``pseudopotential''. 
To recognize this we define a species pseudopotential, $u$, by proposing a generalized Gibbs-Duhem relation with $du = \mathcal{R}dP$. 
Substituting this relation into Eq.~\eqref{eq:Mech_Theory_gen_eqal_area_b}, we have:
\begin{equation}
\label{eq:1componentGD}
    0 = \int_{\mathcal{R}^{\beta}}^{\mathcal{R}^{\alpha}} P(\mathcal{R}) \ d \mathcal{R} -\int_{(P\mathcal{R})^{\beta}}^{(P\mathcal{R})^{\alpha}} d(P\mathcal{R}) = -\int_{u^{\beta}}^{u^{\alpha}} du,
\end{equation}
where the final equality allows us to identify that the pseudopotentials are equal between the two coexisting phases.
The general form of the species pseudopotential is:
\begin{equation}
    u = u^{\rm bulk} - u^{(2,1)}\left(\frac{d\rho}{dz}\right)^2 - u^{(2,2)}\frac{d^2\rho}{dz^2}.
\end{equation}
The precise form of the pseudopotential can be found by using the generalized Gibbs-Duhem relation and matching terms. 
In doing so, we find:
\begin{subequations}
    \begin{align}
        & u^{\rm bulk} = P^{\rm bulk}\mathcal{R} \bigg|_{(P\mathcal{R})^{\beta}}^{(P\mathcal{R})^\alpha} - \int_{\mathcal{R}^\beta}^{\mathcal{R}^\alpha} P^{\rm bulk} d\mathcal{R},\label{eq:u_bulk}\\
        & u^{(2,2)} = \mathcal{R}a,\label{eq:u_22}\\
        & 2 u^{(2,1)} + \frac{d u^{(2,2)}}{d \rho} = \mathcal{R}\left(2b+\frac{d a}{d \rho}\right),\label{eq:u_21_du_22}\\
        & \frac{d u^{(2,1)}}{d \rho} = \mathcal{R}\frac{d b}{d \rho}\label{eq:du_21}.
    \end{align}
\end{subequations}
We can now determine the form of $\mathcal{R}$ in terms of $a$ and $b$ by performing the appropriate derivatives (with respect to density) and combining Eqs.~\eqref{eq:u_22}, \eqref{eq:u_21_du_22}, and \eqref{eq:du_21} to eliminate the species pseudopotential coefficients, resulting in:
\begin{equation}
\label{eq:epsilon_function}
    2b\frac{d\mathcal{R}}{d\rho} = \frac{d}{d\rho} \left(a\frac{d\mathcal{R}}{d\rho}\right),
\end{equation}
which is entirely consistent with the form of $\mathcal{R}$ obtained without invoking the Gibbs-Duhem relation shown in Eq.~\eqref{eq:equalareavariable}.
The individual terms of the pseudopotential can now be determined with:
\begin{subequations}
\label{eq:1component_potential}
\begin{align}
& u^{\rm bulk} = P^{\rm bulk}\mathcal{R} - \int P^{\rm bulk} d\mathcal{R},\\
& u^{(2,1)} = \mathcal{R}b - \frac{1}{2}\frac{d\mathcal{R}}{d\rho}a,\\
& u^{(2,2)} = \mathcal{R}a.
\end{align}
\end{subequations}

The above approach for single-component systems identified the coexistence criteria as equality of pressure and equality of the species pseudopotential (or equivalently, a generalized Maxwell construction on the pressure). 
Again, for a single-component system, we can always immediately identify one of the criteria as equality of the pressure as dynamic stress can always be defined in (quasi) 1D.
With multiple species present, the mechanical balance of each species does not immediately give rise to coexistence criteria. 
Instead, we define an effective body force for each species and must \textit{transform} these mechanical balances into a form where coexistence criteria can be identified. 
Here, for completeness, we show that our body-force-based procedure recovers the above coexistence criteria for single-component systems and identify the relation between $\mathcal{R}(\rho)$ in the original single-component theory and $\mathcal{T}(\rho)$ and $\mathcal{E}(\rho)$ in our multicomponent theory.

We begin by defining the effective body force as:
\begin{subequations}
\label{eq:1component_beff}
\begin{equation}
 0 = \frac{d}{dz}\Sigma \equiv b^{\rm eff},
\end{equation}
with
\begin{align}
    & b^{\rm eff} = b^{(1,1)}\left(\frac{d \rho}{dz}\right) - b^{(3,1)}\left(\frac{d \rho}{dz} \right)^3  - b^{(3,2)}\left(\frac{d^2 \rho}{dz^2} \right)\left(\frac{d \rho}{dz}\right) - b^{(3,3)}\left(\frac{d^3 \rho}{dz^3} \right),\\
    & b^{(1,1)} = -\frac{d P^{\rm bulk}}{d \rho},\\
    & b^{(3,1)} = -\frac{d b}{d \rho},\\
    & b^{(3,2)} = -\left(\frac{d a}{d \rho}+2b\right),\\
    & b^{(3,3)} = -a.
\end{align}
\end{subequations}
Following our multicomponent theory, we define a species pseudopotential as $\mathcal{T} b^{\rm eff} = \partial_z u$ where $\mathcal{T}$ satisfies the following:
\begin{equation}
    2\mathcal{T} b^{(3,1)} - \frac{d}{d\rho}\left(\mathcal{T} b^{(3,2)}\right) + \frac{d^2}{d\rho^2}\left(\mathcal{T} b^{(3,3)}\right) = 0.
\end{equation}
By inserting our expression for the body force coefficients, we can further reduce this to:
\begin{equation}
\label{eq:T_function}
    2b\frac{d\mathcal{T}}{d \rho} = \frac{d}{d\rho}\left(a\frac{d\mathcal{T}}{d\rho}\right).
\end{equation}
Comparing Eqs.~\eqref{eq:T_function} and \eqref{eq:epsilon_function}, we identify $\mathcal{R} = \mathcal{T}$.
This equivalence is consistent with our expectation as both $\mathcal{R}$ and $\mathcal{T}$ allow us to define a pseudopotential beginning from the static mechanical balance. 
In fact, using this equivalence and our expression for the components of the species pseudopotential in terms of the body force coefficients and $\mathcal{T}$:
\begin{subequations}
\label{eq:u_coeff_b}
    \begin{align}
        & u^{\rm bulk} = \int \mathcal{T} b^{(1,1)} d\rho,\\
        & u^{(2,1)} = \frac{1}{2}\left(\mathcal{T}b^{(3,2)} - \frac{d}{d\rho}(\mathcal{T}b^{(3,3)})\right) \\
        & u^{(2,2)} = \mathcal{T} b^{(3,3)},
    \end{align}
\end{subequations}
one can readily verify that Eq.~\eqref{eq:1component_potential} are recovered (within an inconsequential multiplicative constant).

With the species pseudopotential in hand, we now use the generalized Gibbs-Duhem relation, $d\mathcal{G} = \mathcal{E} du$, to arrive at our second and final criterion, which should be equality of the dynamic pressure.
From the main text, $\mathcal{E}$ must satisfy:
\begin{equation}
\label{eq:variable_vector_u}
    2u^{(2,1)}\frac{d \mathcal{E}}{d\rho} = \frac{d}{d\rho}\left(u^{(2,2)}\frac{d \mathcal{E}}{d\rho}  \right).
\end{equation}
By substituting Eq.~\eqref{eq:u_coeff_b} into Eq.~\eqref{eq:variable_vector_u}, we can express the differential equation for $\mathcal{E}$ as:
\begin{equation}
\label{eq:variable_vector_T_a_b}
    \frac{d^2}{d\rho^2}(\mathcal{E}\mathcal{T}a) - \frac{d}{d\rho} \left(2\mathcal{E}\mathcal{T}b+ \mathcal{E}\mathcal{T}\frac{d a}{d \rho}\right) + 2 \mathcal{E}\mathcal{T}\frac{d b}{d\rho} = 0.
\end{equation}
We can now determine the component of the ``global quantity'' $\mathcal{G}$ in terms of the pseudopotential coefficients and $\mathcal{E}$:
\begin{subequations}
\label{eq:G_coeff}
    \begin{align}
        & \mathcal{G}^{\rm bulk} = \mathcal{E}u^{\rm bulk} - \int u^{\rm bulk} d\mathcal{E},\\
        & \mathcal{G}^{(2,1)} = \mathcal{E} u^{(2,1)} -\frac{1}{2} \frac{d \mathcal{E}}{d \rho} u^{(2,2)},\\
        & \mathcal{G}^{(2,2)} = \mathcal{E}u^{(2,2)}.
    \end{align}
\end{subequations}

Intriguingly, $\mathcal{E} = 1/\mathcal{T}$ satisfies Eq.~\eqref{eq:variable_vector_T_a_b}, suggesting that, for a one-component system, $\mathcal{E}$ simply returns us to the mechanical balance. 
Upon substitution of Eq.~\eqref{eq:u_coeff_b} into Eq.~\eqref{eq:G_coeff} and recognizing $\mathcal{E} = 1/\mathcal{T}$ one can straightforwardly verify that $\mathcal{G}$ is indeed the dynamic stress with $\mathcal{G}^{\rm bulk} = -P^{\rm bulk}$, $\mathcal{G}^{(2,1)} = - b$, and $\mathcal{G}^{(2,2)}=- a$ and our theory indeed recovers equality of pressure for single-component systems.

\section{Numerical Details of Phase Diagram Construction}
\subsection{Simulation Details}
In the main text, we determine the complete spatial profiles of the species densities in coexistence, and subsequently the coexistence densities, for the reported phase diagrams by using numerically simulating phenomenological systems of ODEs. 
To do this, we used the numerical path continuation method developed for the binary phase-field-crystal model, as detailed in~\cite{Thiele2019,Frohoff2021}, where the method utilizes the \texttt{MATLAB} continuation and bifurcation package \texttt{PDE2PATH}~\cite{Uecker2021}. 
We implement the dynamics and species pseudopotentials of the MAMB model as the governing equations in the numerical path continuation. 
All simulations are conducted in 1D with periodic boundary conditions under steady-state conditions.

\subsection{Phase Diagram Construction Details}
The theoretical prediction of the phase diagram using our derived coexistence criteria was done using \texttt{Mathematica}. 
In the event that multiple coexistence scenarios solve our criteria for a given set of conditions (this is the case in Fig.~3(c) in the main text which is the pathological scenario where $\partial_B u_A^{\rm bulk}=0$ for the two-component system of species $A$ and $B$), we solely report the solution that aligns with the independent numerical simulation results. 
Assessing which solutions are metastable/globally stable is beyond the scope of the present work.

\section{Path Dependency of Inexact Coexistence Criteria}
\begin{figure}
    \centering
    \includegraphics[width=0.75\linewidth]{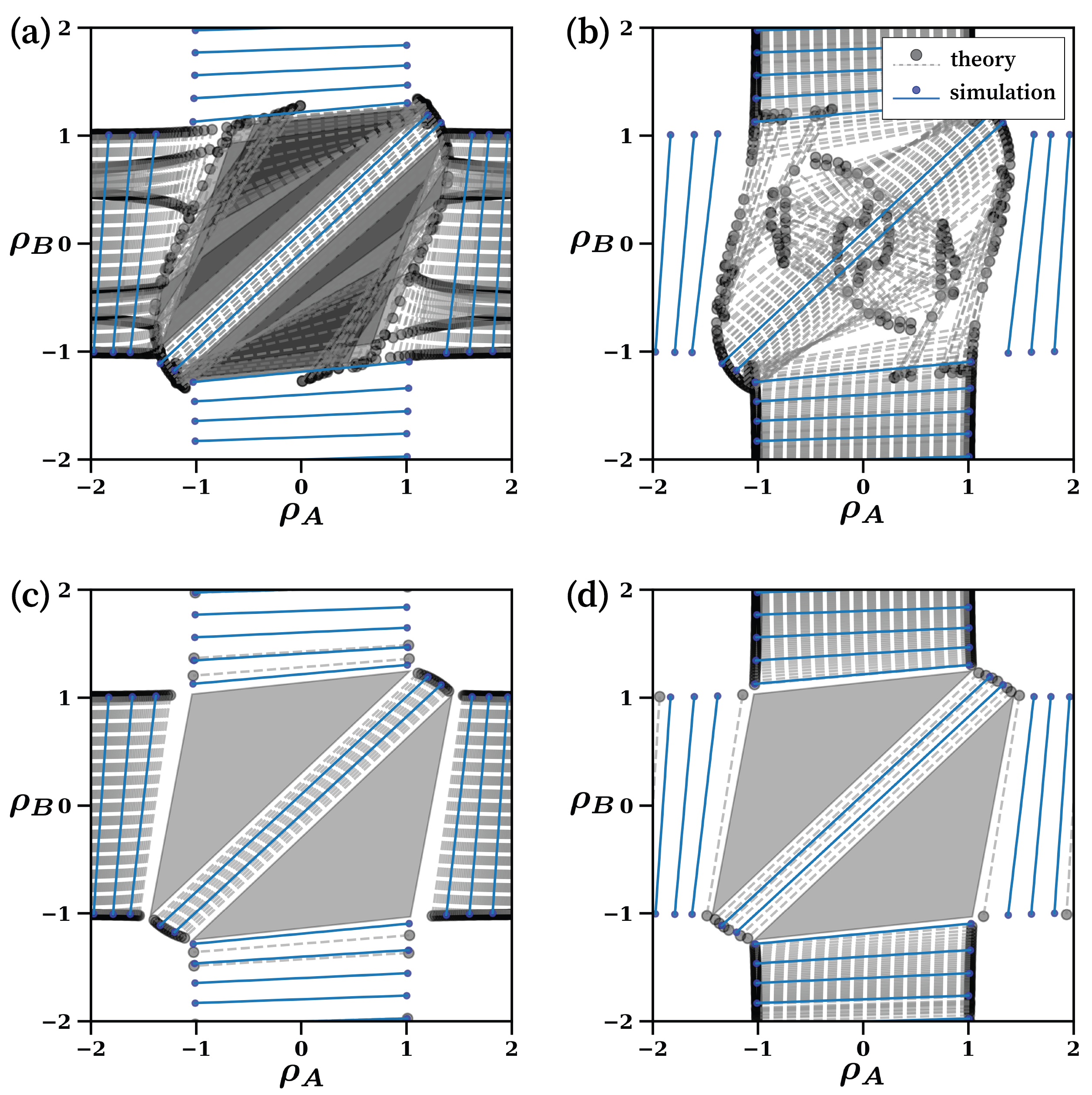}
    \caption{Phase diagram of the NRCH approximated using the equilibrium variable vector. 
    The different integration paths are chosen in each panel to make the approximation. 
    The integration paths are (a) $(\rho_A',\rho_B'):(0,0) \rightarrow (\rho_A,0) \rightarrow (\rho_A,\rho_B)$, (b) $(\rho_A',\rho_B'):(0,0) \rightarrow (0,\rho_B) \rightarrow (\rho_A,\rho_B)$, (c) enforcing $u_A^{\rm bulk} = u_A^{\rm coexist}$, and (d) enforcing $u_B^{\rm bulk} = u_B^{\rm coexist}$. 
    The system parameters are identical to the NRCH model system that satisfies the coupling parameter ratio presented in the main text [Fig.~3(a)].}
    \label{fig:NRCH_eq}
\end{figure}

In this Section, we demonstrate the path dependency of the weighted-area Maxwell construction when approximating $\boldsymbol{\mathcal{E}}$. 
We do so using the NRCH model explored in the main text.
Here we use the equilibrium (incorrect) Maxwell construction vector:
\begin{equation}
    \boldsymbol{\mathcal{E}}=
    \begin{bmatrix} \rho_A & \rho_B\end{bmatrix}^{\rm T}
\end{equation}
where we can explicitly write out the \textit{incorrect} equilibrium generalized Maxwell construction integrals as:
\begin{equation}
    \int_{\rho_A^\alpha}^{\rho_A^\beta} \left(u_A^{\rm bulk}(\rho_A',\rho_B') - u_A^{\rm coexist} \right) d\rho_A' + \int_{\rho_B^\alpha}^{\rho_B^\beta} \left(u_B^{\rm bulk}(\rho_A',\rho_B') - u_B^{\rm coexist} \right) d\rho_B' = 0.
\end{equation}
Since this equilibrium Maxwell construction vector results in path-dependent integrals, we choose four different integration paths to perform the generalized Maxwell construction with the results of each path shown in Fig.~\ref{fig:NRCH_eq}.
We certainly find a strong path dependence and, intriguingly, find that certain paths can lead to accurate phase diagram prediction.

\sloppy The generalized Maxwell construction for the NRCH vanishes, path-independently, when ${\boldsymbol{\mathcal{E}}=\begin{bmatrix} \rho_A & \frac{\chi + \alpha}{\chi - \alpha} \rho_B \end{bmatrix}^{\rm T}}$ as shown in the main text.
Notably, this means one can choose integration paths that mandate $u_A^{\rm bulk} = u_A^{\rm coexist}$ or $u_B^{\rm bulk} = u_B^{\rm coexist}$ such that the generalized Maxwell construction only occurs over one nonzero integrand.
Along these paths, the generalized Maxwell construction is equivalent when using the correct $\boldsymbol{\mathcal{E}}$ and the equilibrium $\boldsymbol{\mathcal{E}}$, meaning the equilibrium Maxwell construction vector can be used to predict the phase diagram [see Fig.~\ref{fig:NRCH_eq}(c) and (d)] if these paths are chosen.
However, if other integration paths are chosen, such as ${(\rho_A',\rho_B'):(0,0) \rightarrow (\rho_A,0) \rightarrow (\rho_A,\rho_B)}$ or ${(\rho_A',\rho_B'):(0,0) \rightarrow (0,\rho_B) \rightarrow (\rho_A,\rho_B)}$, the equilibrium Maxwell construction vector incorrectly predicts the coexistence states [see Fig.~\ref{fig:NRCH_eq}(a) and (b)].

We now examine a system with a set of MAMB parameters that do not result in exact coexistence criteria according to our theory, but can be approximated with the generalized weighted-area construction using $\mathbf{E}=\mathbf{E}^{\rm int}$.
The MAMB functional form and parameter simplification for the two-component system (species $A$ and $B$) are adopted as described in the main text.
We use $\overline{\kappa}_{AA} = \overline{\kappa}_{BB} = \lambda_{AAA} = 2\overline{\kappa}_{AB} = 2\overline{\kappa}_{BA} = 2\lambda_{AAB} = 4\lambda_{ABB}$.
The coexistence criteria are insensitive to the precise values of the parameters (e.g., the value of $\overline{\kappa}_{AA}$), however they depend on the ratios of the parameters.
For this system, the bulk and interfacial Maxwell construction vector differ, $\boldsymbol{\mathcal{E}}^{\rm bulk} \neq \boldsymbol{\mathcal{E}}^{\rm int}$, and thus our theory cannot provide exact coexistence criteria. 
However, as described in the main text, we can still use $\mathbf{E}^{\rm int}$ in the weighted-area construction to estimate the coexistence criteria.
The weighted-area construction is path-dependent as $\boldsymbol{\mathcal{E}}^{\rm bulk} \neq \boldsymbol{\mathcal{E}}^{\rm int}$, however, it is guaranteed to vanish when evaluated along the path corresponding to the spatial species coexistence profiles (the very profiles which we aim to avoid determining).
This has the consequence that integration paths that more closely resemble the coexistence profiles yield more accurate predictions of the coexistence densities.
The interfacial Maxwell construction vector for this system is found to be:
\begin{equation}
    \boldsymbol{\mathcal{E}}^{\rm int} = \begin{bmatrix}
        \exp{\left(2\frac{\lambda_{AAA}}{\overline{\kappa}_{AA}}\rho_A + 2\frac{\overline{\kappa}_{AB}\lambda_{AAA}}{\overline{\kappa}_{AA}^2}\rho_B\right)}& 0 \end{bmatrix}^{\rm T}. 
\end{equation}
The weighted-area Maxwell Construction integrals with $\mathbf{E}^{\rm int}$ for this system are then:
\begin{equation}
    C_A \int_{\rho_A^\alpha}^{\rho_A^\beta} \left(u_A^{\rm bulk} - u_A^{\rm coexist} \right) e^{\left(C_A\rho_A + C_B\rho_B\right)}d\rho_A' + C_B \int_{\rho_B^\alpha}^{\rho_B^\beta} \left(u_A^{\rm bulk} - u_A^{\rm coexist} \right) e^{\left(C_A\rho_A + C_B\rho_B\right)} d\rho_B'  = 0, 
\end{equation}
where $C_A = 2\frac{\lambda_{AAA}}{\overline{\kappa}_{AA}}$ and $C_B = 2\frac{\overline{\kappa}_{AB}\lambda_{AAA}}{\overline{\kappa}_{AA}^2}$.

\begin{figure}
    \centering
    \includegraphics[width=0.6\linewidth]{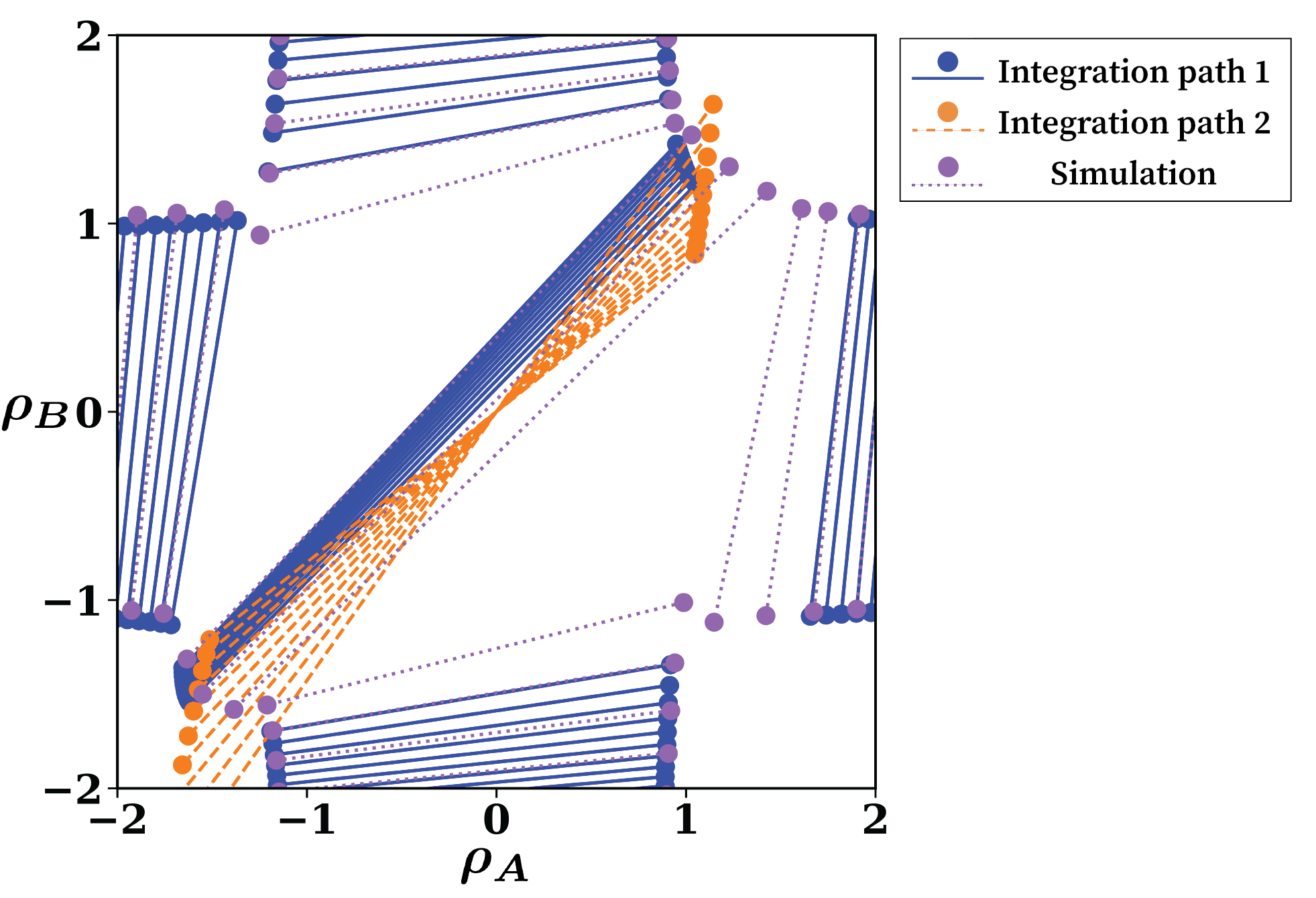}
    \caption{Phase diagram of a MAMB system without exact coexistence criteria. 
    The system parameters are $a= -1$, $\chi = -1$, $\overline{\kappa}_{AA} = \overline{\kappa}_{BB} = \lambda_{AAA} = 0.02$, $\lambda_{AAB} = 0.01$, $\lambda_{ABB} = 0.005$, and $\alpha = \overline{\kappa}_{AB} = \overline{\kappa}_{BA} = 0$.
    The simulation results are compared with the binodal estimated using two different integration paths.
    Integration path 1 is $u_B^{\rm bulk} = u_B^{\rm coexist}$ while integration path 2 is $\rho_A = S\rho_B$, where $S$ is a constant parameter.}
    \label{fig:MAMB_estimate_theory}
\end{figure}

As shown in Fig.~\ref{fig:MAMB_estimate_theory}, we choose two different integration paths for demonstration. 
Integration path 1 is selected with $u_B^{\rm bulk} = u_B^{\rm coexist}$, where we fix the species pseudopotential of $B$ to a constant and solve for coexisting densities.
Integration path 2 is selected with $\rho_A = S\rho_B$, where we vary the constant parameter $S$, and solve for the coexisting densities with $S$ restricting the ratios of species' densities.
Integration path 1 provides high accuracy of the predicted binodals on the edges of the phase diagram where one of $\rho_A$ or $\rho_B$ is relatively constant, while integration path 2 fails to provide accurate predictions.

\section{Approximation of Coexistence Criteria for the NRCH model}
We look to justify approximating the coexistence criteria of the NRCH model by using the weighted-area construction with $\mathbf{E}=\mathbf{E}^{\rm bulk}$ when $\overline{\kappa}_{AB} (\chi - \alpha) \neq \overline{\kappa}_{BA} (\chi + \alpha)$.
This weighted-area construction is equivalent to equating $\mathcal{G}^{\rm bulk} \equiv \boldsymbol{\mathcal{E}}^{\rm bulk} \cdot \mathbf{u}^{\rm bulk} - \int \mathbf{u}^{\rm bulk} \cdot d \boldsymbol{\mathcal{E}}^{\rm bulk}$ across phases.
As detailed in the main text, the difference between $\mathcal{G}^{\rm bulk}$ across phases takes the form:
\begin{equation}
\label{sieq:delta_G}
    \Delta^{\alpha \beta} \mathcal{G}^{\rm bulk} = \left( \frac{\chi + \alpha}{\chi - \alpha} - \frac{\overline{\kappa}_{AB}}{\overline{\kappa}_{BA}} \right) \int_{z^{\alpha}}^{z^{\beta}} \left[ u^{\rm bulk}_B (\boldsymbol{\rho}^{\rm c}) - u^{\rm coexist}_B \right] \partial_z \rho_B^{\rm c} dz,
\end{equation}
where the integral on the right hand side is evaluated along
the coexistence profile, $\boldsymbol{\rho}^{\rm c}(z)$.

\begin{figure}
    \centering
    \includegraphics[width=0.5\linewidth]{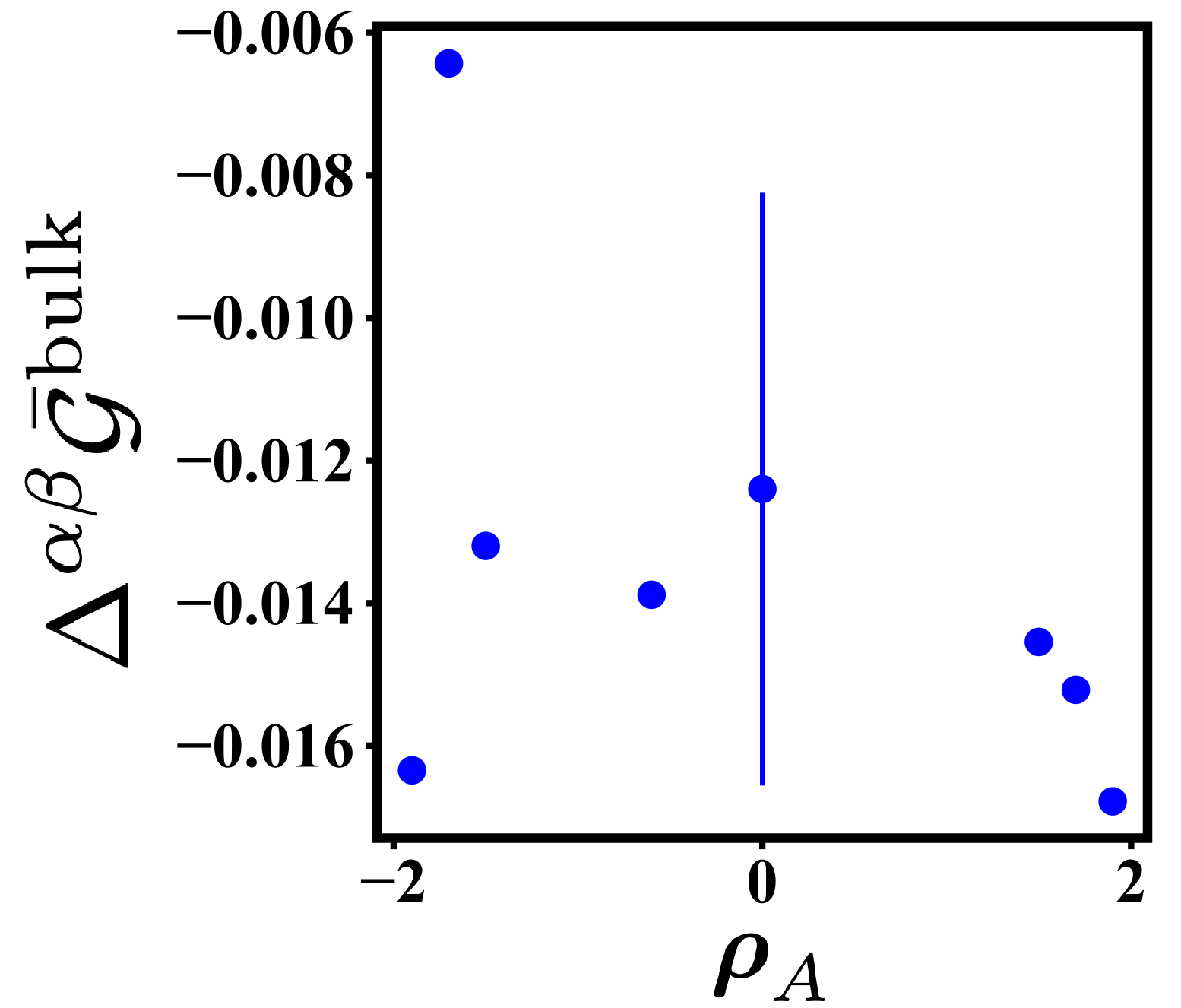}
    \caption{Dimensionless difference of the bulk global quantity between the coexisting phases of an NRCH system breaking the coupling parameter ratio, $\overline{\kappa}_{AB} (\chi - \alpha) \neq \overline{\kappa}_{BA} (\chi + \alpha)$, with respect to the global density of species $A$. 
    The system parameters are described in the main text in Fig.~3(b).
    The error bar shows the standard deviation of $\Delta^{\alpha \beta} \overline{\mathcal{G}}^{\rm bulk}$ with multiple global system densities with the same global $\rho_A$.}
    \label{fig:delta_G}
\end{figure}

Equation~\eqref{sieq:delta_G} provides a raw value for $\Delta^{\alpha \beta} \mathcal{G}^{\rm bulk}$, however it is not clear what constitutes a ``small'' or ``large'' value and therefore when equating $\mathcal{G}^{\rm bulk}$ across phases is a good approximation or not.
To provide a relative measure of the departure from zero of  Eq.~\eqref{sieq:delta_G}, we divide it by the system length in the $z$-direction, $L_z$, to compute a mean of sorts, which we term $\left \langle \Delta^{\alpha \beta} \mathcal{G}^{\rm bulk}_{L_z} \right \rangle$:
\begin{subequations}
\begin{equation}
\label{sieq:delta_G_Lz}
    \left \langle \Delta^{\alpha \beta} \mathcal{G}^{\rm bulk}_{L_z} \right \rangle \equiv \frac{\Delta^{\alpha \beta} \mathcal{G}^{\rm bulk}}{L_z} =  \left( \frac{\chi + \alpha}{\chi - \alpha} - \frac{\overline{\kappa}_{AB}}{\overline{\kappa}_{BA}} \right) \int_{z^{\alpha}}^{z^{\beta}} \frac{\left[ u^{\rm bulk}_B (\boldsymbol{\rho}^{\rm c}) - u^{\rm coexist}_B \right] \partial_z \rho_B^{\rm c}}{L_z} dz,
\end{equation}
with the analogous second moment:
\begin{equation}
\label{sieq:delta_G2_Lz}
    \left \langle \left( \Delta^{\alpha \beta} \mathcal{G}^{\rm bulk}_{L_z} \right)^2 \right \rangle =  \left( \frac{\chi + \alpha}{\chi - \alpha} - \frac{\overline{\kappa}_{AB}}{\overline{\kappa}_{BA}} \right)^2 \int_{z^{\alpha}}^{z^{\beta}} \frac{\left[ u^{\rm bulk}_B (\boldsymbol{\rho}^{\rm c}) - u^{\rm coexist}_B \right]^2 (\partial_z \rho_B^{\rm c})^2}{L_z^2} dz.
\end{equation}
A variance can be defined as ${\rm Var}\left( \Delta^{\alpha \beta} \mathcal{G}^{\rm bulk}_{L_z} \right) \equiv \left \langle \left( \Delta^{\alpha \beta} \mathcal{G}^{\rm bulk}_{L_z} \right)^2 \right \rangle - \left \langle \Delta^{\alpha \beta} \mathcal{G}^{\rm bulk}_{L_z} \right \rangle^2$.
We then define the dimensionless $\Delta^{\alpha \beta} \overline{\mathcal{G}}^{\rm bulk} \equiv \left \langle \Delta^{\alpha \beta} \mathcal{G}^{\rm bulk}_{L_z} \right \rangle / \sqrt{{\rm Var}\left( \Delta^{\alpha \beta} \mathcal{G}^{\rm bulk}_{L_z} \right)}$, which we plot in Fig.~\ref{fig:delta_G}.
\end{subequations}
The consistently low $\Delta^{\alpha \beta} \overline{\mathcal{G}}^{\rm bulk}$ values across all examined global densities suggests that using $\mathbf{E}^{\rm bulk}$ in the weighted-area construction provides a robust approximate form of the final coexistence criterion, as $\mathcal{G}^{\rm bulk}$ is nearly equal in the coexisting phases. 

\section{Completion of System of PDEs to Involution}
In this Section, we complete the system of PDEs resulting from the generalized Gibbs-Duhem relation, $d \mathcal{G}=\sum_i^{n_c} \mathcal{E}_i d u_i$, to involution~\cite{Seiler2010}.
The system of linear PDEs to solve for $\boldsymbol{\mathcal{E}}$ are derived in the main text:
\begin{subequations}
    \label{eq:Esys}
    \begin{align}
    \label{eq:Esys1}
        & \partial_j \mathcal{G}^{\rm bulk} = \sum_i^{n_c} \mathcal{E}_i \partial_j u_i^{\rm bulk}, \\
        \label{eq:Esys2}
        & \partial_k \mathcal{G}_j^{(2, 2)} = \sum_i^{n_c} \partial_k \left( \mathcal{E}_i u_{i j}^{(2,2)} \right), \\
        \label{eq:Esys3}
        & 2 \mathcal{G}_{j k}^{(2, 1)} + \partial_k \mathcal{G}_j^{(2, 2)} = \sum_i^{n_c} \mathcal{E}_i \left( 2 u_{i j k}^{(2,1)} +  \partial_k u_{i j}^{(2,2)} \right), \\
        \label{eq:Esys4}
        & \bigg[\partial_n \mathcal{G}_{j k}^{(2, 1)}\bigg]^{{\rm S}_{n j k}} = \sum_i^{n_c} \mathcal{E}_i \left[  \partial_n u_{i j k}^{(2,1)} \right]^{{\rm S}_{n j k}},
    \end{align}
\end{subequations}
where $\partial_i \equiv \partial / \partial \rho_i$, and each component of the vector $\boldsymbol{\rho}$ are the independent variables. 
Here, the vector $\mathbf{u}^{\rm bulk}(\boldsymbol{\rho})$ and each of the matrices $\mathbf{u}^{(2,1)}(\boldsymbol{\rho})$ and $\mathbf{u}^{(2,2)}(\boldsymbol{\rho})$ are assumed to be known.
The scalar $\mathcal{G}^{\rm bulk}$ and each component of the vectors $\boldsymbol{\mathcal{E}}$ and $\boldsymbol{\mathcal{G}}^{(2, 2)}$ and the matrix $\boldsymbol{\mathcal{G}}^{(2, 1)}$ are the dependent variables to solve for.
Here, the superscript ${\rm S}_{n j k}$ indicates a matrix that is symmetric with respect to exchanging $n$, $j$, and $k$ indices, and results in:
\begin{subequations}
    \label{eq:Esysn}
    \begin{align}
        \label{eq:Esys1n}
        & \partial_j \mathcal{G}^{\rm bulk} = \sum_i^{n_c} \mathcal{E}_i \partial_j u_i^{\rm bulk}, \\
        \label{eq:Esys2n}
        & \partial_k \mathcal{G}_j^{(2, 2)} = \sum_i^{n_c} \partial_k \left( \mathcal{E}_i u_{i j}^{(2,2)} \right), \\
        \label{eq:Esys3n}
        & 2 \mathcal{G}_{j k}^{(2, 1)} + \partial_k \mathcal{G}_j^{(2, 2)} = \sum_i^{n_c} \mathcal{E}_i \left( 2 u_{i j k}^{(2,1)} +  \partial_k u_{i j}^{(2,2)} \right), \\
        \label{eq:Esys4n}
        & \partial_n \mathcal{G}_{j k}^{(2, 1)} + \partial_j \mathcal{G}_{n k}^{(2, 1)} + \partial_k \mathcal{G}_{j n}^{(2, 1)} = \sum_i^{n_c} \mathcal{E}_i\left(  \partial_n u_{i j k}^{(2,1)} + \partial_j u_{i n k}^{(2,1)} + \partial_k u_{i j n}^{(2,1)}\right).
    \end{align}
\end{subequations}
We have $n_c$ independent variables (the number of components of $\boldsymbol{\rho}$), and gain $n_c$ dependent variables each in $\boldsymbol{\mathcal{E}}$ and $\boldsymbol{\mathcal{G}}^{(2, 2)}$, one in $\mathcal{G}^{\rm bulk}$, and $(n_c^2 + n_c)/2$ in $\boldsymbol{\mathcal{G}}^{(2, 1)}$.
Hence, we have in total $m=n_c^2/2 + 5n_c/2 + 1$ dependent variables.
Equation~\eqref{eq:Esys1n} contains $n_c$ equations, Eqs.~\eqref{eq:Esys2n} and \eqref{eq:Esys3n} each contain $n_c^2$ more, and Eq.~\eqref{eq:Esys4n} contains $n_c(n_c+1)(n_c+2)/6$ equations.
Before prolongation, we thus have a system of $t= n_c^3 / 6 + 5 n_c^2/2 + 4 n_c / 3$ equations.

It proves convenient to express all $t$ equations appearing in Eq.~\eqref{eq:Esysn} in vector notation with $\boldsymbol{\Phi} = \mathbf{0}$, where $\boldsymbol{\Phi}$ is defined as:
\begin{equation}
    \boldsymbol{\Phi} = \begin{bmatrix}
        \partial_1 \mathcal{G}^{\rm bulk} - \sum_i^{n_c} \mathcal{E}_i \partial_1 u^{\rm bulk}_i \\
        \vdots \\
        \partial_{n_c} \mathcal{G}^{\rm bulk} - \sum_i^{n_c} \mathcal{E}_i \partial_{n_c} u^{\rm bulk}_i \\
        \partial_1 \mathcal{G}_1^{(2, 2)} - \sum_i^{n_c} \left( \mathcal{E}_i \partial_1 u^{(2,2)}_{i 1} + \partial_1 \mathcal{E}_i u^{(2,2)}_{i 1} \right) \\
        \vdots \\
        \partial_{n_c} \mathcal{G}_{n_c}^{(2, 2)} - \sum_i^{n_c} \left( \mathcal{E}_i \partial_{n_c} u^{(2,2)}_{i n_c} + \partial_n \mathcal{E}_i u^{(2,2)}_{i n_c} \right) \\
        2 \mathcal{G}_{11}^{(2, 1)} + \partial_1 \mathcal{G}_1^{(2, 2)} - \sum_i^{n_c} \mathcal{E}_i \left( 2 u^{(2,1)}_{i 11} + \partial_1 u^{(2,2)}_{i 1} \right) \\
        \vdots \\
        2 \mathcal{G}_{nn}^{(2, 1)} + \partial_{n_c} \mathcal{G}_{n_c}^{(2, 2)} - \sum_i^{n_c} \mathcal{E}_i \left( 2 u^{(2,1)}_{i n_c n_c} + \partial_{n_c} u^{(2,2)}_{i n_c} \right) \\
        \partial_{1} \mathcal{G}_{11}^{(2, 1)} - 2 \mathcal{E}_i \partial_{1} u^{(2,1)}_{i 11} \\
        \vdots \\
        \partial_{n_c} \mathcal{G}_{n_c n_c}^{(2, 1)} - 2 \sum_i^{n_c} \mathcal{E}_i \partial_{n_c} u^{(2,1)}_{i n_c n_c},
    \end{bmatrix}.
\end{equation}

We now consider a system described by two independent variables, $\rho_A$ and $\rho_B$, and hence $n_c=2$, $m=8$, and $t=14$.
Explicitly, our dependent variables are $\mathcal{E}_A$, $\mathcal{E}_B$, $\mathcal{G}^{\rm bulk}$, $\mathcal{G}^{(2, 2)}_A$, $\mathcal{G}^{(2, 2)}_B$, $\mathcal{G}^{(2, 1)}_{A A}$, $\mathcal{G}^{(2, 1)}_{A B}$, and $\mathcal{G}^{(2, 1)}_{B B}$.
Writing out each term in $\boldsymbol{\Phi}$ and simplifying we have:
\begin{equation}
    \label{eq:phifull}
    \boldsymbol{\Phi} = \begin{pmatrix}
        \begin{smallmatrix}
        \partial_A \mathcal{G}^{\rm bulk} - \mathcal{E}_A \partial_A u^{\rm bulk}_A - \mathcal{E}_B \partial_A u^{\rm bulk}_B 
        \end{smallmatrix} \\
        \begin{smallmatrix}
        \partial_B \mathcal{G}^{\rm bulk} - \mathcal{E}_A \partial_B u^{\rm bulk}_A - \mathcal{E}_B \partial_B u^{\rm bulk}_B 
        \end{smallmatrix} \\
        \begin{smallmatrix}
        \partial_A \mathcal{G}_A^{(2, 2)} - \mathcal{E}_A \partial_A u^{(2,2)}_{A A} - \partial_A \mathcal{E}_A u^{(2,2)}_{A A} - \mathcal{E}_B \partial_A u^{(2,2)}_{B A} - \partial_A \mathcal{E}_B u^{(2,2)}_{B A}
        \end{smallmatrix} \\
        \begin{smallmatrix}
        \partial_B \mathcal{G}_A^{(2, 2)} - \mathcal{E}_A \partial_B u^{(2,2)}_{A A} - \partial_B \mathcal{E}_A u^{(2,2)}_{A A} - \mathcal{E}_B \partial_B u^{(2,2)}_{B A} - \partial_B \mathcal{E}_B u^{(2,2)}_{B A}
        \end{smallmatrix} \\
        \begin{smallmatrix}
        \partial_A \mathcal{G}_B^{(2, 2)} - \mathcal{E}_A \partial_A u^{(2,2)}_{A B} - \partial_A \mathcal{E}_A u^{(2,2)}_{A B} - \mathcal{E}_B \partial_A u^{(2,2)}_{B B} - \partial_A \mathcal{E}_B u^{(2,2)}_{B B}
        \end{smallmatrix} \\
        \begin{smallmatrix}
        \partial_B \mathcal{G}_B^{(2, 2)} - \mathcal{E}_A \partial_B u^{(2,2)}_{A B} - \partial_B \mathcal{E}_A u^{(2,2)}_{A B} - \mathcal{E}_B \partial_B u^{(2,2)}_{B B} - \partial_B \mathcal{E}_B u^{(2,2)}_{B B}
        \end{smallmatrix} \\
        \begin{smallmatrix}
        2 \mathcal{G}_{AA}^{(2, 1)} + \partial_A \mathcal{G}_A^{(2, 2)} - \mathcal{E}_A \left( 2 u^{(2,1)}_{A A A} + \partial_A u^{(2,2)}_{A A} \right) - \mathcal{E}_B \left( 2 u^{(2,1)}_{B A A} + \partial_A u^{(2,2)}_{B A} \right)
        \end{smallmatrix} \\
        \begin{smallmatrix}
        2 \mathcal{G}_{AB}^{(2, 1)} + \partial_B \mathcal{G}_A^{(2, 2)} - \mathcal{E}_A \left( 2 u^{(2,1)}_{A A B} + \partial_B u^{(2,2)}_{A A} \right) - \mathcal{E}_B \left( 2 u^{(2,1)}_{B A B} + \partial_B u^{(2,2)}_{B A} \right)
        \end{smallmatrix} \\
        \begin{smallmatrix}
        2 \mathcal{G}_{AB}^{(2, 1)} + \partial_A \mathcal{G}_B^{(2, 2)} - \mathcal{E}_A \left( 2 u^{(2,1)}_{A B A} + \partial_A u^{(2,2)}_{A B} \right) - \mathcal{E}_B \left( 2 u^{(2,1)}_{B B A} + \partial_A u^{(2,2)}_{B B} \right)
        \end{smallmatrix} \\
        \begin{smallmatrix}
        2 \mathcal{G}_{BB}^{(2, 1)} + \partial_B \mathcal{G}_B^{(2, 2)} - \mathcal{E}_A \left( 2 u^{(2,1)}_{A B B} + \partial_B u^{(2,2)}_{A B} \right) - \mathcal{E}_B \left( 2 u^{(2,1)}_{B B B} + \partial_B u^{(2,2)}_{B B} \right) \\
        \end{smallmatrix} \\
        \begin{smallmatrix}
        \partial_{A} \mathcal{G}_{AA}^{(2, 1)} - \mathcal{E}_A \partial_{A} u^{(2,1)}_{A A A} - \mathcal{E}_B \partial_{A} u^{(2,1)}_{B A A}
        \end{smallmatrix} \\
        \begin{smallmatrix}
        2 \partial_{A} \mathcal{G}_{AB}^{(2, 1)} + \partial_{B} \mathcal{G}_{AA}^{(2, 1)} - \mathcal{E}_A \left( 2 \partial_{A} u^{(2,1)}_{A A B} + \partial_{B} u^{(2,1)}_{A A A} \right) - \mathcal{E}_B \left( 2 \partial_{A} u^{(2,1)}_{B A B} + \partial_{B} u^{(2,1)}_{B A A} \right)
        \end{smallmatrix} \\
        \begin{smallmatrix}
        2 \partial_{B} \mathcal{G}_{AB}^{(2, 1)} + \partial_{A} \mathcal{G}_{BB}^{(2, 1)} - \mathcal{E}_A \left( 2 \partial_{B} u^{(2,1)}_{A A B} + \partial_{A} u^{(2,1)}_{A B B} \right) - \mathcal{E}_B \left( 2 \partial_{B} u^{(2,1)}_{B A B} + \partial_{A} u^{(2,1)}_{B B B} \right)
        \end{smallmatrix} \\
        \begin{smallmatrix}
        \partial_{B} \mathcal{G}_{B B}^{(2, 1)} - \mathcal{E}_A \partial_{B} u^{(2,1)}_{A B B} - \mathcal{E}_B \partial_{B} u^{(2,1)}_{B B B}
        \end{smallmatrix}
    \end{pmatrix},
\end{equation}
where we have used the symmetry of Eq.~\eqref{eq:Esys4n} with exchange of the $n,j,k$ indices.

We now look to complete the system of equations in Eq.~\eqref{eq:phifull} to involution following Ref.~\cite{Seiler2010}.
Without loss of generality, we consider the ordering $\rho_A \succ \rho_B$.
We define the order of differentiation of a term to be $q=q_A + q_B$ where $q_A$ and $q_B$ are the order of differentiation with respect to $\rho_A$ and $\rho_B$ within that term, respectively. 
If an equation of order $q$ has any terms (also of order $q$) with $q_A>q_B$ then $\rho_A$ is termed ``multiplicative'' for that equation.
Otherwise, $\rho_A$ is ``non-multiplicative.''
With our choice of $\rho_A \succ \rho_B$,  $\rho_B$ is multiplicative for all equations.
A system of PDEs, $\boldsymbol{\Phi}=\mathbf{0}$, that is locally involutive has the property that the prolongations (i.e.~differentiations) of each equation with respect to its non-multiplicative variables can be expressed as a linear combination of the equations themselves and the prolongations with respect to the multiplicative variables.
If a non-multiplicative prolongation is linearly independent of $\boldsymbol{\Phi}$ and its multiplicative prolongations, the non-multiplicative prolongation is a \textit{compatibility condition}: an additional PDE that a solution to $\boldsymbol{\Phi}=\mathbf{0}$ must satisfy.
A single prolongation of Eq.~\eqref{eq:phifull} results in four second-order compatibility conditions.
A second prolongation results in an additional compatibility condition that is third order. 
A third prolongation reveals that our system of PDEs was involutively complete after two prolongations. 
Our five final compatibility conditions are:
\begin{equation}
    \boldsymbol{\Phi}^{\rm compat} = 
    \resizebox{.9\hsize}{!}{ $
    \begin{pmatrix}
        \begin{smallmatrix}
        \partial^2_{AB} \mathcal{G}^{\rm bulk} - \mathcal{E}_A \partial^2_{AB} u^{\rm bulk}_A - \partial_A \mathcal{E}_A \partial_B u^{\rm bulk}_A - \mathcal{E}_B \partial^2_{AB} u^{\rm bulk}_B - \partial_A \mathcal{E}_B \partial_B u^{\rm bulk}_B
        \end{smallmatrix} \\
        \begin{smallmatrix}
            \partial^2_{AB} \mathcal{G}_A^{(2, 2)} + 2 \partial_A \mathcal{G}_{AB}^{(2, 1)} - \mathcal{E}_A \left(2 \partial_A u^{(2, 1)}_{AAB} + \partial^2_{AB} u^{(2, 2)}_{AA} \right) - \partial_A \mathcal{E}_A \left(2 u^{(2, 1)}_{AAB} + \partial_{B} u^{(2, 2)}_{AA} \right) - \mathcal{E}_B \left(2 \partial_A u^{(2, 1)}_{BAB} + \partial^2_{AB} u^{(2, 2)}_{BA} \right) - \partial_A \mathcal{E}_B \left(2 u^{(2, 1)}_{BAB} + \partial_{B} u^{(2, 2)}_{BA} \right)
        \end{smallmatrix} \\
        \begin{smallmatrix}
            \partial^2_{AB} \mathcal{G}_B^{(2, 2)} + 2 \partial_A \mathcal{G}_{BB}^{(2, 1)}  - \mathcal{E}_A \left(2 \partial_A u^{(2, 1)}_{ABB} + \partial^2_{AB} u^{(2, 2)}_{AB} \right) - \partial_A \mathcal{E}_A \left(2 u^{(2, 1)}_{ABB} + \partial_{B} u^{(2, 2)}_{AB} \right) - \mathcal{E}_B \left(2 \partial_A u^{(2, 1)}_{BBB} + \partial^2_{AB} u^{(2, 2)}_{BB} \right) - \partial_A \mathcal{E}_B \left(2 u^{(2, 1)}_{BBB} + \partial_{B} u^{(2, 2)}_{BB} \right)
        \end{smallmatrix} \\
        \begin{smallmatrix}
        \partial^2_{AB} \mathcal{G}_{BB}^{(2, 1)} - \mathcal{E}_A \partial^2_{AB} u^{(2, 1)}_{ABB} - \partial_A \mathcal{E}_A \partial_B u^{(2, 1)}_{ABB} - \mathcal{E}_B \partial^2_{AB} u^{(2, 1)}_{BBB} - \partial_A \mathcal{E}_B \partial_B u^{(2, 1)}_{BBB}
        \end{smallmatrix} \\
        \begin{smallmatrix}
        \partial^3_{A AB} \mathcal{G}_{BB}^{(2, 1)} - 2 \partial_A \mathcal{E}_A \partial^2_{AB} u^{(2, 1)}_{ABB} - \mathcal{E}_A \partial^3_{AAB} u^{(2, 1)}_{ABB} - \partial^2_{AA} \mathcal{E}_A \partial_B u^{(2, 1)}_{ABB} - 2 \partial_A \mathcal{E}_B \partial^2_{AB} u^{(2, 1)}_{BBB} - \mathcal{E}_B \partial^3_{AAB} u^{(2, 1)}_{BBB} - \partial^2_{AA} \mathcal{E}_B \partial_B u^{(2, 1)}_{BBB}
        \end{smallmatrix}
    \end{pmatrix}. $ }
\end{equation}
Our involutive system of PDEs is now $\tilde{\boldsymbol{\Phi}} = \begin{bmatrix}
    \Phi_1 & \cdots & \Phi_{16} & \Phi^{\rm compat}_1 & \cdots & \Phi^{\rm compat}_5
\end{bmatrix}$:
\begin{equation}
    \label{eq:phiinv}
    \tilde{\boldsymbol{\Phi}} = \resizebox{.9\hsize}{!}{$\begin{pmatrix}
    \begin{smallmatrix}
        \partial_A \mathcal{G}^{\rm bulk} - \mathcal{E}_A \partial_A u^{\rm bulk}_A - \mathcal{E}_B \partial_A u^{\rm bulk}_B 
        \end{smallmatrix} \\
        \begin{smallmatrix}
        \partial_B \mathcal{G}^{\rm bulk} - \mathcal{E}_A \partial_B u^{\rm bulk}_A - \mathcal{E}_B \partial_B u^{\rm bulk}_B 
        \end{smallmatrix} \\
        \begin{smallmatrix}
        \partial_A \mathcal{G}_A^{(2, 2)} - \mathcal{E}_A \partial_A u^{(2,2)}_{A A} - \partial_A \mathcal{E}_A u^{(2,2)}_{A A} - \mathcal{E}_B \partial_A u^{(2,2)}_{B A} - \partial_A \mathcal{E}_B u^{(2,2)}_{B A}
        \end{smallmatrix} \\
        \begin{smallmatrix}
        \partial_B \mathcal{G}_A^{(2, 2)} - \mathcal{E}_A \partial_B u^{(2,2)}_{A A} - \partial_B \mathcal{E}_A u^{(2,2)}_{A A} - \mathcal{E}_B \partial_B u^{(2,2)}_{B A} - \partial_B \mathcal{E}_B u^{(2,2)}_{B A}
        \end{smallmatrix} \\
        \begin{smallmatrix}
        \partial_A \mathcal{G}_B^{(2, 2)} - \mathcal{E}_A \partial_A u^{(2,2)}_{A B} - \partial_A \mathcal{E}_A u^{(2,2)}_{A B} - \mathcal{E}_B \partial_A u^{(2,2)}_{B B} - \partial_A \mathcal{E}_B u^{(2,2)}_{B B}
        \end{smallmatrix} \\
        \begin{smallmatrix}
        \partial_B \mathcal{G}_B^{(2, 2)} - \mathcal{E}_A \partial_B u^{(2,2)}_{A B} - \partial_B \mathcal{E}_A u^{(2,2)}_{A B} - \mathcal{E}_B \partial_B u^{(2,2)}_{B B} - \partial_B \mathcal{E}_B u^{(2,2)}_{B B}
        \end{smallmatrix} \\
        \begin{smallmatrix}
        2 \mathcal{G}_{AA}^{(2, 1)} + \partial_A \mathcal{G}_A^{(2, 2)} - \mathcal{E}_A \left( 2 u^{(2,1)}_{A A A} + \partial_A u^{(2,2)}_{A A} \right) - \mathcal{E}_B \left( 2 u^{(2,1)}_{B A A} + \partial_A u^{(2,2)}_{B A} \right)
        \end{smallmatrix} \\
        \begin{smallmatrix}
        2 \mathcal{G}_{AB}^{(2, 1)} + \partial_B \mathcal{G}_A^{(2, 2)} - \mathcal{E}_A \left( 2 u^{(2,1)}_{A A B} + \partial_B u^{(2,2)}_{A A} \right) - \mathcal{E}_B \left( 2 u^{(2,1)}_{B A B} + \partial_B u^{(2,2)}_{B A} \right)
        \end{smallmatrix} \\
        \begin{smallmatrix}
        2 \mathcal{G}_{AB}^{(2, 1)} + \partial_A \mathcal{G}_B^{(2, 2)} - \mathcal{E}_A \left( 2 u^{(2,1)}_{A B A} + \partial_A u^{(2,2)}_{A B} \right) - \mathcal{E}_B \left( 2 u^{(2,1)}_{B B A} + \partial_A u^{(2,2)}_{B B} \right)
        \end{smallmatrix} \\
        \begin{smallmatrix}
        2 \mathcal{G}_{BB}^{(2, 1)} + \partial_B \mathcal{G}_B^{(2, 2)} - \mathcal{E}_A \left( 2 u^{(2,1)}_{A B B} + \partial_B u^{(2,2)}_{A B} \right) - \mathcal{E}_B \left( 2 u^{(2,1)}_{B B B} + \partial_B u^{(2,2)}_{B B} \right) \\
        \end{smallmatrix} \\
        \begin{smallmatrix}
        \partial_{A} \mathcal{G}_{AA}^{(2, 1)} - \mathcal{E}_A \partial_{A} u^{(2,1)}_{A A A} - \mathcal{E}_B \partial_{A} u^{(2,1)}_{B A A}
        \end{smallmatrix} \\
        \begin{smallmatrix}
        2 \partial_{A} \mathcal{G}_{AB}^{(2, 1)} + \partial_{B} \mathcal{G}_{AA}^{(2, 1)} - \mathcal{E}_A \left( 2 \partial_{A} u^{(2,1)}_{A A B} + \partial_{B} u^{(2,1)}_{A A A} \right) - \mathcal{E}_B \left( 2 \partial_{A} u^{(2,1)}_{B A B} + \partial_{B} u^{(2,1)}_{B A A} \right)
        \end{smallmatrix} \\
        \begin{smallmatrix}
        2 \partial_{B} \mathcal{G}_{AB}^{(2, 1)} + \partial_{A} \mathcal{G}_{BB}^{(2, 1)} - \mathcal{E}_A \left( 2 \partial_{B} u^{(2,1)}_{A A B} + \partial_{A} u^{(2,1)}_{A B B} \right) - \mathcal{E}_B \left( 2 \partial_{B} u^{(2,1)}_{B A B} + \partial_{A} u^{(2,1)}_{B B B} \right)
        \end{smallmatrix} \\
        \begin{smallmatrix}
        \partial_{B} \mathcal{G}_{B B}^{(2, 1)} - \mathcal{E}_A \partial_{B} u^{(2,1)}_{A B B} - \mathcal{E}_B \partial_{B} u^{(2,1)}_{B B B}
        \end{smallmatrix} \\
                \begin{smallmatrix}
            \partial^2_{AB} \mathcal{G}^{\rm bulk} - \mathcal{E}_A \partial^2_{AB} u^{\rm bulk}_A - \partial_A \mathcal{E}_A \partial_B u^{\rm bulk}_A - \mathcal{E}_B \partial^2_{AB} u^{\rm bulk}_B - \partial_A \mathcal{E}_B \partial_B u^{\rm bulk}_B
        \end{smallmatrix} \\
        \begin{smallmatrix}
            \partial^2_{AB} \mathcal{G}_A^{(2, 2)} + 2 \partial_A \mathcal{G}_{AB}^{(2, 1)} - \mathcal{E}_A \left(2 \partial_A u^{(2, 1)}_{AAB} + \partial^2_{AB} u^{(2, 2)}_{AA} \right) - \partial_A \mathcal{E}_A \left(2 u^{(2, 1)}_{AAB} + \partial_{B} u^{(2, 2)}_{AA} \right) - \mathcal{E}_B \left(2 \partial_A u^{(2, 1)}_{BAB} + \partial^2_{AB} u^{(2, 2)}_{BA} \right) - \partial_A \mathcal{E}_B \left(2 u^{(2, 1)}_{BAB} + \partial_{B} u^{(2, 2)}_{BA} \right)
        \end{smallmatrix} \\
        \begin{smallmatrix}
            \partial^2_{AB} \mathcal{G}_B^{(2, 2)} + 2 \partial_A \mathcal{G}_{BB}^{(2, 1)}  - \mathcal{E}_A \left(2 \partial_A u^{(2, 1)}_{ABB} + \partial^2_{AB} u^{(2, 2)}_{AB} \right) - \partial_A \mathcal{E}_A \left(2 u^{(2, 1)}_{ABB} + \partial_{B} u^{(2, 2)}_{AB} \right) - \mathcal{E}_B \left(2 \partial_A u^{(2, 1)}_{BBB} + \partial^2_{AB} u^{(2, 2)}_{BB} \right) - \partial_A \mathcal{E}_B \left(2 u^{(2, 1)}_{BBB} + \partial_{B} u^{(2, 2)}_{BB} \right)
        \end{smallmatrix} \\
        \begin{smallmatrix}
        \partial^2_{AB} \mathcal{G}_{BB}^{(2, 1)} - \mathcal{E}_A \partial^2_{AB} u^{(2, 1)}_{ABB} - \partial_A \mathcal{E}_A \partial_B u^{(2, 1)}_{ABB} - \mathcal{E}_B \partial^2_{AB} u^{(2, 1)}_{BBB} - \partial_A \mathcal{E}_B \partial_B u^{(2, 1)}_{BBB}
        \end{smallmatrix} \\
        \begin{smallmatrix}
        \partial^3_{A AB} \mathcal{G}_{BB}^{(2, 1)} - 2 \partial_A \mathcal{E}_A \partial^2_{AB} u^{(2, 1)}_{ABB} - \mathcal{E}_A \partial^3_{AAB} u^{(2, 1)}_{ABB} - \partial^2_{AA} \mathcal{E}_A \partial_B u^{(2, 1)}_{ABB} - 2 \partial_A \mathcal{E}_B \partial^2_{AB} u^{(2, 1)}_{BBB} - \mathcal{E}_B \partial^3_{AAB} u^{(2, 1)}_{BBB} - \partial^2_{AA} \mathcal{E}_B \partial_B u^{(2, 1)}_{BBB}
    \end{smallmatrix}
    \end{pmatrix}.$}
\end{equation}
The importance of completing the system of PDEs to involution can now be appreciated, as it yields five additional compatibility conditions (when $n_c=2$) that must be satisfied for a well-defined solution to the generalized Gibbs-Duhem relation ($d \mathcal{G} = \mathcal{E}_A d u_A + \mathcal{E}_B d u_B$) to exist.

\section{Derivation of Multicomponent Active Model B+}
In this Section, we show that Multicompnent Active Model B+ (MAMB+) introduced in the main text is the most general multicomponent nonequilibrium field theory for the effective body forces, up to third-order in spatial density gradients and in one spatial dimension.
Considering a spatially odd effective body force, the most general effective body force before simplification is (we adopt indicial notation, where repeated indices imply a summation over that index):
\begin{subequations}
    \label{eqSI:general_MAMB+}
    \begin{align}
        \label{eqSI:general_bieff}
        & b_i^{\rm eff} = \left[ \mathcal{T}_{i j}^{0} \right]^{-1} \left( \partial_z u^0_j + \eta_{j k} \partial_z \rho_k - \theta_{jknm} \partial_z \rho_k \partial_z \rho_n \partial_z \rho_m - \xi_{jkn} \partial^2_{zz} \rho_k \partial_z \rho_n - \omega_{jk} \partial^3_{zzz} \rho_k \right), \\
        \label{eqSI:general_uj}
        & u_j^0 = \frac{\delta \mathcal{F}}{\delta \rho_j} + \tau_j - \lambda_{j k n} \partial_{z} \rho_k \partial_z \rho_n - \pi_{j k} \partial^2_{zz} \rho_k, \\
        \label{eqSI:general_FE}
        & \mathcal{F} = \int_V d \mathbf{r} \left( f^{\rm bulk} + \frac{\kappa_{kn}}{2} \partial_{z} \rho_k \partial_z \rho_n \right), \\
        \label{eqSI:general_deltaFE}
        & \frac{\delta \mathcal{F}}{\delta \rho_j} = \frac{\partial f^{\rm bulk}}{\partial \rho_j} - \frac{1}{2} \frac{\partial \kappa_{jk}}{\partial \rho_n} \partial_{z} \rho_k \partial_z \rho_n - \kappa_{jk} \partial^2_{zz} \rho_k,
    \end{align}
\end{subequations}
where $\left[ \mathcal{T}_{i j}^{0} \right]^{-1}$ is the inverse of the would-be transformation tensor, $\mathcal{T}_{i j}^{0}$, if the non-gradient terms in $b_i^{\rm eff}$ (those other than $\partial_z u_j^0$) were identically zero, $u_j^0$ is the would-be species pseudopotential if the non-gradient terms were all zero, and every component of $\eta_{j k}$, $\theta_{jknm}$, $\xi_{jkn}$, and $\omega_{jk}$ is generally an equation of state that prevents the body force from being expressed as a gradient of $u^0_j$.
As the $k$, $n$, and $m$ indices in $\theta_{jknm}$ are contracted into a symmetric rank 3 tensor, we define $\theta_{jknm}$ to be symmetric with respect to permuting the $k$, $n$, and $m$ indices.
$u^0_j$ is described by $\mathcal{F}$, the would-be free energy functional if the non-variational terms (those other than $\delta \mathcal{F} / \delta \rho_j$) were identically zero, and $\tau_j$, $\lambda_{jkn}$, and $\pi_{jk}$, whose components are generally all equations of state that prevent $u^0_j$ from being the functional derivative of $\mathcal{F}$ with respect to $\rho_j$.
We define $\lambda_{jkn}$ to be symmetric with respect to exchanging the $k$ and $n$ indices as these indices are double contracted into a symmetric rank 2 tensor.
Lastly, the would-be free energy $\mathcal{F}$ is described by the bulk contribution, $f^{\rm bulk}$, and square-gradient contributions coupled through the symmetric, positive-definite matrix $\kappa_{kn}$, whose components are generally equations of state.
Notably, the derivatives of $\kappa_{kn}$, $\partial \kappa_{kn} / \partial \rho_j$, are symmetric with respect to permuting the $j$, $k$, and $n$ indices from the definition of $\kappa_{kn}$ in classical density functional theory~\cite{Hansen2013}.

We now look to eliminate redundant terms in $\eqref{eqSI:general_MAMB+}$.
First, we note $\omega_{jk} \partial^3_{zzz} \rho_k = \partial_z \left( \omega_{jk} \partial^2_{zz} \rho_k \right) - \partial \omega_{jk} / \partial \rho_n \partial^2_{zz} \rho_k \partial_z \rho_n$ and hence we have:
\begin{equation}
\label{eqsi:bieff_mamb+_1}
    b_i^{\rm eff} = \left[ \mathcal{T}_{i j}^{0} \right]^{-1} \bigg( \partial_z \left( u_j^0 - \omega_{jk} \partial^2_{zz} \rho_k \right) + \eta_{j k} \partial_z \rho_k - \theta_{jknm} \partial_z \rho_k \partial_z \rho_n \partial_z \rho_m - \overline{\xi}_{jkn} \partial^2_{zz} \rho_k \partial_z \rho_n \bigg),
\end{equation}
where we have defined $\overline{\xi}_{jkn} \equiv \xi_{jkn} - \partial \omega_{jk} / \partial \rho_n$.
We split $\overline{\xi}_{jkn} = \overline{\xi}_{jkn}^{{\rm S}_{kn}} + \overline{\xi}_{jkn}^{{\rm A}_{kn}}$ into symmetric and antisymmetric parts (with respect to exchanging the $k$ and $n$ indices, indicated by superscipts ${\rm S}_{kn}$ and ${\rm A}_{kn}$, respectively) to obtain:
\begin{subequations}
\label{eqsi:bieff_mamb+_2}
    \begin{multline}
    b_i^{\rm eff} = \left[ \mathcal{T}_{i j}^{0} \right]^{-1} \bigg( \partial_z \left( u_j^0 - \frac{1}{2} \overline{\xi}_{jkn}^{{\rm S}_{kn}} \partial_z \rho_k \partial_z \rho_n - \omega_{jk} \partial^2_{zz} \rho_k \right) + \eta_{j k} \partial_z \rho_k - \overline{\theta}_{jknm} \partial_z \rho_k \partial_z \rho_n \partial_z \rho_m \\ - \overline{\xi}_{jkn}^{{\rm A}_{kn}} \partial^2_{zz} \rho_k \partial_z \rho_n \bigg),
\end{multline}
where:
\begin{equation}
    \overline{\theta}_{jknm} \equiv \theta_{jknm} - \frac{1}{2} \left[\frac{\partial \overline{\xi}_{jkn}^{{\rm S}_{kn}}}{\partial \rho_m}\right]^{{\rm S}_{knm}},
\end{equation}
\end{subequations}
where the ${\rm S}_{knm}$ superscript indicates a quantity that is symmetric under permutation of the $k$, $n$, and $m$ indices.
We note that in moving from Eq.~\eqref{eqsi:bieff_mamb+_1} to Eq.~\eqref{eqsi:bieff_mamb+_2} we have invoked to the product rule to re-express $\overline{\xi}^{{\rm S}_{kn}}_{j k n} \partial^2_{zz} \rho_k \partial_z \rho_n$ as $\frac{1}{2} \partial_z \left( \overline{\xi}^{{\rm S}_{kn}}_{j k n} \partial_{z} \rho_k \partial_z \rho_n \right) - \frac{1}{2} \left[ \partial \overline{\xi}^{{\rm S}_{kn}}_{j k n} / \partial \rho_m \right]^{{\rm S}_{k n m}} \partial_{z} \rho_k \partial_z \rho_n \partial_z \rho_m$.

We now look to extract gradient-like portions, if any, from $\eta_{jk} \partial_z \rho_k$.
The gradient contributions to this term can simply be adsorbed into the definition of the bulk contribution to $u_j^0$: only the non-gradient (or integrable) terms associated with $\eta_{jk} \partial_z \rho_k$ are needed for a minimal theory.
Specifically, if $\partial \eta_{jk} \partial / \rho_n = \partial \eta_{jn} \partial / \rho_k \ \forall k \neq n \in \mathcal{C}$ for a species $j$, one can define $\tau^{(\eta)}_{j} \equiv \int d \rho_k \eta_{jk}$.
This gradient structure is not guaranteed, but we can express each $\eta_{jk}$ as a sum of terms with dependencies on different subsets of species densities:
\begin{equation}
    \eta_{jk} \left( \{ \rho_i \ \forall i \in \mathcal{C} \} \right) = \eta_{jk}^{\rm const} + \sum_{a=1}^{n_c} \sum_{\{\mathcal{S} \subseteq \mathcal{C}, \ |\mathcal{S}|=a\}} \eta_{jk}^{\mathcal{S}} \left( \{ \rho_i \ \forall i \in \mathcal{S} \} \right).
\end{equation}
Here, $\eta_{jk}^{\rm const}$ is a constant with respect to the species densities.
Here, the index $a$ represents the number of species densities that each term will depend on and the second summation corresponds to all possible subsets of species density combinations for $a$ species.
For example, when $a = 1$, there are $n_c$ terms, each solely depending on a single distinct species density.
When $a = n_c$, there is only a single term which depends on all $n_c$ densities. 
In general, the number of terms for a given $a$ will be $n_c!/a!(n_c-a)!$, the number of unique combinations of $a$ species densities.
Differentiating with respect to $\rho_n$ where $n \neq k$ we have:
\begin{equation}
    \frac{\partial \eta_{jk}}{\partial \rho_n} = \sum_{a=1}^{n_c} \sum_{\{\mathcal{S} \subseteq \mathcal{C}, \ |\mathcal{S}|=a, \ n \in \mathcal{S}\}} \frac{\partial \eta_{jk}^{\mathcal{S}}}{\partial \rho_n},
\end{equation}
where the second sum is now only over subsets $\mathcal{S}$ of size $a$ that contain the index $n$.
Splitting $\eta_{jk} = \eta_{jk}^{{\rm I}} + \eta_{jk}^{{\rm N}}$ into integrable (superscript ${\rm I}$) and nonintegrable (superscript ${\rm N}$) parts, we immediately see that every $\eta_{jk}^{\rm const}$ and every $\eta_{jk}^{\mathcal{S}}$ for the corresponding subset of size one $\mathcal{S}=\{ k \}$ (i.e.,~the term in $\eta_{jk}$ that only depends on the associated $\rho_k$) are integrable and therefore contained in $\eta_{jk}^{{\rm I}}$.
We define the sum of these terms to be $\eta_{jk}^{{\rm I},0}$.
The terms where the integrability is in question are thus $\eta_{jk}^{\mathcal{S}} \ \forall \mathcal{S} \neq \{ k \}$.
We define the sum of the largest subset of these terms that satisfy the following relation:
\begin{equation}
    \sum_{\mathcal{S}'} \frac{\partial \eta_{jk}^{\mathcal{S'}}}{\partial \rho_n} = \sum_{\mathcal{S}'} \frac{\partial \eta_{jn}^{\mathcal{S'}}}{\partial \rho_k},
\end{equation}
to be $\eta_{jk}^{{\rm I}'}$, where the sum is over subsets of the species densities $\mathcal{S}'$ that compose the largest subset of terms that satisfy the above symmetry.
We then identify $\eta_{jk}^{{\rm I}} = \eta_{jk}^{{\rm I},0} + \eta_{jk}^{{\rm I}'}$ and hence $\eta_{jk}^{{\rm N}} = \sum_{\mathcal{S}'' \not\subset \mathcal{S}'} \eta_{jk}^{\mathcal{S}''}$ is the sum of all other terms.
Defining $\tau^{(\eta)}_{j} \equiv \int d \rho_k \eta_{jk}^{{\rm I}}$ we can express the effective body forces as:
\begin{subequations}
    \begin{equation}
    \label{eqSI:mostgen_bieff}
    b_i^{\rm eff} = \left[ \mathcal{T}_{i j}^{0} \right]^{-1} \bigg( \partial_z \overline{u}_j^0 + \eta_{j k}^{{\rm N}} \partial_z \rho_k - \overline{\theta}_{jknm} \partial_z \rho_k \partial_z \rho_n \partial_z \rho_m - \overline{\xi}_{jkn}^{{\rm A}_{kn}} \partial^2_{zz} \rho_k \partial_z \rho_n \bigg),
\end{equation}
where:
\begin{equation}
    \overline{u}_j^0 \equiv u_j^0 + \tau^{(\eta)}_j - \frac{1}{2} \overline{\xi}_{jkn}^{{\rm S}_{kn}} \partial_z \rho_k \partial_z \rho_n - \omega_{jk} \partial^2_{zz} \rho_k.
\end{equation}
\end{subequations}
This is the most general form for the effective body forces one can obtain at third-order in spatial gradients.
The non-gradient terms thus correspond to the $\frac{n_c}{2}(n_c^2 - n_c)$ unique components of $\overline{\xi}_{jkn}^{{\rm A}_{kn}}$, the $\frac{n_c}{6}(n_c + 1)(n_c + 2)$ unique components of $\overline{\theta}_{jknm}$, and the $n_c (n_c - 1)$ unique components of $\eta_{j k}^{{\rm N}}$.
We note that for each $j$, there are only $n_c-1$ unique components of $\eta_{j k}^{{\rm N}}$ as we may always redefine $\eta_{j k}^{{\rm I}}$ such that one component $k$ of $\eta_{j k}^{{\rm N}}$ is zero per index $j$.
This ensures that there are no first-order non-gradient terms in the one-component effective body force, as one may always integrate $\tau^{(\eta)} = \int d X \eta(X)$ when $\eta$ is a single-variable function.

One may attempt to manipulate the $\partial_z \rho_k \partial_z \rho_n \partial_z \rho_m$ term by integrating its coefficient, however this will introduce nonintegrable coefficients on $\partial_z \rho_k \partial_z \rho_n \partial_z \rho_m$ in an analogous manner to $\eta_{jk}^{{\rm N}}$.
Additionally, doing so adds terms to the coefficients of $\partial^2_{zz} \rho_k \partial_z \rho_n$ that are symmetric with respect to exchanging $k$ and $n$ and hence they undo the manipulations done to simplify the coefficient of $\partial^2_{zz} \rho_k \partial_z \rho_n$ to be strictly antisymmetric.
In this sense, performing these manipulations does not simplify the effective body forces any further, and consequently we conclude Eq.~\eqref{eqSI:mostgen_bieff} is the most general form of the effective body forces.

We now analyze the form of $\overline{u}_j^0$:
\begin{multline}
    \overline{u}_j^0 = \frac{\delta \mathcal{F}}{\delta \rho_j} + \tau_j + \tau_j^{(\eta)} - \left( \lambda_{j k n} + \frac{1}{2} \overline{\xi}_{jkn}^{{\rm S}_{kn}} \right) \partial_{z} \rho_k \partial_z \rho_n - \left( \pi_{j k} + \omega_{jk} \right) \partial^2_{zz} \rho_k \\ = \frac{\partial f^{\rm bulk}}{\partial \rho_j} + \overline{\tau}_j - \left( \lambda_{j k n} + \frac{1}{2} \overline{\xi}_{jkn}^{{\rm S}_{kn}} + \frac{1}{2} \frac{\partial \kappa_{j k}}{\partial \rho_n} \right) \partial_{z} \rho_k \partial_z \rho_n - \left( \pi_{j k} + \omega_{jk} + \kappa_{jk} \right) \partial^2_{zz} \rho_k,
\end{multline}
where we have defined $\overline{\tau}_j \equiv \tau_j + \tau_j^{(\eta)}$.
We now split $\pi_{j k} + \omega_{jk} + \kappa_{jk}$ into symmetric and antisymmetric (with respect to exchanging the $j$ and $k$ indices) parts, and further split the symmetric part into contributions whose derivatives with respect to $\rho_n$ are symmetric with respect to permuting the $j$, $k$, and $n$ indices.
Defining $\overline{\kappa}_{jk}$ to be the symmetric contributions with respect to exchange of $j$, $k$ indices, whose derivatives with respect to $\rho_n$ are also symmetric with respect to permuting the $j$, $k$, and $n$ indices and $\overline{\pi}_{jk}^{{\rm A'}}$ as all other contributions, we have:
\begin{subequations}
    \begin{equation}
        \overline{u}_j^0 = \frac{\partial f^{\rm bulk}}{\partial \rho_j} + \overline{\tau}_j - \left( \overline{\lambda}_{j k n} + \frac{1}{2} \frac{\partial \overline{\kappa}_{j k}}{\partial \rho_n}\right) \partial_{z} \rho_k \partial_z \rho_n - \left( \overline{\kappa}_{jk} + \overline{\pi}_{jk}^{{\rm A'}} \right) \partial^2_{zz} \rho_k,
    \end{equation}
    where:
    \begin{equation}
        \overline{\lambda}_{j k n} \equiv \lambda_{j k n} + \frac{1}{2} \overline{\xi}_{jkn}^{{\rm S}_{kn}} - \frac{1}{2} \left[ \frac{\partial (\pi_{jk} + \omega_{jk})}{\partial \rho_n} \right]^{{\rm S}_{jkn}}.
    \end{equation}
\end{subequations}

Lastly, we look to integrate $\int d \rho_j \overline{\tau}_j$, however this requires $\partial \overline{\tau}_j / \partial \rho_k = \partial \overline{\tau}_k / \partial \rho_j \ \forall j \neq k$.
Performing an analogous decomposition $\overline{\tau}_j = \overline{\tau}_j^{{\rm I}} + \overline{\tau}_j^{{\rm N}}$ into integrable and nonintegrable contributions as was done for $\eta_{jk}$, we define $\overline{f}^{\rm bulk} \equiv f^{\rm bulk} + \int d \rho_j \overline{\tau}_j^{{\rm I}}$.
We then define the free energy functional $\overline{\mathcal{F}}$:
\begin{subequations}
    \label{eqSI:final_MAMB+}
    \begin{equation}
        \overline{\mathcal{F}} = \int_V d \mathbf{r} \left( \overline{f}^{\rm bulk} + \frac{\overline{\kappa}_{kn}}{2} \partial_{z} \rho_k \partial_z \rho_n \right),
    \end{equation}
    where its functional derivative is:
    \begin{equation}
        \frac{\delta \overline{\mathcal{F}}}{\delta \rho_j} = \frac{\partial \overline{f}^{\rm bulk}}{\partial \rho_j} - \frac{1}{2} \frac{\partial \overline{\kappa}_{jk}}{\partial \rho_n} \partial_{z} \rho_k \partial_z \rho_n - \overline{\kappa}_{jk} \partial^2_{zz} \rho_k.
    \end{equation}
    For the would-be species pseudopotentials we then obtain:
    \begin{equation}
        \overline{u}_j^0 = \frac{\delta \overline{\mathcal{F}}}{\delta \rho_j} + \overline{\tau}_j^{{\rm N}} - \overline{\lambda}_{j k n} \partial_{z} \rho_k \partial_z \rho_n - \overline{\pi}^{\rm A'}_{j k} \partial^2_{zz} \rho_k,
    \end{equation}
    which combined with the expression for the effective body forces:
    \begin{equation}
        b_i^{\rm eff} = \left[ \mathcal{T}_{i j}^{0} \right]^{-1} \bigg( \partial_z \overline{u}_j^0 + \eta_{j k}^{{\rm N}} \partial_z \rho_k - \overline{\theta}_{jknm} \partial_z \rho_k \partial_z \rho_n \partial_z \rho_m - \overline{\xi}_{jkn}^{{\rm A}_{kn}} \partial^2_{zz} \rho_k \partial_z \rho_n \bigg),
    \end{equation}
    is the most general phenomenological multicomponent mechanical balance one can write.
\end{subequations}
The non-variational terms then correspond to the $\frac{1}{2}(n_c^2 - n_c)$ unique components of $\overline{\pi}^{\rm A'}_{j k}$, the $\frac{1}{2}n_c(n_c^2 + n_c)$ unique components of $\overline{\lambda}_{j k n}$, and the $n_c-1$ unique components of $\overline{\tau}_j^{{\rm N}}$, where we again note that $\overline{\tau}_j^{{\rm I}}$ can always be defined such that one component of $\overline{\tau}_j^{{\rm N}}$ is zero.
Removing the overlines, this is the expression found in the main text with the passive transformation tensor, $\left[ \mathcal{T}_{i j}^{0} \right]^{-1} = -\rho_i \delta_{ij}$ (here, repeated indices are not summed).